\documentstyle[preprint,tighten,amssymb,eqsecnum,prd,aps]{revtex}
\makeatletter\def\endmathletters{\global\@ignoretrue}\makeatother
\def\scr{\cal}
\def\slashed{{/}\mskip-10.0mu}
\def\Tr{\mathop{\rm Tr}\nolimits}
\def\tr{\mathop{\rm tr}\nolimits}

\def\Im{\mathop{\rm Im}\nolimits}
\def\D{{\rm d}}
\def\CPN{{\rm CP}^{N-1}}
\def\intm#1{\int {\D^2 #1 \over (2\pi)^2} \,}
\def\intp{\intm{p}}
\def\intk{\intm{k}}
\def\intq{\intm{q}}
\def\goto{\mathop{\;\longrightarrow\;}}
\def\hatvphantom{\vphantom{\widehat{p{+}k}^2}}
\def\Ei{\mathop{\rm Ei}}
\def\arccot{\mathop{\rm arccot}}

\def\arsinh{\mathop{\rm arsinh}\nolimits}
\def\arctanh{\mathop{\rm arctanh}\nolimits}
\def\half{\case1/2}
\def\third{\case1/3}
\def\fourth{\case1/4}
\def\kappa{\varkappa}
\def\princint{\ \raise 0.5 ex\hbox to 0.7em%
{\leaders\hrule height 1pt\hfill}\kern-0.93em\int}
\def\ktaux#1{{\kern#1em\tilde{\kern-#1em\varkappa}}}
\def\kappatilde{%
{\mathchoice{\ktaux{0.2}}{\ktaux{0.2}}{\ktaux{0.14}}{\ktaux{0.1}}}}
\mathchardef\Gamma="0100
\mathchardef\Delta="0101
\mathchardef\Theta="0102
\mathchardef\Lambda="0103
\mathchardef\Xi="0104
\mathchardef\Pi="0105
\mathchardef\Sigma="0106
\mathchardef\Upsilon="0107
\mathchardef\Phi="0108
\mathchardef\Psi="0109
\mathchardef\Omega="010A
\makeatletter
\def\refname{{\normalsize\bf REFERENCES}}
\def\references{%
\ifpreprintsty
\newpage
\hbox to\hsize{\hss\refname\hss}%
\vskip 4truemm
\else
\vskip24pt
\hrule width\hsize\relax
\vskip 1.6cm
\fi
\list{\@biblabel{\arabic{enumiv}}}%
{\labelwidth\WidestRefLabelThusFar  \labelsep4pt %
\leftmargin\labelwidth %
\advance\leftmargin\labelsep %
\ifdim\baselinestretch pt>1 pt %
\parsep  4pt\relax %
\else %
\parsep  0pt\relax %
\fi
\itemsep\parsep %
\usecounter{enumiv}%
\let\p@enumiv\@empty
\def\theenumiv{\arabic{enumiv}}%
}%
\let\newblock\relax %
\sloppy\clubpenalty4000\widowpenalty4000
\sfcode`\.=1000\relax
\ifpreprintsty\else\small\fi
}

\def\thesection{\arabic{section}}
\def\appendix{\par\global\appendixontrue
\setcounter{section}{0}
\setcounter{subsection}{0}
\setcounter{subsubsection}{0}
\def\thesection{\Alph{section}}
\def\thesubsection{\arabic{subsection}}
\def\thesubsubsection{\alph{subsubsection}}
\def\theequation@prefix{\thesection}
\@addtoreset{equation}{section}
\addcontentsline{toc}{section}{\protect\numberline{}{APPENDIXES}}
}
\def\@Dottedtocline#1#2#3#4#5{\ifnum #1>\c@tocdepth \else
  \vskip 5pt plus 2pt
  {\leftskip #2\relax \rightskip \@tocrmarg \parfillskip -\rightskip
    \parindent #2\relax\@afterindenttrue
   \interlinepenalty\@M
   \leavevmode
   \@tempdima #3\relax \advance\leftskip \@tempdima \hbox{}\hskip -\leftskip
    #4\nobreak\leaders\hbox{$\m@th \mkern \@dotsep mu.\mkern \@dotsep
       mu$}\hfill \nobreak \hbox to\@pnumwidth{\hfil\rm #5}\par}\fi}
\def\l@section{\@Dottedtocline{1}{1.0em}{2.0em}}
\def\l@subsection{\@dottedtocline{2}{4.0em}{2.0em}}
\def\l@subsubsection{\@dottedtocline{3}{7.0em}{2.0em}}
\def\@listI{\leftmargin\leftmargini \itemsep\z@ \parsep\z@ \topsep\z@}
\def\@listi{\leftmargin\leftmargini \itemsep\z@ \parsep\z@ \topsep\z@}
\def\@listii{\leftmargin\leftmargini \itemsep\z@ \parsep\z@ \topsep\z@}
\def\@listiii{\leftmargin\leftmargini \itemsep\z@ \parsep\z@ \topsep\z@}
\def\@listiv{\leftmargin\leftmargini \itemsep\z@ \parsep\z@ \topsep\z@}
\newif\if@tbold
\newcommand\tbold{\@tboldtrue\bf}
\def\pbbox#1{\if@tbold\protect\bbox{#1}\else{#1}\fi}
\@tboldfalse
\def\section{\@mainheadtrue
\@startsection {section}{1}{\z@}{0.8cm plus1ex minus
 .2ex}{0.5cm plus1ex minus.2ex}{\reset@font\small\tbold\centering}}
\def\subsection{\@mainheadfalse
\@startsection{subsection}{2}{\z@}{0.8cm plus1ex minus
 .2ex}{0.5cm plus1ex minus.2ex}{\reset@font\small\tbold\centering}}
\def\subsubsection{\@mainheadfalse
\@startsection{subsubsection}{3}{\z@}{.8cm plus1ex minus
 .2ex}{0.5cm plus1ex minus.2ex}{\reset@font\small\it\centering}}
\def\paragraph{\@mainheadfalse
\@startsection{paragraph}{4}{\parindent}{\z@}{-1em}{\reset@font
\normalsize\it}}
\def\subparagraph{\@mainheadfalse
\@startsection{subparagraph}{4}{\parindent}{3.25ex plus1ex minus
 .2ex}{-1em}{\reset@font\normalsize\tbold}}
\def\pacs#1{\par %
\bgroup
\hsize\columnwidth \parindent0pt
\if@twocolumn\else\leftskip=0.10753\textwidth \rightskip\leftskip\fi
\ifdim\prevdepth=-1000pt \prevdepth0pt\fi
\dimen0=-\prevdepth \advance\dimen0 by20pt\nointerlineskip
\vbox to28pt{\small\vrule height\dimen0 width0pt\relax\ifdraft#1\fi\vfill}%
\egroup
}
\makeatother
\begin{document}
\draft

\preprint{IFUP-TH 14/93}

\title{The $\protect\bbox{1/N}$ expansion of two-dimensional spin models}
\author{Massimo Campostrini and Paolo Rossi}
\address{Dipartimento di Fisica dell'Universit\`a and I.N.F.N.,
I-56126 Pisa, Italy}
\maketitle

\begin{abstract}
\vskip -1 truecm
A short review of all available results (perturbative, nonperturbative, and
exact) on $d$-dimensional spin models is presented in order to introduce
the discussion of their $1/N$ expansion at $d=2$, where the models are
asymptotically free.

A general two-dimensional spin model with U$(N)$ invariance, interpolating
between $\CPN$ and ${\rm O}(2N)$ models, is studied in detail in order to
illustrate both the general features of the $1/N$ expansion on the lattice
and the specific techniques devised to extract scaling (field-theoretical)
behavior.

The continuum version of the model is carefully analyzed deriving
quantitative $O(1/N)$ physical predictions in order to establish a
benchmark for lattice computations.

The $1/N$ expansion on the lattice, including second-nearest-neighbor
interactions, is set up by constructing explicitly effective propagators
and vertices, and exhibiting a number of exact results and integral
representations that allow a substantial reduction of the numerical
effort.  The technique of asymptotic expansion of the lattice propagators,
basic to the derivation of analytical results in the scaling domain, is
presented in full detail and applied to the model.  Physical quantities,
like the free energy and different definitions of correlation length, are
evaluated.

The lattice renormalization-group trajectories are identified and
universality among different lattice (and continuum) schemes in the scaling
region is explicitly proven.  As a byproduct, representations of the
$O(1/N)$ contribution to the $\Lambda$-parameter ratios and to the lattice
$\beta$ functions are obtained.

A review of other developments based on the lattice $1/N$ expansion (finite
size scaling, higher orders, fermionic models) is presented.
\end{abstract}
\vskip 1 truecm
\begin{minipage}{\hsize}
\begin{center}
{\it To appear in ``Rivista del Nuovo Cimento''.}
\end{center}
\end{minipage}
\tableofcontents

\bigskip\bigskip

\begin{center}
\begin{minipage}{13 truecm}
{\it ... this (is) the statement of the research, lest human deeds should
in time fade, and great and wonderful works... become unknown...}
\begin{flushright}
[{\sc Herodotus}, Histories, 1, 1]
\end{flushright}
\end{minipage}
\end{center}

\section{Introduction}
\label{introduction}

A better qualitative and quantitative understanding of quantum field
theories requires an improvement of the analytical and numerical methods
of approximation. Lattice field theories are a natural ground of
application of large-scale numerical techniques. More efficient
algorithms and more powerful computing machines lead to an ever
increasing amount of numerical results. These results are however
affected by two major limitations:
\begin{enumerate}
\item
the quality of the information grows as the logarithm
(or, at best, as a small power) of the numerical effort;
\item
the lack of control on the systematic errors possibly
induced by some of the numerical techniques is often tantalizing.
\end{enumerate}

Both limitations are intrinsic to the field-theoretical nature of the
problem addressed: off-critical systems, and even some bulk properties of
critical systems, can usually be studied with great numerical precision
as well as with sensible analytical methods. Systems at criticality, and
the extraction of their scaling properties, however constitute a much
more difficult challenge.

In view of the abovementioned limitations, the parallel development of
more powerful analytical techniques, even with a limited domain of
applicability, is certainly very welcome. On one side it allows the
comparison between numerical an analytical results in a controlled
environment, and therefore it leads to a check of applicability for those
techniques and methods whose reliability cannot be taken for granted {\it
a priori}. On the other side it may improve our understanding of those
properties that cannot be directly tested with present-day numerical
methods and may strengthen some of the theoretical hypotheses that must
unavoidably be used in the field-theoretical interpretation of numerical
data. One can easily understand the need for such theoretical pieces of
evidence when one realizes that, notwithstanding all perturbative results
and the substantial agreement existing in the theoretical physics
community, there is at present no independent nonperturbative proof of
existence of any asymptotically-free quantum field theory, whose
relevance in the construction of models of the fundamental interactions
cannot be overstressed.

In this perspective, we think that the $1/N$ expansion
\cite{Coleman-aspects} (expansion in the number of field components) may be
a rather rewarding instrument of analysis not only in the context of
continuum field theories, where it has long been known as a major source of
nonperturbative information, but also in the case of lattice field
theories, where to the best of our knowledge the $1/N$ expansion is the
only approach leading to some theoretical evidence for the existence of a
continuum limit and of a nonvanishing scaling region, where the
field-theoretical properties of the models can in principle be explicitly
tested with predictable precision.

The conceptual foundations of the power of the $1/N$ expansion are
essentially the following:
\begin{enumerate}
\item
$N$ is an intrinsically adimensional parameter, representing a dependence
whose origin is basically group-theoretical, and leading to well-defined
field representation for all integer values, hence it is not subject to
any kind of renormalization;
\item
$N$ does not depend on any physical scale of the theory,
therefore we may expect it to play no r\^ole in the parametrization of
criticality. As a consequence there is no physical reason not to expect
reasonable convergence properties from an expansion in $1/N$, at least in
well-defined regions of the other physical parameters.
\end{enumerate}

Evidence for a finite radius of convergence of the $1/N$ expansion has been
produced in a number of instances, notably in the proposed exact
$S$-matrices for a number of two-dimensional bosonic and fermionic models.
More generally, the large order behavior of the coefficients of the $1/N$
expansion can be studied by applying inverse-scattering techniques to the
problem of finding instanton solutions of the effective actions
\cite{DeVega,Avan-DeVega-I,Avan-DeVega-II}.  In the case of
${\rm O}(N)$-symmetric $\left(\vec{\phi}^2\right)^2$ theories in less than
four dimensions, the $1/N$ perturbation series can be resummed by a Borel
transformation, and in the two-dimensional nonlinear $\sigma$ model one is
led to conjecture the convergence of the series also for the Green's
functions.

Till now the major domain of application of the $1/N$ expansion has been
in the evaluation of critical exponents for many different classes of
models in the range of dimensions comprised between 2 and 4 (lower and
upper critical dimensions). At the critical dimension the critical
exponents are trivial, but the logarithmic deviations from scaling and
the dynamical mass generation lead to the rich phenomenology
characteristic of asymptotic freedom. This phenomenology can be studied,
on the lattice as well as in the continuum, by applying the $1/N$
expansion in conjunction with proper modifications of the methods usually
adopted in standard renormalizable quantum field theories, notably the
techniques of regularization and renormalization of the physical
parameters. Due to the nonrenormalized character of $N$, the $1/N$
expansion leads to results whose renormalization-group invariance
properties are much more transparent than those of standard perturbation
theory. However, as far as we could check, the $1/N$ expansion commutes
with perturbation theory, and therefore it provides a direct
reinterpretation and an unambiguous resummation of the perturbative
results. Such phenomena as the effects of a change of regularization
scheme, the r\^ole of dimensional transmutation in the parametrization of
renormalization-group invariance, the relationship between dynamical mass
generation and Borel ambiguity in resummations, the interplay between
dimensional regularization, minimal subtraction, and
$\varepsilon$-expansion, the mixing effects and the subtraction of
perturbative tails in the evaluation of quantum expectation values of
composite operators, all find a specific and transparent illustration in
the context of the $1/N$ expansion already at the lowest nontrivial order
of computations
\cite{Campostrini-Rossi-dim,Campostrini-Rossi-condensates}.

The main purpose of the present review is to describe the results that may
be obtained by applying the $1/N$ expansion to the lattice versions of
two-dimensional spin models. Therefore we shall only briefly sketch the
main results presented in the (wide) literature on all the abovementioned
topics, focusing only on those continuum results that are essential in
order to introduce their lattice counterparts.

The subject of lattice $1/N$ expansion has not till now received a
systematic treatment: as a consequence many sections of the present paper
(especially Sect.~\ref{continuum 1/N} and
Sects.\ \ref{lattice 1/N}--\ref{Lambda parameter}) are essentially original
work by the authors.

In order to present a few peculiar techniques and results of the
application of the $1/N$ expansion to renormalizable lattice field
theories, we found it proper to focus on a specific class of models.  These
models should be simple enough to make all calculations as short and
understandable as possible, as well as to make accurate numerical
simulations feasible now or in the near future. However it was necessary
for the completeness of the presentation to deal with a sufficiently rich
phenomenology, notably nontrivial mass spectra, gauge and topological
properties, besides the obvious request of perturbative asymptotic freedom.
After much thought, we decided to study a two-parameter model of
two-dimensional spin fields with U$(N)$ global symmetry \cite{Samuel}.
This model interpolates between the standard ${\rm O}(2N)$
\cite{GellMann-Levy,Stanley-I,Stanley-II} and $\CPN$ models
\cite{Golo-Perelomov,Eichenherr}, and for special values of the parameters
it represents the gauge-fixed bosonized version of the minimal coupling of
massless fermions to $\CPN$ fields
\cite{Abdalla-Abdalla-Alves-Carneiro}.

The present paper is organized as follows:

In Sect.~\ref{renormalizability} we briefly discuss the general results
that have been obtained in the study of $d$-dimensional spin models, with
special emphasis on the topics related to perturbative and nonperturbative
renormalizability.

In Sect.~\ref{spin models} we introduce a class of $1/N$-expandable
two-dimensional spin models, and discuss the qualitative picture of their
properties that one may draw from a large-$N$ analysis.

In Sect.~\ref{exact results} we review a number of exact results,
especially factorized $S$-matrices, that apply to the models under
consideration for peculiar values of the parameters.

In Sect.~\ref{continuum 1/N} we discuss the $1/N$ expansion in the
continuum version of the models, introducing our regularization and
renormalization procedure, defining observables, and extracting some
quantitative $O(1/N)$ physical predictions.

Sect.~\ref{lattice formulation} is dedicated to the presentation of a
number of alternative lattice formulations and motivates our choice of a
lattice action, which will turn out to depend explicitly on an Abelian
vector field and on an extra parameter eventually allowing for a Symanzik
tree-level improvement of the action.

In Sect.~\ref{lattice 1/N} the basic ingredients of the $1/N$ expansion on
the lattice, effective action, propagators, and vertices, are introduced.

Sect.~\ref{integral representations} is devoted to a specific technical
problem, the search for integral representations of the effective lattice
propagators that, in the case of nearest-neighbor interactions, allows
a substantial simplification of the numerical tasks in the evaluation
of the effective Feynman diagrams.

In Sect.~\ref{asymptotic expansions} we introduce the basic ingredient
of all lattice computations in the scaling (field-theoretical) regime,
the asymptotic expansion of the lattice propagators for small values
of the (dynamically generated) mass gap, i.e.\ for values of the
correlation length much bigger than the lattice spacing (we are always
working in the infinite-volume limit).

In Sect.~\ref{asymptotic integral representation} special techniques for
the asymptotic expansion in the case of nearest-neighbor interactions
discussed in Sect.~\ref{integral representations} are presented.

Sect.~\ref{physical quantities} is dedicated to applying the
abovementioned results to the actual $O(1/N)$ evaluation of physical
quantities in the scaling region of our class of models. We discuss the
possible definitions of correlation length, show the universality of the
lattice results, and give a full analysis of the simplest correlation
function, the two-point correlator of the fundamental fields, including
the evaluation of the lattice wavefunction renormalization factor.

In Sect.~\ref{physical standard} the abovementioned lattice contributions
to physical quantities are explicitly evaluated in the case of
nearest-neighbor interactions.

Sect.~\ref{lattice topology} is devoted to the issue of topological
operators on the lattice; different definitions are analyzed and compared
in the context of the $1/N$ expansion of $\CPN$ models.

In Sect.~\ref{Lambda parameter} we rephrase our results in the
language of standard perturbation theory and perturbative renormalization
group. We discuss the evaluation of the ratio of $\Lambda$ parameters in
the context of the $1/N$ expansion and extract an explicit representation
of the $O(1/N)$ contributions to the lattice renormalization-group
$\beta$ function, clarifying some subtleties concerning the
noncommutativity of some limits at the border of the space of parameters,
which however does not affect the physical predictivity of the model.

In Sect.~\ref{finite size scaling} we review some results concerning the
possibility of performing a finite-size-scaling analysis of spin models by
the help of $1/N$-expansion techniques.

In Sect.~\ref{higher lattice 1/N} we analyze the attempts at extracting
physical predictions for models at low $N$ by computing higher orders of
the $1/N$ expansion on finite lattices.

In the same perspective, we discuss in Sect.~\ref{Schwinger-Dyson} an
alternative approach to $1/N$-expandable spin models based on truncated
Schwinger--Dyson equations.

Sect.~\ref{fermionic models} is devoted to a review of the results that can
be obtained by applying the methods discussed here to a wide class of
$1/N$-expandable fermionic lattice models.

Finally in Sect.~\ref{conclusions} we briefly discuss the relevance of our
results and draw our conclusions.

\section{Spin models in $\pbbox{\lowercase{d}}$ dimensions and
renormalizability}
\label{renormalizability}

The renormalization-group properties of two-dimensional spin models,
notably asymptotic freedom, are the foundations of our belief in the
existence of nontrivial renormalized quantum field theories describing the
critical behavior of these models in the coupling domain lying in the
neighborhood of the (trivial) critical coupling $g_c=0$ ($\beta_c=\infty$).
In the case of $1/N$-expandable spin models, the corresponding field
theories are nonlinear $\sigma$ models defined on symmetric spaces. These
in turn may also be thought as special limits of linear $\sigma$ models
(sometimes coupled to gauge fields) enjoying the proper group symmetries.

In order to understand properly the renormalization-group properties of
these models, it is certainly convenient to extend the analysis by treating
the physical dimension $d$ as a continuous parameter in the range $2\le
d\le4$.  It is now possible to compare the $\varepsilon=d-2$ expansion of
the nonlinear models with the $\varepsilon'=4-d$ expansion of the linear
models. These latter theories are known to be superrenormalizable, with an
ultraviolet-stable fixed point at the origin and an infrared-stable fixed
point at ``strong'' couplings when $\varepsilon'>0$. Since the infrared
limit of linear models has the same relevant operator content as the
ultraviolet limit of the nonlinear models, the latter must also be
renormalizable when $\varepsilon<2$ around their nontrivial ultraviolet
fixed point \cite{Brezin-ZinnJustin-breakdown,Amit-Ma-Zia}. As a
consequence, the critical exponents should take the same value when
computed in the linear and nonlinear models at the same dimensionality, and
in particular the critical exponents of the nonlinear $\sigma$ models
should become trivial at $d=4$. As we shall see, this phenomenon is
beautifully illustrated in the $1/N$ expansion.

In the two-dimensional limit $\varepsilon\to0$ however general theorems
\cite{Mermin-Wagner,Coleman} insure the impossibility of spontaneous
breaking of continuous symmetries.  Therefore there is no weak-coupling,
broken-symmetry phase, and $g_c=0$ is an ultraviolet fixed point, around
which logarithmic deviations from scaling are allowed (asymptotic freedom).
Since no massless modes can be present, nonperturbative mass generation
must occur.

This by now standard scenario has been the starting point for most
perturbative studies of nonlinear $\sigma$ models. After Polyakov's
pioneering paper \cite{Polyakov-goldstone}, perturbative ultraviolet
renormalizability was discussed by Brezin, Zinn-Justin, and Le Guillou
\cite{Brezin-ZinnJustin-breakdown,Brezin-ZinnJustin-renormalization,%
Brezin-ZinnJustin-LeGuillou} and by Bardeen, Lee, and Shrock
\cite{Bardeen-Lee-Shrock} for ${\rm O}(N)$ models, and by Valent
\cite{Valent} for $\CPN$ models.  The extension to more general symmetric
spaces was suggested by Eichenherr and Forger
\cite{Eichenherr-Forger,Eichenherr-Forger-II} and discussed by Pisarski
\cite{Pisarski}, by Duane \cite{Duane}, and by Brezin and coworkers
\cite{Brezin-Hikami-ZinnJustin}.  Quite naturally, one adopts dimensional
regularization and evaluates the renormalization-group $\beta$ and $\gamma$
functions in the minimal subtraction ($\overline{\rm MS}$) scheme. However,
a rigorous treatment of these models shows that generic Green's functions
are plagued in two dimensions by severe infrared divergences.  This problem
was first tackled by Jevicki \cite{Jevicki-ground-state} and by Elitzur
\cite{Elitzur}, who showed that two-dimensional Green's functions that are
fully invariant under the symmetry group of the model could be computed (in
low orders of perturbation theory) and found to be infrared-finite. This
property was exploited in
Refs.~\cite{McKane-Stone,Amit-Kotliar,Wolff-O3-mass} in the context of
dimensional regularization, and was given a rigorous proof to all orders of
perturbation theory by David
\cite{David-infrared,David-infrared-letter,David-global-constraint}.
Three-loop calculations for the renormalization-group functions were first
presented by Hikami and Brezin
\cite{Hikami-Brezin} for ${\rm O}(N)$ models, and by Hikami
\cite{Hikami-CPN} for $\CPN$ models. Extensions to more general symmetric
spaces was given in Refs.~\cite{Hikami-3loop,Hikami-sigma}.  Anomalous
dimensions were computed by Wegner and collaborators to three-loop order
\cite{Hof-Wegner}, and later to four-loop order \cite{Wegner-1,Wegner-2}.
The four-loop order $\beta$-function computation was completed in
Refs.~\cite{Bernreuther-Wegner,Wegner}.  It would be beyond the purposes of
the present review to give any detail of the abovementioned computations. A
collection of results is presented in Appendix~\ref{perturbative results}
for easy reference.

In the context of the perturbative approach, another
thoroughly-investigated issue is the classical thermodynamics of the models
at nonzero external magnetic field, which ensures the absence of the
infrared divergences discussed above.  Renormalization and scaling behavior
were discussed in Ref.~\cite{Brezin-ZinnJustin-breakdown} and reconsidered
by Jolicoeur and Niel \cite{Jolicoeur-Niel-mass,Jolicoeur-Niel-magnetic},
who exploited the scaling properties to devise an extrapolation method
allowing for nonperturbative predictions in the limit of vanishing magnetic
field.

When we turn to the approach based on the $1/N$ expansion, willing to
investigate the renormalizability properties of the models for $2\le
d\le4$, we must dramatically change our focus from a situation where the
parameter $\varepsilon$ can be considered infinitesimal and employed as an
ultraviolet regulator to the case where the physical dimensionality is a
fixed finite parameter. Nothing prevents us in principle from using some of
the dimensional regularization techniques, and this is one of the basic
ingredients in the study of $d$-dimensional models around criticality and
in the evaluation of critical exponents in the $1/N$ expansion. An
impressive series of results were obtained by Abe
\cite{Abe-II,Abe-eta-II,Abe-Hikami-scaling}, Brezin and Wallace
\cite{Brezin-Wallace}, and Ma \cite{Ma-exponents,Ma-largeN}, and more
recently with improved techniques by Vasilev and coworkers
\cite{Vasilev-Pismak-Khonkonen-indices,Vasilev-Pismak-Khonkonen-exponents,%
Vasilev-Pismak-Khonkonen-bootstrap,Vasilev-Nalimov-CPN}. In ${\rm O}(N)$
models, the critical exponent $\eta$ is by now known to $O(1/N^3)$, while
the critical exponent $\nu$ is known to $O(1/N^2)$; $O(1/N)$ results are
available for $\CPN$ models.  Following the procedure indicated in
Ref.~\cite{Hikami-Brezin}, these results are also the starting point for an
evaluation of the renormalization-group $\beta$ and $\gamma$ functions (in
dimensional regularization and minimal subtraction) at the same orders of
the $1/N$ expansion. All the results for critical exponents confirm the
abovementioned observations about universality between nonlinear and linear
models and triviality in $d=4$.

We want to stress that, as far as the present evidence
goes, the $\varepsilon$ and $1/N$ expansions appear to be strictly
commuting when applied to the evaluation of physical quantities, such as
the critical exponents.

The problem of renormalizability in the framework of the $1/N$ expansion
was studied by Symanzik \cite{Symanzik-ON,Symanzik-Pphi2} and by Arefeva
and collaborators. In
Refs.~\cite{Arefeva-I,Arefeva-II,Arefeva-III,Arefeva-IV,Arefeva-AP} the
ultraviolet renormalizability of the three-dimensional ${\rm O}(N)$ models
in both the symmetric and the broken-symmetry phase was shown to all orders
of $1/N$ by applying dimensional regularization.  The ultraviolet
renormalizability of $\CPN$ models when $d=2,3$ was shown in
Ref.~\cite{Arefeva-Azakov} by similar methods. However, in order to prove
the existence of a renormalized critical theory free of infrared
divergences, it was originally necessary to give up dimensional
regularization and attack the problem from the point of view of BPHZL
renormalization, which was done for three-dimensional ${\rm O}(N)$ models
in Ref.~\cite{Arefeva-Nissimov-Pacheva}. Subsequently the result was
generalized to all $2<d<4$ by the introduction of analytic regularization
\cite{Vasilev-Nalimov-dimension}.

A different view of renormalizability for asymptotically-free
$1/N$-expandable field theories has been put forward by Rim and Weisberger
\cite{Rim-Weisberger}. The essential, if subtle, equivalence of this point
of view with more standard dimensional regularization approaches has
however been exposed in Ref.~\cite{Campostrini-Rossi-dim}.

A very important issue in the context of the $1/N$ expansion of $\CPN$
models is the relevance of classical instanton configurations, that appear
to be nonperturbative in the expansion parameter $1/N$, and therefore might
in principle invalidate conclusions obtained in a purely perturbative
context. The problem was however solved by Jevicki
\cite{Jevicki-instantons}, who showed that, at the quantum effective-action
level, instantons, instead of being stationary points, appear in the form
of poles. One may then demonstrate that the $1/N$ expansion and the
semiclassical method correspond to two alternative contour integrations of
the functional integral. Further insight on the r\^ole of instantons was
obtained by David \cite{David}, who discussed also the problem of
summability of the instanton contributions, computed in
Refs.~\cite{Fateev-Frolov-Schwarz,Berg-Luscher-CPN}. The quantum statistics
of $\CPN$ models was studied by Affleck \cite{Affleck-I,Affleck-II} also in
connection with the topological properties ($\theta$-dependence) of the
models, and later analyzed and reviewed by Actor \cite{Actor}.  An
extension to $\CPN$ models coupled to fermions was discussed in
Ref.~\cite{Davis-Matheson-chiral}.  For sake of completeness, we also
mention that a different nonperturbative approach to $\CPN$ models, based
on the r\^ole of ``torons'' (classical solutions with fractional
topological charge) has been put forward in recent years by Zhitnitsky
\cite{Zhitnitsky-CPN} and found to agree with large-$N$ predictions.

A more general nonperturbative issue that may be addressed in the context
of the $1/N$ expansion is the existence and the r\^ole of infrared
renormalons \cite{Parisi-infrared}, appearing as singularities on the
positive real axis of the Borel transform in massless ultraviolet-free
theories, and related to the appearance of nonperturbative expectation
values. These in turn are the basic ingredients in the operator product
expansion approach advocated by Shifman, Vainshtein and Zakharov in order
to describe large-distance effects in asymptotically free theories
\cite{Shifman-Vainshtein-Zakharov}.

David showed that, in the context of the $1/N$ expansion of ${\rm O}(N)$
models, nonperturbative terms can be organized in an operator expansion,
but they have infrared renormalons \cite{David-renormalons}; these
renormalons cancel against the corresponding renormalons appearing in the
coefficients of the operator product expansion when Green's functions
(involving only zero-dimension operators) are computed. According to the
same author \cite{David-ambiguity,David-OPE}, only in well-definite
instances (e.g.\ the topological charge density, and other quantities with
a direct physical meaning) nonperturbative expectation values can be
defined unambiguously.  In any case, it is possible to show that, in each
order in $1/N$, the ${\rm O}(N)$ two-dimensional $S$-matrix amplitudes can
be written as series in powers of the dynamically-generated mass times a
convergent perturbative series \cite{Brunelli-Gomes}.  For the partially
different point of view supported by the ITEP group, one should see
Refs.~\cite{Novikov-Shifman-Vainshtein-Zakharov-sigma,%
Novikov-Shifman-Vainshtein-Zakharov-OPE}, where the issue of the operator
product expansion in the context of the $1/N$ expansion of ${\rm O}(N)$
models is also discussed.

The subject of the operator product expansion and renormalizability for
critical ${\rm O}(N)$ models in dimension $2<d<4$
(where nontrivial criticality exists) has been thoroughly
investigated in recent years by Lang and Ruhl
\cite{Lang-Ruhl-field,Lang-Ruhl-tensor,Lang-Ruhl-OPE,Lang-Ruhl-ancestor,%
Lang-Ruhl-quasiprimary,Lang-Ruhl-fusion}.

Finally we should mention that the possibility of including
nonperturbative effects directly into the perturbative expansion has been
explored by Davis and Nahm, who discussed both ${\rm O}(N)$ models
\cite{Davis-Nahm-sigma} and  $\CPN$ models \cite{Davis-Nahm-CPN}, showing
that proper normal-ordering may lead automatically to the inclusion of a
nonperturbative mass gap in the perturbative series (and confinement in
$\CPN$ models), and the result of this procedure commutes with the $1/N$
expansion \cite{Davis-Mayger}. In a related development
\cite{DelrioGatzelurrutia-Davis} the vacuum structure of the ${\rm O}(N)$
model is studied by a variational technique and agreement with conventional
large-$N$ results is found.

\section{$\pbbox{1/N}$-expandable two-dimensional spin models}
\label{spin models}

In order to achieve some generality, we shall investigate the
properties of a two-parameter class of $1/N$-expandable spin models,
described by the continuum action
\begin{equation}
S = N \int\D^2 x \left\{\beta_{\rm v}\,\partial_\mu\bar z\partial_\mu z +
\beta_{\rm g}\,\overline{D_\mu z}D_\mu z \right\},
\label{S-cont}
\end{equation}
where $z$ is an $N$-component complex field subject to the constraint
\begin{mathletters}
\label{constraints}
\begin{equation}
\bar z z = 1
\end{equation}
and a covariant derivative $D_\mu = \partial_\mu + i A_\mu$ has been
defined in terms of the composite gauge fields
\begin{equation}
A_\mu = \half i \left\{\bar z \partial_\mu z - \partial_\mu\bar z\,z
\right\} = i \bar z \partial_\mu z.
\end{equation}
\end{mathletters}
This action was introduced first by Samuel \cite{Samuel} and it is an
interpolating action between pure $\CPN$ models ($\beta_{\rm v}=0$) and
U$(N)$ vector models ($\beta_{\rm g}=0$), which in turn are nothing but
${\rm O}(2N)$ vector models.

We notice that pure $\CPN$ models enjoy a ${\rm U}(1)$ gauge
invariance related to the local transformations
\begin{mathletters}
\begin{eqnarray}
z(x) &\to& e^{i\lambda(x)} z(x), \\
\bar z(x) &\to& e^{-i\lambda(x)}\bar z(x), \\
A_\mu(x) &\to& A_\mu(x) - \partial_\mu\lambda(x).
\end{eqnarray}
\end{mathletters}
This invariance will play an important r\^ole in determining the structure
of the effective vertices.

We can introduce the (rescaled) weak coupling parameter $f$ by the
following change of variables:
\begin{mathletters}
\label{kappa-f}
\begin{eqnarray}
{1\over2f} &=& \beta_{\rm v} + \beta_{\rm g}, \\
\kappa f &=& {\beta_{\rm v}\over\beta_{\rm g}}\,.
\end{eqnarray}
\end{mathletters}
The coupling constant $\kappa$, as defined by Eqs.~(\ref{kappa-f}), enjoys
the property of non-re\-nor\-ma\-li\-za\-tion, i.e. the
renormalization-group trajectories in the $(\beta_{\rm g},\beta_{\rm v})$
plane (in the continuum version of the model) are just the curves of
constant $\kappa$ and correspond to physically different theories. This
property will emerge rather clearly from the discussion of the $1/N$
expansion. The renormalization-group trajectories are plotted in
Fig.~\ref{kappa-plot} for several values of $\varkappa$.

The $1/N$ expansion is achieved as usual by implementing the constraints
(\ref{constraints}) by Lagrange-multiplier fields $\alpha$ and
$\theta'_\mu$. The manipulations are quite standard and we obtain
\begin{eqnarray}
S &=& {N\over2f} \int\D^2x \left\{\partial_\mu\bar z\partial_\mu z
+ {1\over1+\kappa f}\left(\bar z\partial_\mu z\right)^2
+ i \alpha \left(\bar z z - 1\right) + {1\over1+\kappa f}
  \left(\theta'_\mu - i \bar z\partial_\mu z\right)^2 \right\}
\nonumber \\
&=& {N\over2f} \int\D^2x \left\{{\kappa f\over1+\kappa f}
  \partial_\mu\bar z\partial_\mu z + {1\over1+\kappa f}
  \left|\left(\partial_\mu+i\theta'_\mu\right)z\right|^2
+ i\alpha\left(\bar z z - 1\right)\right\}.
\end{eqnarray}
We can now perform the Gaussian integration over the $z$ fields and
obtain the effective action
\begin{equation}
S_{\rm eff} = N \Tr\ln\left\{
  {\kappa f\over1+\kappa f} \left(-\partial_\mu\partial_\mu\right)
+ {1\over1+\kappa f} \left(-\overline{D}_\mu D_\mu\right)
    + i \alpha \right\} + {N\over2f}(-i\alpha),
\end{equation}
where now $D_\mu = \partial_\mu + i\theta'_\mu$.  Finally, rescaling the
multiplier field $\theta'_\mu$ to $\theta_\mu = \theta'_\mu/(1+\kappa f)$
and introducing the vacuum expectation value of the $\alpha$ field,
$\alpha(x) = \left<\alpha\right> + \alpha_{\rm q}(x)$, $\left<\alpha\right>
= -i m_0^2$, we obtain the following form of the effective action:
\begin{equation}
S_{\rm eff} = N \Tr\ln\left\{
   -\partial_\mu\partial_\mu
   - i\left\{\partial_\mu,\theta_\mu\right\} + m_0^2
   + i\alpha_{\rm q}\right\}
+ {N\over2f}\left\{-m_0^2-i\alpha_{\rm q}
+ (1+\kappa f)\theta_\mu\theta_\mu\right\}.
\label{Seff-cont}
\end{equation}
In the large-$N$ limit, the value of $m_0^2$ is determined, as a function
of $f$ only, by the saddle-point condition (gap equation)
\begin{equation}
{1\over2f} = \intp {1\over p^2 + m_0^2}.
\label{gap-cont}
\end{equation}
Eq.~(\ref{gap-cont}) is in need of ultraviolet regularization; we
shall come to this point in Sect.~\ref{continuum 1/N}.

By taking the second functional derivative of the effective action around
the saddle point we may now obtain the propagators of the quantum
fluctuations associated with the fields $\alpha_{\rm q}$ and
$\theta_\mu$; both are $O(1/N)$ quantities that can be expressed by the
functions
\begin{mathletters}
\label{Delta-cont}
\begin{eqnarray}
\Delta^{-1}_{(\alpha)}(p) &=& \intq {1\over q^2 + m_0^2} \,
  {1\over (p+q)^2 + m_0^2} = {1\over2\pi p^2\xi}\ln{\xi+1\over\xi-1}
\,,\\
\Delta^{-1}_{(\theta)\,\mu\nu}(p) &=& \left(\kappa+{1\over f}\right)
  \delta_{\mu\nu} - \intq {(p_\mu+2q_\mu)(p_\nu+2q_\nu)\over
  [q^2 + m_0^2][(p+q)^2 + m_0^2]}  \nonumber \\
&=& \kappa\delta_{\mu\nu} + {1\over2\pi}
  \left(\xi\ln{\xi+1\over\xi-1} - 2\right)
  \left(\delta_{\mu\nu} - {p_\mu p_\nu\over p^2}\right),
\end{eqnarray}
\end{mathletters}
where $\xi = \sqrt{1+4 m_0^2/p^2}$. The effective propagators in $d$
dimensions are presented for reference in Appendix~\ref{d propagators}.

It is now crucial to observe that all the high-order effective vertices
resulting from Eq.~(\ref{Seff-cont}) by taking higher functional
derivatives are completely unaffected by the value of $\kappa$. As a
consequence, the vertices share the gauge properties enjoyed by the
$\CPN$ model, and in particular transversality. This fact in turn implies
the possibility of replacing the propagator (\ref{Delta-cont}b)
\begin{mathletters}
\label{Delta-theta}
\begin{equation}
\Delta_{(\theta)\,\mu\nu}(p) =\Delta_{(\theta)}(p)
\left(\delta_{\mu\nu} - {p_\mu p_\nu\over p^2}\right)
+ {1\over\kappa}\,{p_\mu p_\nu\over p^2}\,,
\end{equation}
where
\begin{equation}
\Delta^{-1}_{(\theta)}(p) = {1\over2\pi}
  \left(\xi\ln{\xi+1\over\xi-1} - 2 + 2\pi\kappa\right),
\end{equation}
\end{mathletters}
with its transverse part, when computing expectations of gauge-invariant
operators. Eq.~(\ref{Delta-theta}) shows that $\kappa$ is a ``physical''
parameter, related to the ratio of the mass of the Lagrangian field $z$
($m_0$ in the large-$N$ limit) to the mass $m_\theta$ of the propagating
field $\theta_\mu$. The exact relationship between the two masses is
expressed in the large-$N$ limit by the equation
\begin{equation}
\sqrt{{4m_0^2\over m_\theta^2} - 1}\;
\arccot\sqrt{{4m_0^2\over m_\theta^2} - 1}
= 1 - \pi\kappa, \qquad 0 < \kappa < {1\over\pi} \,.
\label{mu-cont}
\end{equation}
This is the basic physical reason why $\kappa$ is not subject to
renormalization. $m_\theta(\kappa)$ is plotted in Fig.~\ref{mu-plot}.

Let us now briefly describe the classes of different physical theories
parametrized by $\kappa$.
\begin{enumerate}
\item
When $\kappa=0$ we get $m_\theta=0$: $\theta_\mu$ becomes a dynamical gauge
field giving rise to a linear confining potential between $z$ and $\bar z$,
and the physical states are the bound states that are singlets under gauge
transformations. This is the well-known physical picture of the $\CPN$
models \cite{Luscher-secret,DAdda-Luscher-DiVecchia-1/N,Witten-CPN}.
\item
For very small values of $\kappa$ the above picture is substantially
unchanged: in the absence of a gauge symmetry the $z$ fields are not
automatically confined by Elitzur's theorem; however, their mass is so
much higher than that of their bound states to make them effectively
disappear from the physical spectrum.
\item
With growing $\kappa$ an inversion occurs and the $z$ fields become the
fundamental states of the model, while the mass of the bound states
becomes bigger and bigger, when measured in units of $m_0$.
\item
At $\kappa=1/\pi$ we meet a threshold: $m_\theta=2m_0$ and the Yukawa
potential that was the remnant of the linear confining potential
completely disappears. This is a quite interesting model: it is easy
to get convinced that the corresponding action is nothing but the
effective action resulting from the functional integration over a set
of $N$-component massless fermion fields minimally coupled to the
``gauge'' field $A_\mu$ (cf.\ e.g.\
Refs.~\cite{Abdalla-Abdalla-Alves-Carneiro,Abdalla-LimaSantos-CPN}).
This model has no quantum anomaly, and its factorized $S$-matrix is
therefore known: the physical states are in the fundamental
representation of U$(N)$ and the bound states have disappeared, as
expected.
\item
When $\kappa>1/\pi$, there are no bound states, and the models interpolate
smoothly from ${\rm U}(N)$ to ${\rm O}(2N)$ symmetry. In particular, for
integer values of $n=\pi\kappa$ the models describe the minimal
gauge-invariant coupling of $n$ ``flavors'' of massless fermions
\cite{DAdda-DiVecchia-Luscher}.  This picture gives further support to the
notion that $\kappa$ is a physical parameter not subject to
renormalization.
\item
Finally, when $\kappa\to\infty$ the effective field $\theta_\mu$ completely
decouples and we are left, as expected, with the well-known ${\rm O}(2N)$
nonlinear sigma model, possessing a factorized $S$-matrix for the
fundamental $2N$ real fields and showing absence of bound states.
\end{enumerate}

\section{Review of exact results}
\label{exact results}

In Sect.~\ref{spin models} we mentioned that, for special values of the
parameter $\kappa$, a number of exact results are available, especially
concerning exact factorized $S$-matrices and bound-state spectra.  We
must however keep in mind that these results have been obtained in a rather
indirect way, by applying such methods as analytic $S$-matrix theory or
Bethe Ansatz. As a consequence, the $1/N$ expansion offers the possibility
of verifying the applicability of the abovementioned methods to the models
at hand and therefore the correctness of the physical interpretation.  For
future reference and comparison, we would like to present here a short
review of these exact results.

We start from the observation that the key ingredient for the possibility
of constructing an exact $S$-matrix is the assumption of factorization of
multiparticle amplitudes into two-particle amplitudes, i.e.\ absence of
particle production \cite{Iagolnitzer}. This property is in turn related to
the existence of higher-order conservation laws.  Consider the standard
Noether current $j_\mu$ associated with the global symmetry, and notice
that the nonlocal charge \cite{Luscher-Pohlmeyer}
\begin{equation}
Q^{({\rm nl})} = \half\int\D x_1\D x_2 \,\varepsilon(x_1{-}x_2)
\left[j_0(t,x_1)\,j_0(t,x_2)\right] + \int\D x\,j_1(t,x),
\end{equation}
where $\varepsilon(x)$ is the sign function, is classically conserved owing
to the equations of motion.  At the quantum level one may show that this
conservation law is in general spoiled by anomalies generated by the
renormalization process. These anomalies are nonperturbative, but their
coefficients can be calculated in perturbation theory. A detailed analysis
shows that in ${\rm O}(2N)$ models the anomaly is actually absent
\cite{Luscher-charge}, while in the pure $\CPN$ case one finds
\cite{Abdalla-Abdalla-Gomes-I,Abdalla-Abdalla-Gomes-III}
\begin{equation}
{\D Q^{({\rm nl})}\over\D t} = {N\over\pi}\int\varepsilon^{\mu\nu}
\partial_\mu\theta_\nu(t,x)\,\D x.
\label{dQ-nl}
\end{equation}
However, when we consider the inclusion of minimally-coupled massless
fermions, we realize that the nonlocal charge is classically conserved only
if we include a contribution from the fermionic axial vector current. The
axial current in turn is known to possess a quantum anomaly, opposite in
sign to the r.h.s.\ of Eq.~(\ref{dQ-nl}). As a consequence, the modified
nonlocal charge has no net quantum anomaly and the corresponding model
turns out to possess a factorized $S$-matrix
\cite{Abdalla-Abdalla-Gomes-III,Abdalla-Abdalla-Gomes-II}. In our language,
these results imply the possibility of finding explicit $S$-matrices in the
cases $\kappa=\infty$ and $\kappa=1/\pi$ respectively.

Without belaboring on the techniques used in order to solve the
factorization equations, we only recall that the Fock space is decomposed
irreducibly into subspaces labeled by a definite particle number $n$.
States are identified by sets of particle momenta $\{P_r\}$, and
factorization is expressed by
\begin{equation}
\bigl<P'\;({\rm out})\bigm|P\;({\rm in})\bigr> =
\Bigl<P'\;({\rm in})\Bigm|\prod_{1\le r\le s\le n}S(P'_r,P_s)
\Bigm|P\;({\rm in})\Bigr>,
\end{equation}
where $S(P'_r,P_s)$ is the two-particle $S$-matrix.  $S$ is best expressed
in terms of rapidity variables $\theta_r=\arctanh(P_{r,1}/P_{r,0})$.
Imposing ${\rm O}(2N)$ symmetry \cite{Zamolodchikov-Zamolodchikov-ON} one
may assume
\begin{eqnarray}
\bigl<&&\theta'_1,i; \theta'_2,j\;({\rm out})\bigm|
\theta_1,k; \theta_2,l\;({\rm in})\bigr> \nonumber \\
= \delta&&(\theta_1-\theta'_1)\,\delta(\theta_2-\theta'_2)
 [\delta_{ij}\delta_{kl}\,S_1(\theta) + \delta_{ik}\delta_{jl}\,S_2(\theta)
+ \delta_{il}\delta_{jk}\,S_3(\theta)] \nonumber \\
+\,\delta&&(\theta_1-\theta'_2)\,\delta(\theta_2-\theta'_1)
[\delta_{ij}\delta_{kl}\,S_1(-\theta) + \delta_{il}\delta_{jk}\,S_2(-\theta)
+ \delta_{ik}\delta_{jl}\,S_3(-\theta)]
\end{eqnarray}
where $\theta = \theta_1-\theta_2$. Solving the constraints one finds
\begin{mathletters}
\begin{eqnarray}
S_3(\theta) &=& -{i\pi\over(N-1)\theta}\,S_2(\theta), \\
S_1(\theta) &=& -{i\pi\over(N-1)(i\pi-\theta)}\,S_2(\theta),
\end{eqnarray}
\end{mathletters}
and, assuming minimality in the number of poles and zeros in the physical
sheet,
\begin{equation}
S_2(\theta) = R(\theta)\,R(i\pi-\theta),
\end{equation}
where
\begin{equation}
R(\theta) = {\displaystyle
\Gamma\!\left({1\over2(N-1)} - {i\theta\over2\pi}\right)
\Gamma\!\left({1\over2} - {i\theta\over2\pi}\right) \over \displaystyle
\Gamma\!\left({1\over2} - {1\over2(N-1)} - {i\theta\over2\pi}\right)
\Gamma\!\left(- {i\theta\over2\pi}\right)} \,.
\end{equation}
The result shows no bound state poles \cite{Banks-Zaks}, and was checked in
a $1/N$ expansion up to second order \cite{Berg-Karowski-Kurak-Weisz}.

$\CPN$ models with minimally-coupled fermions in turn correspond to
${\rm SU}(N)$ symmetry, and the form of the factorized $S$-matrix is
\cite{Abdalla-LimaSantos-CPN,Berg-Karowski-Weisz-Kurak,Koberle-Kurak-CPN}
\begin{mathletters}
\begin{eqnarray}
\bigl<\theta'_1,i; \theta'_2,j\;({\rm out})\bigm|
\theta_1,k; \theta_2,l\;({\rm in})\bigr>
= \delta(\theta_1'-\theta_1)\,\delta(\theta'_2-\theta_2)
&&[\delta_{ik}\delta_{jl}\,u_1(\theta) + \delta_{jk}\delta_{il}\,u_2(\theta)]
\nonumber \\ -\, \delta(\theta'_1-\theta_2)\,\delta(\theta'_2-\theta_1)
&&[\delta_{jk}\delta_{il}\,u_1(\theta) + \delta_{ik}\delta_{jl}\,u_2(\theta)]
, \\ \bigl<\theta'_1,i; \bar\theta'_2,j\;({\rm out})\bigm|
\theta_1,k; \bar\theta_2,l\;({\rm in})\bigr>
= \delta(\theta_1'-\theta_1)\,\delta(\theta'_2-\theta_2)
&&[\delta_{ik}\delta_{jl}\,t_1(\theta) + \delta_{kl}\delta_{ij}\,t_2(\theta)]
\nonumber \\ +\, \delta(\theta_1'-\theta_2)\,\delta(\theta'_2-\theta_1)
&&[\delta_{ik}\delta_{jl}\,r_1(\theta) + \delta_{kl}\delta_{ij}\,r_2(\theta)],
\end{eqnarray}
\end{mathletters}
where the bar indicates antiparticles. The constraints imply
\begin{mathletters}
\begin{eqnarray}
u_2(\theta) &=& -{2i\pi\over N\theta}\,u_1(\theta), \\
r_1(\theta) &=& r_2(\theta) = 0, \\
t_2(\theta) &=& -{2i\pi\over N(i\pi-\theta)}\,t_1(\theta).
\end{eqnarray}
Moreover, the crossing symmetry requires
\begin{eqnarray}
t_1(\theta) &=& u_1(i\pi-\theta), \\
t_2(\theta) &=& u_2(i\pi-\theta).
\end{eqnarray}
\end{mathletters}
Finally, from minimality one obtains
\begin{equation}
t_1(\theta) = {\displaystyle
\Gamma\!\left({1\over2} - {i\theta\over2\pi}\right)
\Gamma\!\left({1\over2} + {1\over N} + {i\theta\over2\pi}\right)
\over \displaystyle \Gamma\!\left({1\over2} + {i\theta\over2\pi}\right)
\Gamma\!\left({1\over2} + {1\over N} - {i\theta\over2\pi}\right)} \,.
\end{equation}
There are no bound state poles: the bosons interact repulsively and the
fermions are screened by a ``secret'' long-range force, while the gauge
field loses the zero mass pole. This result was again checked in the $1/N$
expansion.

We must notice that these integrable models can sometimes be solved by the
Bethe Ansatz and quantum inverse-scattering methods, always reproducing the
abovementioned results.

For completeness we mention that the factorized $S$-matrix approach allows
in some special cases the determination of exact form factors. In
particular, in ${\rm O}(2N)$ models it is possible to evaluate the matrix
elements of the Noether current between the two-particle state and the
vacuum.  The result was checked in the $1/N$ expansion to $O(1/N)$ by the
use of the (explicitly known) spectral representation of the propagator
$\Delta_{(\alpha)}(p^2)$ \cite{Karowski-Weisz}. One can write
\begin{equation}
\Delta_{(\alpha)}(p^2) = 4\pi m_0^2\,{\sinh\psi\over\psi}\,,\qquad
\sinh^2\half\psi =  {p^2\over4m_0^2}\,.
\end{equation}
Therefore
\begin{eqnarray}
\Delta_{(\alpha)}(p^2) &=& 4\pi m_0^2\left[1
  + {p^2\over2\pi} \int_{4m_0^2}^\infty \D\mu^2
    {\rho_{(\alpha)}(\mu^2)\over\mu^2(p^2+\mu^2)}\right], \\
\rho_{(\alpha)}(\mu^2) &=& 2\pi\,{\sinh\phi\over\phi^2+\pi^2}\,,
\qquad \cosh^2\half\phi = {\mu^2\over4m_0^2}\,.
\nonumber
\end{eqnarray}
Likewise, when $\kappa=1/\pi$ we have
\begin{eqnarray}
\Delta_{(\theta)}(p^2) &=& 2\pi\,{\tanh\half\psi\over\psi}
  = \int_{4m_0^2}^\infty \D\mu^2\,{\rho_{(\theta)}(\mu^2)\over p^2+\mu^2}, \\
&&\rho_{(\theta)}(\mu^2) = 2\pi\,{\coth\half\phi\over\phi^2+\pi^2}\,.
\nonumber
\end{eqnarray}

A last very important exact result that can be obtained from the analysis
of the $S$-matrices for integrable models is the analytic determination of
the so-called mass -- $\Lambda$-parameter ratio, where the $\Lambda$
parameter is defined in standard perturbation theory ($\overline{\rm MS}$
scheme) in terms of the subtraction scale $\mu$ and the universal
coefficients $b_0$ and $b_1$ of the renormalization-group $\beta$ function:
\begin{equation}
\Lambda_{\overline{\rm MS}} = \mu\exp\Biggl[
-\int{\D g'\over\beta(g')}\Biggr] \cong
\mu(b_0g)^{-b_1/b_0^2}\exp\left[-{1\over b_0g}\right].
\end{equation}
The seminal result was obtained by Hasenfratz and collaborators in the case
of ${\rm O}(2N)$ models
\cite{Hasenfratz-Maggiore-Niedermayer,Hasenfratz-Niedermayer-ON}:
\begin{equation}
m = \left(8\over e\right)^{\textstyle{1\over2(N-1)}}\,
{1\over\displaystyle\Gamma\!\left(1+{1\over2(N-1)}\right)}\,
\label{Lambda-ON}
\Lambda_{\overline{\rm MS}}\,.
\end{equation}
The analysis can be repeated in the $\kappa=1/\pi$ model; the final result
is
\begin{equation}
m = \left(2\over e\right)^{\textstyle{1\over N}}\,
{1\over\displaystyle\Gamma\!\left(1+{1\over N}\right)}\,
\Lambda_{\overline{\rm MS}}\,.
\label{Lambda-1/pi}
\end{equation}
As we shall see in the next Section, these results can also be explicitly
verified in the context of the $1/N$ expansion.

\section{The $\pbbox{1/N}$ expansion in the continuum:
regularization and renormalization}
\label{continuum 1/N}

The Feynman rules for the $1/N$ expansion of the model can be easily
derived by introducing external currents coupled to the $z$ fields before
performing the functional integration. They are summarized in
Fig.~\ref{Feynman-rules-cont}.
The $1/N$ expansion is an expansion in the loops of the effective fields
$\alpha_{\rm q}$ and $\theta_\mu$.  In the graphical representation, a
closed loop of the $z$ fields stands for an effective vertex of the
expansion; effective vertices can be obtained by taking functional
derivatives of the effective action and carry a factor of $N$.

The effective vertices actually amount to one-loop integrals over the
fundamental field propagators. In two dimensions, all the effective
vertices may be in principle computed analytically, but the computation
may become very cumbersome in the case of exceptional configurations of
momenta, which are often those relevant to the actual computations one
would like to perform. For a discussion of this technical problem, cf.\
Ref.~\cite{Campostrini-Rossi-CPN}.

While no regularization is needed in the evaluation of the effective
vertices, it becomes unavoidable when one wants to compute the Green's
functions of the physical fields. In our choice of regularization we were
not guided by the usual requirements of Poincar\'e invariance, gauge
invariance, and consistency up to all perturbative orders that made
dimensional regularization a favorite tool of quantum field theory.
Having in mind our final purpose of performing explicit lattice
computations, we rather focused on the requests of computational ease and
transparency in the regularization mechanism, which is often obscured in
dimensional regularization by the interplay of ultraviolet and infrared
singularities.

Relaxing the consistency request down to one-loop consistency, we
found that the simplest and most transparent scheme was a kind of
sharp-momentum (SM) cutoff. The regularization procedure (roughly
formulated for ${\rm O}(N)$ models in Ref.~\cite{Orloff-Brout} and
discussed more precisely in Ref.~\cite{Campostrini-Rossi-dim}) starts from
the observation that, since the $\Delta$ propagators are finite, the
$O(1/N)$ (one-loop) Feynman integrals appearing in the computation of the
$n$-point Green's functions are integrals of regular functions. They can
therefore be regularized by subtracting explicitly the highest powers of
the integration variable appearing in the Taylor expansion of the
integrand. The lower limit for the integration of the subtraction terms is
arbitrary, and we are therefore introducing a dependence on the cutoff
value $M^2$. When the integrals are finite, this dependence disappears when
taking the limit $M^2\to\infty$.

Let us however consider the regularized gap equation
\begin{equation}
{1\over2f} = \intp\,{1\over p^2+m_0^2} - \int_{M^2}^\infty{\D p^2\over4\pi}
  \,{1\over p^2} = {1\over4\pi}\ln{M^2\over m_0^2}\,.
\label{ren-gap-cont}
\end{equation}
Eq.~(\ref{ren-gap-cont}) allows us to eliminate the dependence on $M^2$
of any superficially divergent diagram in favor of an explicit dependence
on the coupling constant, which in turn will be reabsorbed in the
renormalization-group-invariant definition of the physical mass and in
the wavefunction renormalization.

A first obvious analogy with the lattice formulation lies in the fact
that we are working with a ``bare'' coupling constant which can be varied
together with the cutoff while keeping all physical quantities constant.
As we shall show in Sect.~\ref{asymptotic expansions}, the analogy can be
made much more stringent by finding the relationship between the SM
cutoff and the lattice cutoff, which will also lead to a natural
regularization of the infrared singularities appearing in massless
lattice integrals.

It is important to observe that our regularization procedure differs by
$O(1/M^2)$ terms from a na\"\i ve sharp-momentum cutoff (i.e.\ simply
setting the upper integration limit to $M^2$).  This difference is
irrelevant in the continuum, but it will become crucial when discussing the
relationship with the lattice scheme.

The connection between SM regularization and dimensional regularization
(and renormalization) has been explored in some detail in the asymptotic
regime of large Euclidean momenta, where perturbation theory holds
\cite{Campostrini-Rossi-dim}. We shall not belabor on this point in the
present paper.

\subsection{The free energy}

SM regularization leads to a simple parametrization of the perturbative
tails that contribute to the nonscaling part of physical quantities. We can
therefore evaluate the scaling contributions to the free energy to $O(1/N)$
in our models, thus also offering a first explicit example of our
computational techniques. The free energy is the sum of the connected
vacuum diagrams of the effective theory.  In lowest orders only the trivial
Gaussian integrations over the $z$, $\alpha$ and $\theta_\mu$ fields do
contribute.  Subtracting the perturbative tail and keeping only the scaling
part, according to the rules of SM regularization, leads to the following
expression for the first two nontrivial contributions:
\begin{equation}
F = N \Tr\ln{p^2+m_0^2\over p^2} - {N\over2f}\,m_0^2
+ {1\over2}\Tr\ln{\Delta^{(0)}_{(\alpha)}(p)\over\Delta_{(\alpha)}(p)}
+ {1\over2}\Tr\ln{\Delta^{(0)}_{(\theta)}(p)\over\Delta_{(\theta)}(p)}
+ O\!\left(1\over N\right),
\label{F-cont}
\end{equation}
where
\begin{mathletters}
\label{Delta-0}
\begin{eqnarray}
\Delta_{(\alpha)}(p) &\displaystyle\goto_{p\to\infty}&
{2\pi p^2\over\ln(p^2/m_0^2)}
\equiv\Delta^{(0)}_{(\alpha)}(p), \\
\Delta_{(\theta)}(p) &\displaystyle\goto_{p\to\infty}&
{2\pi\over\ln(p^2/m_0^2) - 2 + 2\pi\kappa}
\equiv\Delta^{(0)}_{(\theta)}(p).
\end{eqnarray}
\end{mathletters}
According to our rules, the regularized expression of the free energy is
therefore
\begin{eqnarray}
F &=& N\,{m_0^2\over4\pi}
+ {1\over2} \intp \left[\ln\ln{\xi+1\over\xi-1}
- \ln\ln{p^2\over m_0^2}\right] - {1\over2} \int_{M^2}^\infty
  {\D p^2\over4\pi}\,{2m_0^2\over p^2}\,{1\over\ln(p^2/m_0^2)} \nonumber\\
&&\quad+\;{1\over2} \intp\left[\ln\!\left(\ln{\xi+1\over\xi-1} +
    {2\pi\kappa-2\over\xi}\right) - \ln\!\left(\ln{p^2\over m_0^2}
    + 2\pi\kappa - 2\right)\right] \nonumber\\
&&\quad-\;{1\over2} \int_{M^2}^\infty
    {\D p^2\over4\pi}\,{2m_0^2\over p^2}\,
    {3-2\pi\kappa\over\ln(p^2/m_0^2)+2\pi\kappa-2}
    + O\!\left(1\over N\right) \nonumber\\
&=& N\,{m_0^2\over4\pi} + {m_0^2\over 4\pi}\left[\ln\ln{M^2\over m_0^2}
+ (3-2\pi\kappa)\ln\!\left(\ln{M^2\over m_0^2}+ 2\pi\kappa - 2\right)
+ c_F(\kappa)\right] + O\!\left(1\over N\right), \nonumber\\
\label{F-reg-cont}
\end{eqnarray}
where $c_F(\kappa)$ is a numerical constant which can be computed to all
desired numerical accuracy; it is plotted in Fig.~\ref{c-Fm}.  Notable
special cases are
\begin{mathletters}
\begin{eqnarray}
&&c_F(0) \cong 1.18887122, \\
&&c_F\!\left(1\over\pi\right) = 2\gamma_{\rm E}.
\end{eqnarray}
\end{mathletters}
The issue of the evaluation of dimensionless SM-regulated one-loop
continuum integrals is discussed in Appendix~\ref{continuum integrals}.

$c_F(\kappa)$ may also be evaluated in the context of a $1/\kappa$
expansion of Eq.~(\ref{F-reg-cont}). The result is
\begin{equation}
c_F(\kappa) = (2\pi\kappa-3)\ln2\pi\kappa + \gamma_{\rm E} + \ln4 - 3
   + {5\over2\pi}\,{1\over\kappa} + O\!\left(1\over\kappa^2\right).
\label{cF-1k}
\end{equation}
The large-$\kappa$ limit of Eq.~(\ref{F-reg-cont}) is
\cite{Biscari-Campostrini-Rossi}
\begin{equation}
F \cong N\,{m_0^2\over4\pi}
+ {m_0^2\over 4\pi}\left[\ln\ln{M^2\over m_0^2}
+ \gamma_{\rm E} - \ln{M^2\over4m_0^2} - 1\right].
\end{equation}

Substituting Eq.~(\ref{ren-gap-cont}) in (\ref{F-reg-cont}) we obtain
\begin{equation}
F = N\,{m_0^2\over4\pi} + {m_0^2\over 4\pi}\left[\ln{2\pi\over f}
+ (3-2\pi\kappa)\ln\!\left({2\pi\over f} + 2\pi\kappa - 2\right)
+ c_F(\kappa) \right] + O\!\left(1\over N\right).
\label{F-ren-cont}
\end{equation}
Now, exploiting the renormalization-group invariance of the scaling part
of the free energy, we are ready to obtain the renormalization-group
$\beta$ function:
\begin{eqnarray}
\beta(f) = -2\left(\partial\ln F\over\partial f\right)^{-1} =
-{f^2\over\pi}\left[1 + {1\over N}\,{f\over2\pi}
\left(1 + {3-2\pi\kappa\over1+f(\kappa-1/\pi)}\right)
+ O\!\left(1\over N^2\right)\right].
\label{beta-cont}
\end{eqnarray}
It is easy to check that the known universal coefficients are correctly
reproduced. In particular, when $\kappa=1/\pi$
\begin{equation}
\beta(f) = -{f^2\over\pi}\left[1 + {1\over N}\,{f\over\pi}\right]
+ O\!\left(1\over N^2\right)
\end{equation}
and when $\kappa\to\infty$
\begin{equation}
\beta(f) = -{f^2\over\pi}\left(1 - {1\over N}\right)
\left[1 + {1\over N}\,{f\over2\pi}\right] + O\!\left(1\over N^2\right),
\end{equation}
as expected for ${\rm O}(2N)$ nonlinear sigma models.

\subsection{The two-point function: regularization}

We now focus our attention on the properties of the invariant
two-point correlation function of the fields in the fundamental
representation. Let us define
\begin{equation}
G(p) = {1\over2f}\int\D^2x\,e^{ipx}\left<\bar z(x) z(0)\right>
\equiv {1\over p^2+m_0^2} - {1\over N}\,{1\over p^2+m_0^2}
\Sigma_1(p)\,{1\over p^2+m_0^2}
+ O\!\left(1\over N^2\right)\,.
\label{G-cont}
\end{equation}
The $O(1/N)$ contributions to the two-point function are drawn in
Fig.~\ref{two-point}.  We obtain
\begin{eqnarray}
\Sigma_1(p) &=& \intk{\Delta_{(\alpha)}(k)\over(p+k)^2+m_0^2}
-\Delta_{(\alpha)}(0) \intk {\D^2q\over(2\pi)^2}\,{1\over(q^2+m_0^2)^2}\,
  {\Delta_{(\alpha)}(k)\over(q+k)^2+m_0^2} \nonumber \\
&-&\, \intk {\Delta_{(\theta)\,\mu\nu}(k)\over(p+k)^2+m_0^2}\,
  (2p_\mu+k_\mu)(2p_\nu+k_\nu) \nonumber \\
&+&\,\Delta_{(\alpha)}(0) \intk {\D^2q\over(2\pi)^2}\,
  {1\over(q^2+m_0^2)^2}\,\Delta_{(\theta)\,\mu\nu}(k)\,
  {(2q_\mu+k_\mu)(2q_\nu+k_\nu)\over(q+k)^2+m_0^2} \,.
\label{Sigma1-int}
\end{eqnarray}
By straightforward manipulations (essentially replacing zero-momentum
insertions of the $\alpha_{\rm q}$ field with derivatives with respect to
$m_0^2$), Eq.~(\ref{Sigma1-int}) can be recast into the form
\begin{eqnarray}
\Sigma_1(p) &=& \intk{\Delta_{(\alpha)}(k)\over(p+k)^2+m_0^2}
+ {1\over2}\Delta_{(\alpha)}(0) \intk\Delta_{(\alpha)}(k)\,
  {\partial\over\partial m_0^2}\,\Delta^{-1}_{(\alpha)}(k) \nonumber \\
&+&\, \intk\Delta_{(\theta)\,\mu\nu}(k) \left[\delta_{\mu\nu}
  - {(2p_\mu+k_\mu)(2p_\nu+k_\nu)\over(p+k)^2+m_0^2}\right] \nonumber \\
&+&\, {1\over2}\Delta_{(\alpha)}(0) \intk\Delta_{(\theta)\,\mu\nu}(k)
  \,{\partial\over\partial m_0^2}\,\Delta^{-1}_{(\theta)\,\mu\nu}(k).
\end{eqnarray}
Furthermore, explicit knowledge of the propagators allows us to make use
of the identities
\begin{mathletters}
\label{Delta-derivatives}
\begin{eqnarray}
{\partial\over\partial m_0^2}\,\Delta^{-1}_{(\alpha)}(p) &=&
-{2\over p^2\xi^2}\left[\Delta^{-1}_{(\alpha)}(p) +
  {1\over4\pi m_0^2}\right], \\
{\partial\over\partial m_0^2}\,\Delta^{-1}_{(\theta)}(p) &=&
{2\over p^2\xi^2}\left[\Delta^{-1}_{(\theta)}(p) - \kappa -
  {p^2\over4\pi m_0^2}\right].
\end{eqnarray}
\end{mathletters}
As a consequence, we obtain the representation
\begin{eqnarray}
\Sigma_1(p) &=& \intk{\Delta_{(\alpha)}(k)\over(p+k)^2+m_0^2}
- \intk {\Delta_{(\alpha)}(k)\over k^2+4 m_0^2} \nonumber \\
&+&\, \intk\Delta_{(\theta)}(k) \left\{1 - {4 p^2 k^2 - 4 (p \cdot k)^2
  \over k^2[(p+k)^2+m_0^2]}\right\} - \intk\Delta_{(\theta)}(k)\,
  {k^2 + 4\pi\kappa m_0^2 \over k^2 + 4m_0^2} \nonumber \\
&+&\, {1\over\kappa} \intk {p^2+m_0^2\over k^2}
  \left[1 - {p^2+m_0^2\over(p+k)^2+m_0^2}\right].
\label{Sigma1-cont}
\end{eqnarray}
The last term in Eq.~(\ref{Sigma1-cont}) reflects the dependence on the
longitudinal degrees of freedom of the field. Its contribution can be
computed in closed form (after regularization) and it amounts to
\begin{equation}
{1\over\kappa}\,(p^2+m_0^2)
\left[{1\over4\pi}\ln{M^2\over m_0^2}
	- {1\over2\pi}\ln{p^2+m_0^2\over m_0^2}\right].
\label{last-Sigma1-cont}
\end{equation}
Eq.~(\ref{last-Sigma1-cont}) is singular in the $\kappa\to0$ limit, when
the model becomes gauge-invariant and the contribution of the
longitudinal degrees of freedom becomes gauge-dependent.

The regularized version of Eq.~(\ref{Sigma1-cont}) is obtained by
applying the SM scheme prescriptions:
\begin{eqnarray}
\Sigma_1^{\rm reg}(p) &=& \Sigma_1(p)
   - \int_{M^2}^\infty {\D k^2\over4\pi}
\left[{2\pi\over\ln(k^2/m_0^2)} + {2\pi(3 - 2\pi\kappa) \over
    \ln(k^2/m_0^2) + 2\pi\kappa - 2}\right] {2m_0^2\over k^2}
\nonumber \\
&-&\, \int_{M^2}^\infty {\D k^2\over4\pi}\left[{2\pi\over\ln(k^2/m_0^2)}
  - {2\pi\cdot 2 \over \ln(k^2/m_0^2) + 2\pi\kappa - 2} + {1\over\kappa}
  \right] {p^2+m_0^2\over k^2} \,.\quad
\label{Sigma1-cont-reg}
\end{eqnarray}
Eq.~(\ref{Sigma1-cont-reg}) implies the possibility of parametrizing
$\Sigma_1^{\rm reg}(p)$ in the form
\begin{eqnarray}
\Sigma_1^{\rm reg}(p) &=& \Sigma_1^{\rm fin}(p) + m_0^2\left[
\ln\ln{M^2\over m_0^2} + (3-2\pi\kappa)\ln\!\left(\ln{M^2\over m_0^2}
  + 2\pi\kappa - 2\right) + c_m(\kappa)\right] \nonumber \\
&+&\, (p^2+m_0^2)\left[{1\over2} \ln\ln{M^2\over m_0^2}
  - \ln\!\left(\ln{M^2\over m_0^2} + 2\pi\kappa - 2\right)
  + {1\over4\pi\kappa}\ln{M^2\over m_0^2}\right],
\nonumber \\
\label{Sigma1-cont-fin}
\end{eqnarray}
where $\Sigma_1^{\rm fin}(p)$ is a regular, $M$-independent function of
$p^2$ (and $\kappa$) subject to the normalization condition
\begin{equation}
\Sigma_1^{\rm fin}(i m_0) = 0,
\label{Sigma1-norm}
\end{equation}
and $c_m(\kappa)$ is a numerically computable constant.

\subsection{Mass and wavefunction renormalization}

The interpretation of Eqs.~(\ref{Sigma1-cont-fin}) and
(\ref{Sigma1-norm}) becomes straightforward in the context of
renormalization, when we identify $G(p)$ with the bare two-point function
and write
\begin{equation}
{1\over p^2+m_0^2 + \displaystyle {1\over N}\,\Sigma_1^{\rm reg}(p)} =
{Z\over p^2+m^2 + \displaystyle {1\over N}\,\Sigma_1^{\rm fin}(p)}\,.
\end{equation}
Eq.~(\ref{Sigma1-norm}) is then the on-shell renormalization condition
and
\begin{mathletters}
\label{m-Z-cont}
\begin{eqnarray}
m^2 &=& m_0^2 + {1\over N}\,m_0^2\left[\ln{2\pi\over f} +
  (3-2\pi\kappa)\ln\!\left({2\pi\over f} + 2\pi\kappa - 2\right)
  + c_m(\kappa)\right], \\
Z &=& 1 - {1\over N}\left[{1\over2} \ln{2\pi\over f}
  - \ln\!\left({2\pi\over f} + 2\pi\kappa - 2\right)
  + {1\over2\kappa f}\right].
\end{eqnarray}
\end{mathletters}
In particular Eq.~(\ref{Sigma1-norm}) insures that the mass gap $m^2$
is identified as the pole of the two-point function and its $O(1/N)$
contribution $m_1^2$ satisfies the condition
\begin{equation}
m_1^2 = \Sigma_1^{\rm reg}(i m_0).
\label{m1-def}
\end{equation}
Applying Eq.~(\ref{m1-def}) directly to Eq.~(\ref{Sigma1-cont-reg}) we
obtain the representation
\begin{eqnarray}
m_1^2 &=& \intp\left\{\Delta_{(\alpha)}(p)\,{\xi-1\over p^2\xi^2} +
 \Delta_{(\theta)}(p)\left[(\xi-1)
  + (\pi\kappa-1) \left({1\over\xi^2}-1\right)\right]\right\}
\nonumber \\
&-&\,\int_{M^2}^\infty{\D p^2\over4\pi}\left[{2\pi\over\ln(p^2/m_0^2)}
  + {2\pi(3-2\pi\kappa)\over\ln(p^2/m_0^2)+2\pi\kappa-2}\right]
  {2m_0^2\over p^2}\,,
\label{m1-cont}
\end{eqnarray}
allowing a direct determination of $c_m(\kappa)$, which is plotted
in Fig.~\ref{c-Fm}.

In the $\kappa\to0$ limit, $c_m(\kappa)$ is an infrared-divergent
quantity, which shows that in the $\CPN$ models no single-particle mass
for the fundamental fields can be consistently defined
\cite{Campostrini-Rossi-CPN}. This is a reflection of the gauge
properties of the model and is further evidence for confinement. More
generally, it is pleasant to notice that $m_1^2$ does not depend on the
longitudinal degrees of freedom of the vector field, as a consequence of
gauge-invariance of the couplings and the on-shell condition. This allows
a direct physical interpretation of Eq.~(\ref{m1-cont}) also in those
instances (e.g.\ the $\kappa=1/\pi$ model) where we are dealing with a
gauge-fixed version of a theory enjoying gauge symmetry properties.
Actually we can compute exactly
\begin{equation}
c_m(1/\pi) = 2\gamma_{\rm E}.
\label{cm-1/pi}
\end{equation}
Eq.~(\ref{cm-1/pi}) can be shown to be consistent with the $1/N$ expansion
of the exact result (\ref{Lambda-1/pi}).  The $1/\kappa$ expansion of
Eq.~(\ref{m1-cont}) leads to
\begin{equation}
c_m(\kappa) = (2\pi\kappa-3)\ln2\pi\kappa + \gamma_{\rm E} + \ln4 - 2
   + {5\over2\pi}\,{1\over\kappa} + O\!\left(1\over\kappa^2\right).
\label{cm-1/k}
\end{equation}
The large-$\kappa$
limit of Eq.~(\ref{m1-cont}) is
\cite{Biscari-Campostrini-Rossi,Flyvbjerg-scaling,Flyvbjerg}
\begin{equation}
m_1^2 \goto_{\kappa\to\infty} m_0^2\left[\ln{2\pi\over f}
+ \gamma_{\rm E} - {2\pi\over f} + \ln4\right],
\end{equation}
consistently with the exact result (\ref{Lambda-ON}).  In the same limit
one may also show that
\begin{equation}
\lim_{p^2+m_0^2\to0^+}{\partial\Sigma_1(p)\over\partial p^2}
\goto_{\kappa\to\infty} {1\over2}
\left[\ln{2\pi\over f} + \gamma_E - \ln{\pi\over2} - 1\right],
\end{equation}
and the on-shell renormalization condition can be imposed on $Z$.

It is pleasant to notice that Eq.~(\ref{m-Z-cont}a) allows an
independent determination of the renormalization-group $\beta$ function
and the result is completely consistent with Eq.~(\ref{beta-cont}). As a
consequence the adimensional ratio
\begin{equation}
{F\over m^2} = {1\over4\pi}\left[N + c_F(\kappa) - c_m(\kappa)\right]
+ O\!\left(1\over N\right)
\label{F/m-cont}
\end{equation}
is universal and scheme-independent.
It is interesting to notice that one obtains from Eqs.~(\ref{cF-1k})  and
(\ref{cm-1/k}) the relationship
\begin{equation}
c_F(\kappa) - c_m(\kappa) = -1 + O\!\left(1\over\kappa^2\right).
\end{equation}

An alternative renormalization-group-invariant definition of the
correlation length can be defined starting from the second moment of the
two-point correlation function
\begin{equation}
\left<x^2\right> = {\int\D^2 x \,\fourth x^2\left<\bar z(x)z(0)\right>
\over \int\D^2 x\left<\bar z(x)z(0)\right>}\,.
\label{second-moment}
\end{equation}
In momentum space this definition leads to the relationship
\begin{equation}
m^2_R \equiv {1\over\left<x^2\right>} \cong
{m_0^2 + \displaystyle{1\over N}\,\Sigma_1(0) \over
 1 + \displaystyle{1\over N}\,\Sigma'_1(0)}
\cong m_0^2 + {1\over N}\left(\Sigma_1(0) - m_0^2 \Sigma'_1(0)\right),
\label{m-R-Sigma}
\end{equation}
where
\begin{equation}
\Sigma'_1(p) = {\partial\Sigma_1(p)\over\partial p^2}\,.
\end{equation}

Substituting Eq.~(\ref{Sigma1-cont-fin}) in Eq.~(\ref{m-R-Sigma}) and
comparing with Eq.~(\ref{m-Z-cont}a), we can easily show that
\begin{equation}
m^2_R \equiv m^2 + \delta m^2_R =
m^2 + {1\over N}\left(\Sigma_1^{\rm fin}(0) - m_0^2
\Sigma^{\prime\,\rm fin}_1(0)\right) + O\!\left(1\over N^2\right).
\label{m-R-Sigma-fin}
\end{equation}
Keeping also in mind Eq.~(\ref{Sigma1-norm}), we come to the conclusion
that $\delta m^2_R$ is amenable to a (typically small) calculable constant,
which can be interpreted as a universal scheme-independent adimensional
ratio.  The $O(1/N)$ contribution to $\delta m^2_R$ is plotted as a
function of $\kappa$ in Fig.~\ref{mR}.

Eq.~(\ref{m-R-Sigma}) becomes singular in the $\kappa\to0$ limit, when
only gauge-invariant correlations can be sensibly defined. The strategy
for such computations is discussed in Ref.~\cite{Campostrini-Rossi-CPN}
and is not especially relevant to our present analysis. We only mention
that gauge invariance is obtained at the price of introducing ``strings''
connecting the $z$ fields, and defining
\begin{equation}
G_{\scr C}(x,y) = {1\over2f} \left<\bar z(x)
\exp\left\{i\int_x^0\D t_\mu \theta_\mu(t)\right\}z(0)\right>.
\end{equation}
String renormalization is required. It is then possible to define
$\left<x^2\right>_{\scr C}$ in analogy with Eq.~(\ref{second-moment}).

We stress that the gauge dependence of the definition
(\ref{second-moment}) makes the observable $m_R^2$ dependent on the
longitudinal degrees of freedom of the vector field. Therefore it has no
direct physical interpretation in those special cases when we want to
recover the gauge-invariant properties of a gauge-fixed model.  Similar
considerations hold for the so-called magnetic susceptibility
\begin{equation}
\chi = \int\D^2 x \left<\bar z(x)z(0)\right> = 2f G(0) \cong
2f\left({1\over m_0^2} - {1\over N}\,{\Sigma_1(0)\over m_0^4}\right).
\label{mag-susc}
\end{equation}
The value of $\Sigma_1^{\rm reg}(0)$ can be computed analytically for
special values of $\kappa$; ignoring the contributions of the longitudinal
degrees of freedom we have (cf.\ Ref.~\cite{Biscari-Campostrini-Rossi})
\begin{mathletters}
\begin{eqnarray}
\left.\Sigma_1^{\rm reg}(0)\over m_0^2\right|_{\kappa=1/\pi}
&=& {3\over2}\left[\ln{2\pi\over f} + \gamma_E - c_1\right], \\
\left.\Sigma_1^{\rm reg}(0)\over m_0^2\right|_{\kappa=\infty}
&=& {3\over2}\left[\ln{2\pi\over f} + \gamma_E - c_1\right]
    -{2\pi\over f} + \ln 4,
\end{eqnarray}
\end{mathletters}
where
\begin{equation}
c_1 = \ln {\Gamma({1\over3})\,\Gamma({7\over6})
     \over \Gamma({2\over3})\,\Gamma({5\over6})}
\cong 0.4861007.
\label{c1}
\end{equation}

We can adopt a wavefunction renormalization condition slightly
different from Eq.~(\ref{m-Z-cont}b), defining the renormalization
constant $\tilde Z$ by
\begin{equation}
\tilde Z = {\chi m_R^2 \over 2f} = 1 - {1\over N} \Sigma_1'(0) +
O\!\left(1\over N^2\right).
\label{Z-tilde}
\end{equation}
It is easy to check that the ratio of the two definitions is a
$\beta$-independent constant:
\begin{equation}
{\tilde Z\over Z} =  1 - {1\over N}\,\Sigma^{\prime\,\rm fin}_1(0)
+ O\!\left(1\over N^2\right).
\end{equation}

Finally it is possible to compute the renormalization-group function
$\gamma(f)$, after its definition
\begin{equation}
\gamma(f) = -\beta(f)\,{\partial\over\partial f}\ln[2fZ(f)].
\label{gamma-SM}
\end{equation}
Ignoring the contributions from the longitudinal degrees of freedom of the
vector field, we obtain
\begin{equation}
\gamma(f) = {f\over\pi}\left(1-{1\over2N}\right)\left[1
+ {1\over N}\,{f\over\pi}\left(1+{1\over1+f(\kappa-1/\pi)}\right)\right]
+ O\!\left(1\over N^2\right),
\end{equation}
and, specifically for $O(2N)$ models,
\begin{equation}
\gamma(f) \approx {f\over\pi}\left(1-{1\over2N}\right)
\left[1 + {1\over N}\,{f\over2\pi}\right].
\end{equation}
Here and in the following, when checking agreement with the expressions
presented in Appendix~\ref{perturbative results}, one must perform an
appropriate change of variables, whose form may be extracted by comparing
SM and $\overline{\rm MS}$ $\beta$ functions.

\subsection{Correlations of composite operators}

Another very important class of correlation functions is obtained by
considering the Green's function of the (gauge-invariant) composite
operators
\begin{eqnarray}
P_{ij}(x) = \bar z_i(x)z_j(x).
\label{P-def}
\end{eqnarray}
We have shown in Ref.~\cite{Campostrini-Rossi-CPN} that, for $\CPN$
models ($\kappa=0$)
\begin{equation}
G_{ij,kl}(x-y) \equiv \left<P_{ij}(x) P_{kl}(y)\right>_{\rm conn}
= {B(x-y) \over N(N+1)}
\left(\delta_{il}\delta_{jk} - {1\over N}\,\delta_{ij}\delta_{kl}\right),
\end{equation}
where
\begin{equation}
B(x-y) = {N\left<\bar z_i(x)z_j(x) z_i(y)\bar z_j(y)\right> - 1 \over N-1}
= {4f^2\over N}\left(1 + {1\over N}\right)
  \left(\Delta_0^{-1} +\Delta_1^{-1}\right),
\end{equation}
and in turn $\Delta_0^{-1} = N\Delta_{(\alpha)}^{-1}$ and
$\Delta_1^{-1}$ is expressed to $O(1/N)$ by the sum of Feynman
diagrams drawn in Fig.~\ref{Delta1}.
More generally, $\Delta_1^{-1}$ is the sum of all the diagrams such that
the external $\alpha$ legs emerge from the same effective vertex
(generalized tadpole contributions).

This analysis applies almost literally to the more general case presented
here, and we can quote the final result which is just a very slight
generalization of Eq.~(8.20) in Ref.~\cite{Campostrini-Rossi-CPN}:
\begin{eqnarray}
\Delta_1^{-1}(p) &=& -\intk\Delta_{(\alpha)}(k)
\Biggl[V_4^{(a)}(p,k) + V_4^{(a)}(p,-k) + V_4^{(b)}(p,k)
+ {1\over k^2+4m_0^2}\,{\partial\over\partial m_0^2}
 \Delta_{(\alpha)}^{-1}(p)\Biggr] \nonumber \\
&-&\,\intk\Delta_{(\theta)}(k)
\Biggl\{(k^2+4m_0^2)\left[V_4^{(a)}(p,k) + V_4^{(a)}(p,-k)\right]
- 4\left[V_3(p,k) + V_3(p,-k)\right]
\nonumber \\ &&\qquad\qquad+\;(k^2+4m_0^2+2p^2)V_4^{(b)}(p,k)
+ {k^2+4\pi\kappa m_0^2\over k^2+4m_0^2}\,
{\partial\over\partial m_0^2}\Delta_{(\alpha)}^{-1}(p)\Biggr\}.
\label{Delta-1-comp}
\end{eqnarray}
The effective vertices entering Eq.~(\ref{Delta-1-comp}) are drawn in
Fig.~\ref{V3-V4}.
The formal definitions of $V_3$, $V_4^{(a)}$, and $V_4^{(b)}$, together
with their explicit expressions in terms of elementary functions, can be
found in Appendix~\ref{effective vertices}.  We stress that algebraic
manipulations, based on the gauge invariance of the effective vertices,
lead to the possibility of replacing all explicit dependence on the mixed
vector-scalar vertices appearing in Fig.~\ref{Delta1} with purely scalar
vertices.

Regularization and renormalization are straightforward along the lines
presented in Ref.~\cite{Campostrini-Rossi-CPN}. One may analyze the large
$k$ behavior of the effective vertices and find that the ultraviolet
divergence of Eq.~(\ref{Delta-1-comp}) is regulated by the SM counterterm
\begin{eqnarray}
&&\int_{M^2}^\infty {\D k^2\over4\pi}\,
   {\Delta_{(\alpha)}^{(0)}(k)\over k^4}
\left[4\Delta^{-1}_{(\alpha)}(p) - 2m_0^2\,{\partial\over\partial m_0^2}
\Delta^{-1}_{(\alpha)}(p)\right] \nonumber \\
-\;&&\int_{M^2}^\infty {\D k^2\over4\pi}\,
   {\Delta_{(\theta)}^{(0)}(k)\over k^2}
\left[4\Delta^{-1}_{(\alpha)}(p) + 2(3-2\pi\kappa)m_0^2\,
{\partial\over\partial m_0^2}\Delta^{-1}_{(\alpha)}(p)\right].
\label{Delta1-counter}
\end{eqnarray}
Eq.~(\ref{Delta1-counter}) shows that renormalized Green's functions are
obtained by mass and wavefunction renormalization, and in particular mass
renormalization is once more consistent with Eq.~(\ref{F-reg-cont}) and
Eq.~(\ref{m-Z-cont}a), while wavefunction renormalization is obtained by
defining
\begin{equation}
Z_P \cong 1 - {1\over N}\left[2\ln{2\pi\over f}
- 2 \ln\!\left({2\pi\over f} + 2\pi\kappa - 2\right) + c_Z(\kappa)\right].
\label{Z-P}
\end{equation}
We can extract from Eq.~(\ref{Z-P}) the anomalous dimension of the
composite field $P_{ij}$ in the SM regularization scheme:
\begin{equation}
\gamma_P(f) = -\beta(f)\,{\partial\over\partial f}
\ln\left[4f^2Z_P\right] = {2f\over\pi}\left[1 +
{1\over N}\,{f\over2\pi}\left(1 + {1\over1+f(\kappa-1/\pi)}
\right)\right] + O\!\left(1\over N^2\right).
\label{gamma-P}
\end{equation}
We notice that, in contrast with Eq.~(\ref{m-Z-cont}b) and consistently
with the gauge properties of $P_{ij}$, $Z_P$ is independent of the
longitudinal degrees of freedom of $\theta_\mu$.

A magnetic susceptibility and a second moment of the correlation function
can be defined for the field $P_{ij}$. These quantities can be shown to
satisfy all the renormalization-group requirements. In particular
\begin{equation}
\left<x^2\right>_P = {\int\D^2 x \,\fourth x^2
   \left<\tr\{P(x)P(0)\}\right>
\over \int\D^2 x\left<\tr\{P(x)P(0)\}\right>}
\label{second-moment-P}
\end{equation}
can be used as an alternative definition of the correlation length.
In the large-$N$ limit,
\begin{equation}
\left<x^2\right>_P = {1\over6m_0^2} + O\!\left(1\over N\right).
\end{equation}

In analogy with Eq.~(\ref{Z-tilde}), we can obtain a computationally
convenient definition of renormalization constant $\tilde Z_P$ by the
prescription
\begin{equation}
\tilde Z_P = {2\pi\over 3(2f)^2}\,{\chi_P\over\left<x^2\right>_P} =
{2\pi\over 3}\left.\left[{\partial\over\partial p^2}
   \left(\Delta_{(\alpha)}(p)
    - {1\over N}\,\Delta^2_{(\alpha)}(p)\Delta^{-1}_1(p)\right)
    \right]^{-1}\right|_{p^2=0} + O\!\left(1\over N^2\right),
\label{Z-P-tilde}
\end{equation}
where
\begin{equation}
\chi_P = \int\D^2 x \left<\tr\{P(x)P(0)\}\right>
\label{chi-P}
\end{equation}
and we fixed the normalization by noticing that
\begin{equation}
\left.\left[{\partial\over\partial p^2}\Delta_{(\alpha)}(p)
    \right]^{-1}\right|_{p^2=0} = {3\over2\pi}\,.
\end{equation}
In order to regularize Eq.~(\ref{Z-P-tilde}), we apply
Eq.~(\ref{Delta1-counter}) and recognize that the structure of the
counterterms of $\Delta^2_{(\alpha)}
\Delta^{-1}_1$ is simply
\begin{eqnarray}
&&\int_{M^2}^\infty {\D k^2\over4\pi}\left(4\,{\Delta_{(\alpha)}^{(0)}(k)
    \over k^4} - 4\,{\Delta_{(\theta)}^{(0)}(k)\over k^2}\right)
   \Delta_{(\alpha)}(p) \nonumber \\
+\;&&\int_{M^2}^\infty
  {\D k^2\over4\pi}\left(2\,{\Delta_{(\alpha)}^{(0)}(k)
  \over k^4} +2(3-2\pi\kappa){\Delta_{(\theta)}^{(0)}(k)\over k^2}\right)
    m_0^2\,{\partial\over\partial m_0^2}\Delta_{(\alpha)}(p).
\end{eqnarray}
Now, noticing that
\begin{equation}
\left.{\partial\over\partial p^2}
  \left[m_0^2\,{\partial\over\partial m_0^2}
   \Delta_{(\alpha)}(p)\right]\right|_{p^2=0} = 0,
\end{equation}
we immediately check that Eq.~(\ref{Z-P-tilde}) is consistent with the
parametrization (\ref{Z-P}); $c_Z(\kappa)$ can be computed numerically
and is plotted in Fig.~\ref{c-Z}. It is worth noticing that $c_Z(\kappa)$
is finite for $\kappa\to0$.

\subsection{Wilson loops and static potential}

Having exhausted the discussion of the correlations of fundamental fields
that may be relevant to a $O(1/N)$ analysis, we would like to consider
also the properties of (gauge-invariant) correlations of the vector field
$\theta_\mu$. These correlations are most naturally expressed in terms of
expectation values of the Wilson loops. The vector field being Abelian,
no path ordering is required and the general definition, for arbitrary
loops $\scr C$, is
\begin{eqnarray}
L({\scr C}) &=& \left<\exp\left\{i\oint_{\scr C}\D t_\mu\,\theta_\mu(t)
\right\}\right> \nonumber \\
&=& 1 - {1\over2N}\oint_{\scr C}\D t_\mu\oint_{\scr C}
   \D t'_\nu\intk e^{ik\cdot(t-t')}
\Delta_{(\theta)\,\mu\nu}(k) + O\!\left(1\over N^2\right).
\label{loop}
\end{eqnarray}
We defined $L({\scr C})$ having in mind the interpretation of our
Lagrangian as an effective theory for an underlying gauge-invariant model
where the gauge field is $\theta_\mu$, {\it not\/} the original vector
field $\theta'_\mu$.

For our purposes we shall only consider long rectangular loops of size $R
\times T$, and consider the limit $T\to\infty$. The quantity
\begin{eqnarray}
V(R) = -\lim_{T\to\infty}{1\over T}\ln L(R,T)
\label{V-cont}
\end{eqnarray}
can be interpreted as the interaction potential generated by vector
fields between two static sources. This quantity is relevant to the
discussion of the nonrelativistic bound-state spectrum, a reasonable
approximation in the large-$N$ limit.  One can easily show that $V(R)$
does not depend on the longitudinal components of $\theta_\mu$, and
within the $1/N$ expansion one may generalize the result of
Ref.~\cite{Campostrini-Rossi-CPN} to
\begin{equation}
N V(R) \cong -2\int_0^\infty \cos kR\;\Delta_{(\theta)}(k)\,{\D k\over2\pi} +
   2\int_0^\infty \left[\Delta_{(\theta)}(k)-\Delta_{(\theta)}^{(0)}(k)
   \right]\,{\D k\over2\pi}\,,
\end{equation}
where $\Delta_{(\theta)}^{(0)}(k)$ is defined in Eq.~(\ref{Delta-0}b); it
has been introduced in the context of SM regularization of the loop
ultraviolet singularities giving rise to the so-called ``perimeter
term''.  Rotating the integration contour in the complex $k$ plane to
$k=ix+\varepsilon$ we finally obtain the following representation of the
static potential, for $0\le\kappa\le1/\pi$:
\begin{eqnarray}
N\,{V(R)\over m_0} &\cong& 2\int_0^\infty \left[\Delta_{(\theta)}(k)
  -\Delta_{(\theta)}^{(0)}(k)\right]\,{\D k\over2\pi m_0}
-\pi\,{m_\theta\over m_0}\,{1-m_\theta^2/(4m_0^2) \over m_\theta^2/(4m_0^2)
  - \pi\kappa}\,e^{-m_\theta R} \nonumber \\
&-&\,\int_{2m_0}^\infty e^{-xR}\,{2\pi\xi'\over\left[\xi'
    \displaystyle\ln{1+\xi'\over1-\xi'} + 2\pi\kappa - 2\right]^2
  + \pi^2{\xi'}^2}\,{\D x\over m_0} \,,
\label{potential-cont}
\end{eqnarray}
where $\xi' = \sqrt{1-4 m_0^2/x^2}$ and we have defined the mass $m_\theta$
according to Eq.~(\ref{mu-cont}).
In Fig.~\ref{potential-fig} we have drawn the function $NV(R)/m_0$ for a
few different values of $\kappa$.

Eq.~(\ref{potential-cont}), considered as a function of $\kappa$,
interpolates between the linear confining potential at $\kappa=0$ and the
limiting case $\kappa=1/\pi$, when $\lim_{R\to\infty}V(R) = 0$ and the
bound state spectrum finally disappears, while $m_\theta=2m_0$.  The
qualitative discussion of the properties of the models as a function of
$\kappa$, presented in Sect.~\ref{spin models}, is essentially based upon
an analysis of this static potential, plus the information coming from
integrability at $\kappa=1/\pi$.

A more quantitative discussion of the bound state spectrum can be obtained,
for sufficiently large $N$, by considering the semiclassical approximation,
i.e.\ by solving the Schr\"odinger equation in the asymptotic potential
$V_{\rm as}(R)$. $V_{\rm as}(R)$ is obtained by removing from
Eq.~(\ref{potential-cont}) all the contributions not affecting leading
order predictions, and it can be represented in the form
\begin{equation}
V_{\rm as}(R,\kappa) = {6\pi m_0\over N}\,
A\!\left(m_\theta\over m_0\right)\left(1-e^{-m_\theta R}\right),
\label{V-as}
\end{equation}
where
\begin{equation}
A(x) = {x\over6}\,{1-\fourth x^2 \over
\fourth x^2 - 1 + \sqrt{4/x^2-1}\,\arccot\sqrt{4/x^2-1}}\,,
\label{A-def}
\end{equation}
enjoying the property
\begin{equation}
\lim_{x\to0} x A(x) = 1.
\end{equation}
The analysis of the resulting Schr\"odinger equation is presented in some
detail in Appendix~\ref{Schrodinger equation}. In the case $\kappa=0$
($\CPN$ models), the analysis of the Schr\"odinger equation in the linear
potential was presented in Refs.~\cite{Witten-CPN,%
Haber-Hinchliffe-Rabinovici} and in more detail in Ref.~\cite{Davis-Mayger}.

\subsection{Topological charge and susceptibility}

Before concluding this Section, we must mention that, in the special case
$\kappa=0$, the $1/N$ expansion can also be applied to the problem of
the so-called topological susceptibility
\begin{equation}
\chi_t = \int\D^2 x\left<q(x)\,q(0)\right>,
\label{chit}
\end{equation}
where
\begin{equation}
q(x) = {i\over2\pi}\,\varepsilon_{\mu\nu}\,\overline{D_\mu z}\,D_\nu z
     = {1\over2\pi}\,\varepsilon_{\mu\nu}\partial_\mu \theta_\nu
\label{qt}
\end{equation}
is the topological charge density.  One can easily show that
\begin{equation}
\chi_t = \lim_{p^2\to0}\,{1\over(2\pi)^2}\,p^2\tilde\Delta_{(\theta)}(p),
\label{chidef}
\end{equation}
where $\tilde\Delta_{(\theta)}(p)$ is the full propagator of
the quantum field $\theta_\mu$. $\chi_t$ is trivially zero when
$\kappa\ne0$.

In the large $N$ limit, Eq.~(\ref{Delta-theta}b) implies the simple
relationship \cite{Luscher-secret}
\begin{equation}
\chi_t = {3\over\pi N}\,m_0^2 + O\!\left(1\over N^2\right).
\end{equation}
The computation of the $1/N^2$ corrections to the inverse vector field
propagator was performed in Ref.~\cite{Campostrini-Rossi-CPN-topology}.
They can be obtained from the diagrams of Fig.~\ref{topology-diagrams}.
The vector and mixed scalar-vector vertices can be replaced by combination
of scalar vertices; the result is
\begin{eqnarray}
\Delta^{-1}_{1\,(\theta)}(p) &=& \intk\Delta_{(\alpha)}(k)\,W_1(k,p) -
\intk\Delta_{(\theta)}(k)\,W_2(k,p) \nonumber \\
&-&\,\intk{\Delta_{(\alpha)}(k) + k^2\Delta_{(\theta)}(k)
\over k^2+4m_0^2}\,{\partial\over\partial m_0^2}\Delta_{(\theta)}^{-1}(p),
\label{Delta1lf}
\end{eqnarray}
where
\begin{mathletters}
\begin{eqnarray}
W_1(k,p) &=& -(p^2+4m_0^2)[V_4^{(a)}(k,p) + V_4^{(a)}(k,-p)] \nonumber \\
&-&\,(p^2+2k^2+4m_0^2)V_4^{(b)}(k,p) + 4[V_3(k,p)+V_3(k,-p)], \\
W_2(k,p) &=& (p^2+4m_0^2)(k^2+4m_0^2)[V_4^{(a)}(k,p) + V_4^{(a)}(k,-p)]
\nonumber \\
&+&\,(p^2+2k^2+4m_0^2)(k^2+2p^2+4m_0^2)V_4^{(b)}(k,p) \nonumber \\
&-&\,4(p^2+k^2+4m_0^2)[V_3(k,p)+V_3(k,-p)] \nonumber \\
&-&\,2\,{k^2p^2\over(k\cdot p)^2} [\Delta_{(\alpha)}(k+p)\,Z_+^2(k,p)
    + \Delta_{(\alpha)}(k-p)\,Z_-^2(k,p)],
\label{W-12}
\end{eqnarray}
\end{mathletters}
and in turn
\begin{equation}
Z_\pm(k,p) = (p^2+k^2\pm 4m_0^2)V_3(\mp k,p) -\Delta_{(\alpha)}^{-1}(k) -
\Delta_{(\alpha)}^{-1}(p).
\end{equation}
A property of $W_2(k,p)$ relevant to the computation of higher-order
corrections to the slope of the linear static potential is
\begin{equation}
\lim_{\stackrel{\scriptstyle p\to0}{\scriptstyle k\to0}}
{W_2(k,p)\over k^2p^2} = {4\over45\pi}\,{1\over m_0^6}\,.
\label{W2-zero}
\end{equation}
Ultraviolet regularization in the SM scheme is straightforward and can be
proven to be consistent with the expected renormalizability properties of
the model: only mass and $z$-field wavefunction renormalization are
required.

The $1/N$-expandable, dimensionless ratio $R=\chi_t\left<x^2\right>_P$ is
computed; it takes the scheme-independent value
\begin{equation}
R = {1\over2\pi N}\left[1+{c_R\over N} + O\!\left(1\over N^2\right)\right],
\label{Rres}
\end{equation}
where $c_R \cong -0.380088$.

\section{Lattice formulation of two-dimensional spin models}
\label{lattice formulation}

It is well known that infinitely many lattice actions share the same
na{\"\i}ve continuum limit. However, from the point of view of numerical
simulations, the choice of a lattice action is a quite important topic,
because one is trying to optimize the speed of the computation and the
width of the scaling region. In the lattice $1/N$ expansion, which we are
discussing, another relevant criterion of choice is the possibility of
performing analytic calculations as far as possible, in order to keep under
control and possibly minimize the number of the altogether unavoidable
numerical integrations.

In a formal approach (and for one-coupling models), one might follow the
original suggestion by Stone \cite{Stone}, and adopt a lattice action
defined by the kernel of the heat equation on the manifold where the
fundamental degrees of freedom are defined. This choice has the advantage
of corresponding to the fixed-point (continuum limit) action in the
exactly-solvable one-dimensional case \cite{Rossi-Brihaye}. However for our
purposes it is definitely more convenient to consider actions that are
polynomial in the fields.

In passing we mention that a lattice Hamiltonian formulation was introduced
by Hamer, Kogut, and Susskind \cite{Hamer-Kogut-Susskind}, and studied in
the large-$N$ limit in Refs.~\cite{Srednicki,Banks,Guha-Sakita}.  $1/N$
correction have not however been evaluated.

We shall discuss a number of possibilities, and finally focus on those
formulations that meet the abovementioned criterion.

\subsection{Nearest-neighbor quadratic actions}

The lattice counterpart of Eq.~(\ref{S-cont}) satisfying the request that
only nearest-neighbor interactions be involved and no term higher than
quadratic in any given field be present is \cite{Samuel}
\begin{equation}
S^{(1)} = N \sum_{n,\mu} \left\{
\beta_{\rm v} \left[ \vphantom{\bar\lambda}
         2 - \bar z_{n+\mu} z_n - \bar z_n z_{n+\mu} \right] +
\beta_{\rm g} \left[ 2 - \bar z_{n+\mu} \lambda_{n,\mu} z_n -
        \bar z_n \bar \lambda_{n,\mu}z_{n+\mu} \right] \right\} ,
\label{latt-S1}
\end{equation}
where the $N$-component complex field $z_n$ satisfies the constraint
\begin{mathletters}
\label{constraints-latt}
\begin{equation}
\bar z_n z_n = 1
\end{equation}
and we have introduced explicitly in the action a ${\rm U}(1)$ gauge field
$\lambda_{n,\mu}$ satisfying
\begin{equation}
\bar \lambda_{n,\mu} \lambda_{n,\mu} = 1 \,.
\end{equation}
\end{mathletters}
This form of the action was introduced in the literature and its large $N$
features were discussed in detail in Ref.~\cite{Samuel}.  Recently the case
$\beta_{\rm v}=0$ ($\CPN$ models) of Eq.~(\ref{latt-S1}) has been the
starting point of many numerical simulations at small and intermediate
values of $N$ \cite{Campostrini-Rossi-Vicari-I,%
Campostrini-Rossi-Vicari-II,Irving-Michael-potential,Vicari}.

\subsection{Nearest-neighbor quartic actions}

It is possible to write down a gauge-invariant lattice action without
explicitly introducing a ${\rm U}(1)$ gauge field
\cite{Rabinovici-Samuel,DiVecchia-Holtkamp-Musto-Nicodemi-Pettorino}:
\begin{equation}
S^{(1)}_{\rm g4} = N \beta_{\rm g} \sum_{n,\mu}
\left[1- \left|\bar z_{n+\mu} z_n\right|^2\right].
\label{latt-S1-4g}
\end{equation}
This used to be a favorite version of lattice $\CPN$ models,
especially because of the property that the $N=2$ action is completely
equivalent to the popular standard lattice action of the ${\rm SU}(2)
\approx {\rm O}(3)$ nonlinear $\sigma$ model.  $S^{(1)}_{\rm g4}$ is
sometimes referred to as the ``adjoint'' form of lattice $\CPN$ models, and
is characterized by possessing a large-$N$ first-order phase transition,
that does not however show up for any finite value of $N$.

The mixed action obtained by combining $S^{(1)}$ (for $\beta_{\rm g}=0$)
and $S^{(1)}_{\rm g4}$ is also solvable in the large-$N$ limit
\cite{Samuel}.  More generally, the qualitative features of the phase
diagrams for mixed actions can be explored, for finite $N$ and in the $1/N$
expansion, by the mean-field technique
\cite{Duane-Green,DiVecchia-Musto-Nicodemi-Pettorino-Rossi,Ruhl-mean},
which is an increasingly accurate description of the lattice models when
one considers higher and higher space dimensionality.

However large-$N$ studies and numerical simulations have shown that the
approach to scaling of $S^{(1)}_{\rm g4}$ is very slow even for quite large
correlation lengths, and the situation is even worse for what concerns
asymptotic scaling, since the large-$N$ $\beta$ function gets contributions
to all loops, in contrast with the continuum and the action (\ref{latt-S1})
(cf.\ Ref.~\cite{DiVecchia-Musto-Nicodemi-Pettorino-Rossi}). We shall
therefore avoid any further effort concerning Eq.~(\ref{latt-S1-4g}) and
its $1/N$ expansion.

\subsection{Next-to-nearest-neighbor actions}

It is obviously possible to formulate the models in terms of
next-to-nearest-neighbor interactions:
\begin{eqnarray}
S^{(2)} = N \sum_{n,\mu} \Bigl\{
\fourth &&\beta_{\rm v} \left[ \vphantom{\bar\lambda}
         2 - \bar z_{n+2\mu} z_n - \bar z_n z_{n+2\mu} \right]
\nonumber \\ + \;
\fourth &&\beta_{\rm g} \left[ 2 -
    \bar z_{n+2\mu} \lambda_{n+\mu,\mu} \lambda_{n,\mu} z_n -
    \bar z_n \bar\lambda_{n,\mu} \bar\lambda_{n+\mu,\mu}z_{n+2\mu}
\right] \Bigr\}
\label{latt-S2}
\end{eqnarray}
and similarly
\begin{equation}
S^{(2)}_{\rm g4} = \fourth N \beta_{\rm g} \sum_{n,\mu}
\left[1 - \left|\bar z_{n+2\mu} z_n\right|^2\right].
\label{latt-S2-4g}
\end{equation}
More generally, any combination in the form
\begin{equation}
c_1S^{(1)} + c_2S^{(2)}, \qquad c_1+c_2 = 1, \quad c_1>0
\end{equation}
is a lattice action belonging to the same universality class. Amongst
these combinations, a special r\^ole is played by the choice
\begin{eqnarray}
S^{\rm Sym} = \case4/3 S^{(1)} - \third S^{(2)},
\label{latt-S-Sym}
\end{eqnarray}
corresponding to the so-called Symanzik tree-improved version of the model
\cite{Symanzik-sigma,Musto-Nicodemi-Pettorino}.  The choice $S^{\rm Sym}$
corresponds to forcing a short-distance (ultraviolet) behavior markedly
more similar to the behavior of the continuum action. As a byproduct, the
leading logarithmic dependence on the correlation length of the deviations
from scaling is removed and the scaling region is therefore enlarged. This
makes Eq.~(\ref{latt-S-Sym}) a natural candidate for numerical simulations
of the models.

In our discussion of the $1/N$ expansion, we shall try to carry the
analysis of the general case $c_1S^{(1)} + c_2S^{(2)}$ as far as
possible, and we shall concentrate on $S^{(1)}$ when its peculiar
analytic properties will become crucial to the development of our study.

\section{$\pbbox{1/N}$ expansion on the lattice:
effective action, propagators and vertices}
\label{lattice 1/N}

In the context of the $1/N$ expansion, the crucial property that selects
quadratic actions is the possibility of performing the exact Gaussian
integration over the $z$ fields at the only price of introducing a scalar
Lagrange multiplier for the constraint (\ref{constraints-latt}a).

Let us introduce the matrices
\begin{mathletters}
\label{matrices}
\begin{eqnarray}
{\scr M}^{(1)}_{mn} &=& \beta_{\rm v} \sum_\mu
\left( 2 \delta_{m,n} - \delta_{m+\mu,n} - \delta_{m,n+\mu} \right)
\nonumber \\  &+&\, \beta_{\rm g} \sum_\mu
\left( 2 \delta_{m,n} - \lambda_{m,\mu}\delta_{m+\mu,n} -
    \bar\lambda_{m,\mu}\delta_{m,n+\mu} \right), \\
{\scr M}^{(2)}_{mn} &=& \fourth \beta_{\rm v} \sum_\mu
\left( 2 \delta_{m,n} - \delta_{m+2\mu,n} - \delta_{m,n+2\mu} \right)
\nonumber \\ &+&\, \fourth \beta_{\rm g} \sum_\mu \left(
  2 \delta_{m,n} - \lambda_{m,\mu}\lambda_{m+\mu,\mu}\delta_{m+2\mu,n} -
  \bar\lambda_{m,\mu}\bar\lambda_{m-\mu,\mu}\delta_{m,n+2\mu} \right),
\end{eqnarray}
\end{mathletters}
and obtain the effective action in the form
\begin{equation}
S_{\rm eff} = N \Tr\ln \left[c_1 {\scr M}^{(1)}_{mn} +
c_2 {\scr M}^{(2)}_{mn} + {i\over2f}\,\alpha_n\delta_{m,n}\right]
+ {N\over2f}\,\sum_n(-i\alpha_n).
\label{S-eff-1}
\end{equation}
In order to perform the expansion, we must solve also the constraint
involving the $\lambda_{n,\mu}$ fields, which can be done by setting
\begin{equation}
\lambda_{n,\mu} = \exp\left(i\theta'_{n,\mu}\right),
\end{equation}
where $\theta'_{n,\mu}$ is an (unconstrained) real field essentially
playing the r\^ole of $\theta'_\mu$ in the continuum version.

Assuming translation invariance of the classical fields $\alpha_n$ and
$\theta'_{n,\mu}$, we can write down the effective action in momentum
space:
\begin{eqnarray}
S_{\rm eff} = N \intp \Biggl\{\ln\Biggl[{i\over2f}\,\alpha +
    c_1&&\left(\beta_{\rm v}\sum_\mu4\sin^2{p_\mu\over2} +
        \beta_{\rm g}\sum_\mu4\sin^2{p_\mu+\theta'_\mu\over2}\right)
\nonumber \\
+\; c_2&&\left(\beta_{\rm v}\sum_\mu\sin^2 p_\mu +
   \beta_{\rm g}\sum_\mu\sin^2 \left(p_\mu+\theta'_\mu\right)\right)\Biggr]
- {i\over2f}\,\alpha \Biggr\}.\nonumber \\
\label{S-eff}
\end{eqnarray}
A solution of the saddle-point equation is easily found to be
\begin{equation}
\theta'_\mu = 0 ,\qquad \alpha = -im_0^2,
\label{saddle-point}
\end{equation}
where $m_0^2$ is defined by the mass gap equation
\begin{equation}
{1\over2f} = \intp {1\over c_1 \sum_\mu4\sin^2 \half p_\mu
    + c_2 \sum_\mu\sin^2 p_\mu + m_0^2}
\equiv \intp {1\over {\bar p}^2 + m_0^2} \,;
\label{mass-gap}
\end{equation}
we adopted the standard notations
\begin{mathletters}
\label{hat-def}
\begin{eqnarray}
&\displaystyle
\hat p_\mu \equiv 2 \sin \half p_\mu, \qquad
\hat p^2 \equiv \sum_\mu \hat p_\mu^2, \qquad
\hat p^4 \equiv \sum_\mu \hat p_\mu^4, \\
&\displaystyle
{\bar p}^2 \equiv c_1\sum_\mu4\sin^2 \half p_\mu
   + c_2 \sum_\mu\sin^2 p_\mu
= \sum_\mu \hat p_\mu^2\left(c_1 + c_2 \cos^2 \half p_\mu\right)
= \hat p^2 - \fourth c_2\hat p^4, \\
&\displaystyle \intp \equiv
\int_{-\pi}^\pi{\D p_1\over 2\pi} \int_{-\pi}^\pi{\D p_2\over 2\pi} \,.
\end{eqnarray}
\end{mathletters}

Going back to coordinate space, taking the second functional derivatives
with respect to the fields $\alpha_n$ and $\theta'_{n,\mu}$, and
evaluating them at the saddle point defined by Eq.~(\ref{saddle-point}),
we obtain a representation of the lattice propagators of the effective
fields:
\begin{mathletters}
\label{propagator-alpha-theta}
\begin{eqnarray}
\Delta^{-1}_{(\alpha)}(k) &=& \intp
{1\over \overline{p{+}\half k}^2 + m_0^2}\,
{1\over \overline{p{-}\half k}^2 + m_0^2}\,, \\
\Delta^{-1}_{(\theta)\,\mu\nu}(k) &=& (1 + \kappa f)\,\delta_{\mu\nu}
    \intp {2c_1 \cos p_\mu + c_2 \cos 2p_\mu (1 + \cos k_\mu) \over
           {\bar p}^2 + m_0^2} \nonumber \\
&-&\,\intp \exp\left[\half i (k_\mu-k_\nu)\right] \nonumber \\
&&\quad\times \;
{\left(2c_1 \sin p_\mu + c_2 \sin 2p_\mu \cos \half k_\mu\right)
 \left(2c_1 \sin p_\nu + c_2 \sin 2p_\nu \cos \half k_\nu\right) \over
 \left(\overline{p{+}\half k}^2 + m_0^2\right)
 \left(\overline{p{-}\half k}^2 + m_0^2\right)} \,,
\end{eqnarray}
\end{mathletters}
where the $\theta'_\mu$ field has been rescaled to
$\theta_\mu = \theta'_\mu/(1 + \kappa f)$.

The effective vertices are obtained by taking higher order functional
derivatives of the lattice effective action. They are in the form of
one-loop integrals of the fundamental fields propagators. Computations
can be cumbersome, but they are conceptually straightforward. The Feynman
rules needed for the computations are summarized in
Fig.~\ref{Feynman-rules-latt}. We used the notation
\begin{mathletters}
\begin{eqnarray}
V_{3\mu}(p,k) &=&
2 c_1 \sin p_\mu + c_2 \sin 2p_\mu \cos \half k_\mu, \\
V_{4\mu}(p,k) &=&
c_1 \cos p_\mu + c_2 \cos 2p_\mu \cos^2 \half k_\mu.
\end{eqnarray}
\end{mathletters}
As discussed in Sect.~\ref{spin models},
the vertices are gauge-invariant for all values of $\kappa$. As a
consequence, the following relationships hold:
\begin{mathletters}
\label{gauge-lattice}
\begin{eqnarray}
\sum_\mu \hat k_\mu V_{3\mu}(p,k) &=&
\overline{p{+}\half k}^2 - \overline{p{-}\half k}^2, \\
2\hat k_\mu V_{4\mu}(p,k) &=&
V_{3\mu}(p{+}\half k, k) - V_{3\mu}(p{-}\half k, k).
\end{eqnarray}
\end{mathletters}

The notations we have introduced allow the following representation
of the effective vector propagator
\begin{eqnarray}
\Delta^{-1}_{(\theta)\,\mu\nu}(k) &=& 2(1 + \kappa f)\,\delta_{\mu\nu}
    \intp {V_{4\mu}(p,k) \over
           {\bar p}^2 + m_0^2} \nonumber \\
&-&\, \exp\left[\half i (k_\mu-k_\nu)\right] \intp
{ V_{3\mu}(p{+}\half k,k) V_{3\nu}(p{+}\half k,k)\over
 \left(\bar p^2 + m_0^2\right)
 \left(\overline{p{+}k}^2 + m_0^2\right)} \,.\quad
\label{effective-vector}
\end{eqnarray}
By applying Eqs.~(\ref{gauge-lattice}), we may check that
\begin{eqnarray}
\sum_\mu &&\exp\left(-\half i k_\mu\right) \hat k_\mu \left\{
\vphantom{1\over\left(\overline{p{+}k}^2 + m_0^2\right)}
2\,\delta_{\mu\nu} \intp {V_{4\mu}(p,k) \over {\bar p}^2 + m_0^2}
\right. \nonumber \\ &&\qquad-\;
 \left. \exp\left[\half i (k_\mu-k_\nu)\right] \intp
{ V_{3\mu}(p{+}\half k,k) V_{3\nu}(p{+}\half k,k)\over
 \left(\bar p^2 + m_0^2\right)
 \left(\overline{p{+}k}^2 + m_0^2\right)}\right\} = 0,
\end{eqnarray}
as an effect of gauge invariance: when $\kappa = 0$ the inverse
propagator $\Delta^{-1}_{(\theta)\,\mu\nu}(k)$ is transverse on the
lattice, i.e.\ it vanishes when contracted with $e^{-i k_\mu} - 1$ or
$e^{i k_\nu} - 1$. It is therefore convenient to restate
Eq.~(\ref{effective-vector}) in a form reminiscent of
Eq.~(\ref{Delta-cont}b):
\begin{equation}
\Delta^{-1}_{(\theta)\,\mu\nu}(k) \equiv
\Delta^{-1}_{(\theta)\,\mu}(k)\,\kappa\,\delta_{\mu\nu} +
\Delta^{-1}_{(\theta)}(k)\,\delta_{\mu\nu}^t(k),
\end{equation}
where
\begin{equation}
\delta_{\mu\nu}^t(k) \equiv \delta_{\mu\nu} -
\exp\left[\half i (k_\mu-k_\nu)\right]
{\hat k_\mu\hat k_\nu \over \hat k^2},
\end{equation}
\begin{mathletters}
\begin{eqnarray}
  &&\qquad{1\over2f}\,\Delta^{-1}_{(\theta)\,\mu}(k) =
\intp {V_{4\mu}(p,k) \over {\bar p}^2 + m_0^2}\,, \\
&&\Delta^{-1}_{(\theta)}(k) =
2 \sum_\mu\intp {V_{4\mu}(p,k) \over {\bar p}^2 + m_0^2} -
\sum_\mu\intp { \left[V_{3\mu}(p{+}\half k,k)\right]^2\over
 \left(\bar p^2 + m_0^2\right)
 \left(\overline{p{+}k}^2 + m_0^2\right)} \,, \qquad
\end{eqnarray}
\end{mathletters}
enjoying the important property
\begin{equation}
\Delta^{-1}_{(\theta)}(0) = 0.
\end{equation}
It is worth noticing that
\begin{equation}
\Delta^{-1}_{(\theta)\,\mu}(k) \goto_{c_2\to0} 1-{f\over2}\intp
{\hat p^2 \over \hat p^2 + m_0^2} =
1-{f\over2} + {m_0^2\over4}\,,
\end{equation}
independently of $k$ and $\mu$.  However even in this simplified case
one cannot trivially identify $\kappa$ with its continuum counterpart,
even in the scaling regime, and, for finite values of $f$,
$\kappa_{\rm latt}$ must be tuned in order to recover an assigned
value of $\kappa_{\rm cont}$.

\section{Integral representations of lattice propagators}
\label{integral representations}

We mentioned in Sect.~\ref{lattice formulation} that some powerful
analytical techniques can be applied in some specific version of the
model. In this Section we introduce the first of these techniques, the
use of integral representations of lattice propagators. We could only
apply this technique to the case $c_2 = 0$ (nearest-neighbor
interactions) which we shall discuss in great detail.

When $c_2 = 0$ dramatic simplifications occur already in
Eqs.~(\ref{propagator-alpha-theta}). By simple manipulations and use
of the gap equation
\begin{equation}
{1\over2f} = \intp {1\over \hat p^2 + m_0^2}
\label{nn-mass-gap}
\end{equation}
we obtain
\begin{mathletters}
\label{delta-def}
\begin{eqnarray}
\Delta^{-1}_{(\alpha)}(k) &\displaystyle\goto_{c_2\to0}&
\intp {1\over \hat p^2 + m_0^2 \hatvphantom}
   \,{1\over \widehat{p{+}k}^2 + m_0^2}\,, \\
\Delta^{-1}_{(\theta)}(k) &\displaystyle\goto_{c_2\to0}& \left[
{4 + m_0^2 \over 2f} - 1 - \intp {\widehat{2p{+}k}^2 \over
  \Bigl(\hat p^2 + m_0^2\Bigr) \Bigl(\widehat{p{+}k}^2 + m_0^2\Bigr)}
\right]\,.
\end{eqnarray}
\end{mathletters}

The first analytical result concerns Eq.~(\ref{nn-mass-gap}), that can be
cast into the form
\begin{equation}
{1\over2f} = {1\over2\pi}\,{1\over1+m_0^2/4}\,
K\!\left(1\over1+m_0^2/4\right),
\label{nn-mg-ellipt}
\end{equation}
where $K$ is the complete elliptic integral of the first kind.  It is also
possible to evaluate in closed form the inverse propagators along the
principal diagonal of the momentum lattice, i.e.\ when $k_1 = k_2
\equiv l$.  We change variables to $q_1 = \half(p_1+p_2)$ and
$q_2 = \half(p_1-p_2)$, and notice that, for any periodic function
$f(p_1,p_2)$ we have
\begin{equation}
\int_{-\pi}^\pi{\D p_1\over2\pi}\int_{-\pi}^\pi{\D p_2\over2\pi}\,f(p_1,p_2)
= \int_{-\pi}^\pi{\D q_1\over2\pi}\int_{-\pi}^\pi{\D q_2\over2\pi}
\,f(q_1{+}q_2,q_1{-}q_2).
\end{equation}
In the case $k_1 = k_2 = l$ in particular
\begin{eqnarray}
\hat p^2 &=& 4\bigl(1-\cos q_1\cos q_2\bigr), \\
\widehat{p{+}k}^2 &=& 4\bigl(1-\cos (q_1+l)\cos q_2\bigr).
\end{eqnarray}
Therefore the $q$-integrations of Eqs.~(\ref{delta-def}) can be performed
in terms of the standard complete elliptic integrals $K$, $E$, and $\Pi$
(we use the conventions of Ref.~\cite{Gradstein}); the result is
\begin{mathletters}
\label{Delta-diagonal}
\begin{eqnarray}
\Delta^{-1}_{(\alpha)}(l,l) &=&
{1\over8\pi\left(1+m_0^2/4\right)^2}\,
\Pi\!\left({\cos^2\half l\over\left(1+m_0^2/4\right)^2},
   {1\over1+m_0^2/4}\right), \\
\Delta^{-1}_{(\theta)}(l,l) &=&
-{1\over\pi}\,E\!\left({1\over1+m_0^2/4}\right) +
\left(1-{1\over\cos^2\half l}\right){1\over\pi}\,
K\!\left({1\over1+m_0^2/4}\right) \nonumber \\ &+&\,
\left({1\over\cos^2\half l} - {1\over\left(1+m_0^2/4\right)^2}\right)
{1\over\pi}\,\Pi\!\left({\cos^2\half l\over\left(1+m_0^2/4\right)^2},
   {1\over1+m_0^2/4}\right).
\end{eqnarray}
\end{mathletters}

In the general case, we had to resort to an integral representation of the
inverse propagators.  Let us make use of the standard Feynman parameters to
write
\begin{equation}
{1\over \hat p^2 + m_0^2 \hatvphantom}\,
{1\over \widehat{p{+}k}^2 + m_0^2} = \int_0^1\D x\,{1\over
\left[\hat p^2(1-x) + \widehat{p{+}k}^2 x + m_0^2\right]^2}
\label{Feynman-repr}
\end{equation}
and notice that, via trigonometric identities,
\begin{equation}
\bigl(\widehat{p_\mu{+}k_\mu}\bigr)^2 =
\hat p_\mu^2\left(1-\half \hat k_\mu^2\right)
+ \hat k_\mu^2 + 2 \sin p_\mu \sin k_\mu \,.
\end{equation}
Let us now change variables to $q_\mu = p_\mu + z_\mu$, where $z_\mu$ is
defined by the relationships
\begin{equation}
\sin z_\mu = {x \sin k_\mu \over \sqrt{1-x(1-x)\hat k_\mu^2}}\,, \quad
\cos z_\mu = {1 - \half x \hat k_\mu^2
    \over \sqrt{1-x(1-x)\hat k_\mu^2}}\,.
\label{zmu-def}
\end{equation}
Eq.~(\ref{zmu-def}) implies the identities
\begin{mathletters}
\label{q-identities}
\begin{eqnarray}
&&\hat p_\mu^2 (1-x) + \bigl(\widehat{p_\mu{+}k_\mu}\bigr)^2 x =
\hat q_\mu^2 \sqrt{1-x(1-x)\hat k_\mu^2} + 2
- 2 \sqrt{1-x(1-x)\hat k_\mu^2}\,, \\
&&\sin\left(p_\mu + \half k_\mu\right) =
{\sin q_\mu \cos \half k_\mu + (1-2x)\cos q_\mu \sin\half k_\mu
\over \sqrt{1-x(1-x)\hat k_\mu^2}}\,.
\end{eqnarray}
\end{mathletters}
Substituting Eqs.~(\ref{Feynman-repr}) and (\ref{q-identities}) in the
relevant one-loop integrals we obtain
\begin{mathletters}
\label{Delta-def}
\begin{eqnarray}
\Delta^{-1}_{(\alpha)} &=& \int_0^1\D x \intq
{1\over\left[4+m_0^2-\sum_\mu a_\mu\cos q_\mu\right]^2} \,, \\
\Delta^{-1}_{(\theta)} &=& {4+m_0^2\over2f} - 1 - \int_0^1\D x \intq
{c - \sum_\mu b_\mu \sin^2 q_\mu \over
   \left[4+m_0^2-\sum_\mu a_\mu\cos q_\mu\right]^2} \,,
\end{eqnarray}
\end{mathletters}
where in order to simplify the notation we have defined the auxiliary
variables
\begin{mathletters}
\label{ABC}
\begin{eqnarray}
a_\mu &=& 2 \sqrt{1-x(1-x)\hat k_\mu^2}\,, \\
b_\mu &=& {(1-2x)^2\hat k_\mu^2 + \hat k_\mu^2 - 4 \over
   1-x(1-x)\hat k_\mu^2}\,, \\
c &=& \sum_\mu{(1-2x)^2 \hat k_\mu^2 \over 1-x(1-x)\hat k_\mu^2}\,.
\end{eqnarray}
\end{mathletters}
We can now exploit the relationships
\begin{mathletters}
\begin{eqnarray}
b_\mu &=& a^2_\mu\,{\D\over\D x}\,{2(1-2x)\over a^2_\mu}\,, \\
c  &=& -2(1-2x)\sum_\mu{1\over a_\mu}\,{\D  a_\mu\over\D x}\,,
\end{eqnarray}
\end{mathletters}
to perform an integration by parts in the variable $x$ in
Eq.~(\ref{Delta-def}b) and obtain the more convenient representation
\begin{eqnarray}
\Delta^{-1}_{(\theta)} &=& \int_0^1\D x\,2(1-2x)
\sum_\mu\left\{{1\over a_\mu}\,{\D  a_\mu\over\D x}\intq
{1\over\left[4+m_0^2-\sum_\mu a_\mu\cos q_\mu\right]^2}\right. \nonumber \\
&&\qquad\qquad\left.-\;{1\over a^2_\mu}\,{\D\over\D x}\intq
{a^2_\mu\sin^2q_\mu\over\left[4+m_0^2-\sum_\mu a_\mu\cos q_\mu\right]^2}
\right\}.
\end{eqnarray}
Trivial algebraic manipulations (involving repeated use of integration by
parts in the momentum variables) lead finally to the form
\begin{equation}
\Delta^{-1}_{(\theta)} = \hat k^2 \int_0^1\D x\,{(1-2x)^2\over a_1a_2}\,
\intq {4\cos q_1\cos q_2 \over
\left[4+m_0^2-\sum_\mu a_\mu\cos q_\mu\right]^2}\,.
\label{Delta-new}
\end{equation}
Starting from Eq.~(\ref{Delta-def}a) and (\ref{Delta-new}) we can now
perform the momentum integrations.  One momentum component can be
integrated easily, thanks to the relationship
\begin{equation}
\int_{-\pi}^\pi {\D\theta\over2\pi}\,{1\over b-a\cos\theta}
= {1\over\sqrt{b^2-a^2}}
\end{equation}
and its parametric derivatives.  As a consequence we obtain
\begin{mathletters}
\label{Delta-comp}
\begin{eqnarray}
\Delta^{-1}_{(\alpha)} &=& \int_0^1\D x
\int_{-\pi}^\pi {\D q_1\over2\pi}\, {4+m_0^2 - a_1\cos q_1 \over
\left[\left(4+m_0^2 - a_1\cos q_1\right)^2 - a_2^2\right]^{3/2}}\,, \\
\Delta^{-1}_{(\theta)} &=& \hat k^2 \int_0^1\D x\,{(1-2x)^2\over a_1}\,
\int_{-\pi}^\pi {\D q_1\over2\pi}\, {4\cos q_1 \over
\left[\left(4+m_0^2 - a_1\cos q_1\right)^2 - a_2^2\right]^{3/2}}\,.
\end{eqnarray}
\end{mathletters}
The $q_1$ integrations in Eqs.~(\ref{Delta-comp}) are easily reducible to
standard elliptic integrals \cite{Gradstein}.  Without belaboring on the
straightforward algebraic tricks involved in the derivation, we present the
final result in the following form:
\begin{mathletters}
\label{Delta-repr}
\begin{eqnarray}
\Delta^{-1}_{(\alpha)} &=&
{4+m_0^2\over4\pi}\int_0^1\D x \, {1\over(a_1a_2)^{3/2}}
\,{\zeta^3 E(\zeta) \over 1-\zeta^2} \,, \\
\Delta^{-1}_{(\theta)} &=&
\hat k^2\,{4+m_0^2\over\pi}\int_0^1\D x \,
{(1-2x)^2\over(a_1a_2)^{5/2}} \left[{\zeta^3 E(\zeta) \over 1-\zeta^2}
    + 2\zeta\bigl(E(\zeta)-K(\zeta)\bigr)\right],
\end{eqnarray}
\end{mathletters}
where
\begin{equation}
\zeta = \sqrt{4 a_1 a_2 \over (4+m_0^2)^2 - (a_1-a_2)^2} \,.
\end{equation}
Eqs.~(\ref{Delta-repr}) lead to an enormous computational gain: from the
numerical point of view, the values of the elliptic integrals can be
routinely generated with high accuracy, and therefore the number of
numerical integrations is reduced from two to one; moreover, the $x$
integration is much more regular then the original momentum integration: if
we exploit the explicit expressions of $\Delta^{-1}_{(\alpha)}$ and
$\Delta^{-1}_{(\theta)}$ on the principal diagonal (\ref{Delta-diagonal})
to further regularize the integrand, a moderate-size Gauss-Legendre
integration suffices to produce very accurate results. From the point of
view of analytic manipulations, we can exploit the knowledge of the
asymptotic expansions of the elliptic integrals to simplify dramatically
the otherwise quite complicated problem of generating an asymptotic
expansion of the propagators in the small $m_0$ (scaling) region. This
expansion in turn is the essential ingredient in quantitative computations
for lattice models in the scaling (i.e.\ field-theoretical) regime. We
shall discuss this point in the next Section.  Let us only recall here the
relevant expansion formulae of the complete elliptic integrals, in the
regime $k \approx 1$, $k'=\sqrt{1-k^2} \ll 1$:
\begin{mathletters}
\label{ellipt-expansion}
\begin{eqnarray}
E(k) &\approx& 1 + {{k'}^2\over2}
  \left(\ln{4\over k'} - {1\over2}\right), \\
K(k) &\approx& \ln{4\over k'} + {{k'}^2\over4}
  \left(\ln{4\over k'} - 1\right), \\
\Pi(n,k) &\approx& {1\over1-n}\left(\ln{4\over k'}
+ {\sqrt{n}\over2}\ln{1-\sqrt{n}\over1+\sqrt{n}}\right) \nonumber \\
&+&\,{k^{\prime2}\over4(1-n)^2}\left((1+n)\ln{4\over k'}
+ \sqrt{n}\ln{1-\sqrt{n}\over1+\sqrt{n}}\right).
\end{eqnarray}
\end{mathletters}

\section{Asymptotic expansions in the scaling region}
\label{asymptotic expansions}

We are interested in evaluating physical quantities in a lattice model
possessing a nontrivial (continuum) field theory limit when the
ultraviolet regulator ($1/a$, where $a$ is the lattice spacing) is sent
to infinity while properly tuning the lattice coupling $f$ according to
the renormalization-group equation of the model. In such a model, lattice
expectation values for finite $a$ and $f(a)$ will necessarily receive
contributions from the irrelevant operators included in the lattice
definition of a physical quantity. In order to isolate the contribution
that will survive in the continuum limit, i.e.\ the scaling part of the
expectation value, we must be able to perform an expansion in the powers
of the lattice spacing. Within the $1/N$ expansion of lattice spin
models, there is another dimensionful parameter that can be employed to
classify the relevance of the contributions to any given expectation
value: the bare large-$N$ vacuum expectation value $m_0^2$ (``large-$N$
mass''). Since the continuum limit is the limit of infinite correlation
length, it is also the limit of vanishing $m_0^2$: contributions carrying
higher powers of $m_0^2$ with respect to the scaling contribution will be
{\it irrelevant}.  Actually, since the physical value of the mass in
these asymptotically-free models is not strictly zero, we must be careful
not to set $m_0^2=0$ and to get our results in the form of asymptotic
expansions in which $1 \gg |1/\ln(m_0^2 a^2)| \gg m_0^2 a^2$.

The detailed analysis of the procedures for the evaluation of physical
quantities in the scaling region, and the related problem of regularization
of the infrared divergences generated by the na\"\i ve $m_0^2\to0$ limit,
will be discussed in Sect.~\ref{physical quantities}.  Preliminary to such
a discussion is however the asymptotic expansion for small values of
$m_0^2$ of the basic ingredients in the evaluation of physical quantities,
i.e.\ the effective propagators themselves.  Due to their original
definition as one-loop integrals over Feynman propagators, the inverse
effective propagators $\Delta^{-1}_{(\alpha)}$ and
$\Delta^{-1}_{(\theta)\,\mu\nu}$ will have formal asymptotic expansions in
the form
\begin{equation}
\sum_{n=0}^\infty \left[\tilde A_n(k) +
    \tilde B_n(k) \ln m_0^2\right] \left(m_0^2\right)^n .
\label{-expansion}
\end{equation}
We may however recognize from Eq.~(\ref{mass-gap}) that the coupling
constant itself admits such an asymptotic expansion in the form
\begin{equation}
\beta \equiv {1\over2f} = \sum_{n=0}^\infty \left(\alpha_n +
    \beta_n \ln m_0^2\right) \left(m_0^2\right)^n ,
\label{beta-asympt}
\end{equation}
and Eq.~(\ref{beta-asympt}) can in principle be inverted to
\begin{equation}
\ln m_0^2 = \sum_{n=0}^\infty \left(\gamma_n +
    \delta_n \beta\right) \left(m_0^2\right)^n .
\end{equation}
As a consequence, a perfectly acceptable form of the asymptotic expansion
of the inverse propagators is:
\begin{mathletters}
\label{props-expansion}
\begin{eqnarray}
\Delta^{-1}_{(\alpha)}(k) &=& \sum_{n=0}^\infty
\left[A^{(\alpha)}_n(k) + \beta B^{(\alpha)}_n(k) \right]
\left(m_0^2\right)^n , \\
\Delta^{-1}_{(\theta)}(k) &=& \sum_{n=0}^\infty
\left[A^{(\theta)}_n(k) + \beta B^{(\theta)}_n(k) \right]
\left(m_0^2\right)^n .
\end{eqnarray}
\end{mathletters}
The expansion (\ref{props-expansion}), in comparison with
(\ref{-expansion}), involves no extra effort, but it shows the great
advantage of expressing final results in a form that makes direct contact
with the lattice weak-coupling expansion of the models, and allows
explicit checks of commutativity between weak-coupling and $1/N$
expansion.

We obviously aim at a systematic way of generating the coefficients $A_n$
and $B_n$. We shall discuss here a general technique that does not rely
upon any specific feature of the one-loop integrals involved, and in the
next Section we shall present a different approach specific to the case
discussed in Sect.~\ref{integral representations} when an integral
representation is available.

\subsection{General technique}

The general technique is worth a detailed analysis, because it will be
very useful also in the evaluation of expectation values of physical
quantities.  It will also allow us to show the connection between
sharp-momentum regularization and lattice formulation. It is therefore
convenient for our purposes to discuss the case of a one-loop lattice
integral with arbitrary propagators and vertices. The general form of the
lattice integral is
\begin{equation}
I(k; m_0a) = \int_0^\pi {\D^d p \over (2\pi)^d}\, F(k; m_0a, p),
\label{I-form}
\end{equation}
where $k$ is the external momentum (or a collection of external momenta)
and we use the notation
\begin{equation}
\int_0^b {\D^d p \over (2\pi)^d} \equiv
\prod_{i=1}^d\int_{-b}^b {\D p_i \over 2\pi}, \qquad
\int_a^b {\D^d p \over (2\pi)^d} \equiv
\prod_{i=1}^d\int_{-b}^b {\D p_i \over 2\pi} -
\prod_{i=1}^d\int_{-a}^a {\D p_i \over 2\pi}\,.
\end{equation}

When comparing with the continuum formulation, we must keep in mind that
powers of $a$ have been included in the definitions of $F$ and $I$ in
order to make them dimensionless. Formally we can rewrite
Eq.~(\ref{I-form}) in the form
\begin{eqnarray}
I(k; m_0a) &=& \int_0^\pi {\D^d p \over (2\pi)^d}\,F(k; m_0a, p) +
a^d \int_0^\infty {\D^d p \over (2\pi)^d}\,F(k; m_0a, pa)\nonumber \\
&-&\,a^d \int_0^\infty {\D^d p \over (2\pi)^d}\,
  {\rm T}F(k; m_0a, pa),
\label{I-formal}
\end{eqnarray}
where ${\rm T}F$ is the Taylor series expansion of the function $F$ in
powers of $m_0a$ around $m_0a=0$.  In order to turn Eq.~(\ref{I-formal})
into a mathematically rigorous statement, we must show that, order by
order in $a$, all ultraviolet and infrared divergences are explicitly
canceled. Let us therefore introduce the truncated Taylor series
expansion ${\rm T}_J$ (including all the powers of $m_0$ up to $m_0^J$)
and notice that
\begin{eqnarray}
&&\!\!\int_0^\pi {\D^d p \over (2\pi)^d}\,F(k; m_0a, p) +
a^d \int_0^\infty {\D^d p \over (2\pi)^d}\,F(k; m_0a, pa)
- a^d \int_0^\infty {\D^d p \over (2\pi)^d}\,
  {\rm T}_J F(k; m_0a, pa) \nonumber \\
&=&\left[\int_0^\pi {\D^d p \over (2\pi)^d}\,F(k; m_0a, p)
  -a^d \int_0^{\pi/a} {\D^d p \over (2\pi)^d}\,
  {\rm T}_J F(k; m_0a, pa)\right]\nonumber \\
&+&\,a^d \int_0^{\pi/a} {\D^d p \over (2\pi)^d}\,F(k; m_0a, pa)
  +  a^d \int_{\pi/a}^\infty {\D^d p \over (2\pi)^d}\,
     \left(1-{\rm T}_J\right) F(k; m_0a, pa) \nonumber \\
&=&\int_0^\pi {\D^d p \over (2\pi)^d}\,\left(1-{\rm T}_J\right)
  F(k; m_0a, p)
+ a^d \int_{\pi/a}^\infty {\D^d p \over (2\pi)^d}\,
     \left(1-{\rm T}_J\right) F(k; m_0a, pa) + I(k;m_0a) \nonumber \\
&=&I(k;m_0a) + O(a^{J+1}).
\label{I-messy}
\end{eqnarray}
Indeed the first contribution to Eq.~(\ref{I-messy}) is trivially
$O(a^{J+1})$, and the same can be shown to hold for the second
contribution by noticing that
\begin{eqnarray}
F(k; m_0a, pa) &=& \sum_j a^j F_j(k; m_0, p), \nonumber \\
F_j(k; \lambda m_0, \lambda p) &=& \lambda^j F_j(k; m_0, p);
\end{eqnarray}
therefore we have
\begin{equation}
a^d \int_{\pi/a}^\infty {\D^d p \over (2\pi)^d}\,
     \left(1-{\rm T}_J\right) F_j(k; m_0, p) = O(a^{J+1}).
\end{equation}

Eq.~(\ref{I-messy}) paves the way to two major strategies in the
regularization of lattice integrals. The first strategy, till now most
popular, stems from the observation that the last term in the l.h.s.\ of
Eq.~(\ref{I-messy}) vanishes in dimensional regularization.  One may
therefore adopt the definition
\begin{equation}
I(k;m_0a) = \lim_{d\to2}
\left[\int_0^\pi {\D^d p \over (2\pi)^d}\, F(k; m_0a, p)
+ a^d \int_0^\infty {\D^d p \over (2\pi)^d}\,F(k; m_0a, pa)\right].
\label{I-definition}
\end{equation}
When expanding in powers of $a$, the first contribution corresponds to
infrared-divergent massless lattice integrals and the second to
ultraviolet-divergent massive continuum integrals; in dimensional
regularization infrared and ultraviolet poles cancel exactly and we are
left with a finite result. However one may also notice that the last term
in the l.h.s.\ of Eq.~(\ref{I-messy}) is reminiscent of the structure of
counterterms in the sharp-momentum regularization scheme. More precisely,
we can decompose the expansion ${\rm T}_J F$ into contributions which are
ultraviolet divergent (relevant), contributions which are infrared
divergent (irrelevant), and contributions which are ultraviolet and
infrared divergent (marginal). Symbolically we may write
\begin{equation}
{\rm T}_J = {\rm T}_J^{\rm(UV)} + {\rm T}_J^{\rm(IR)} +  {\rm T}_J^{(0)}.
\end{equation}
We can introduce an arbitrary cutoff $M^2$ and split the counterterm
${\rm T}_J^{(0)}$ into two separate integrals. Finally we can express
the original integral in terms of two separate contributions whose
singularities are now independently regularized:
\begin{eqnarray}
I(k;m_0a) &=& \left[\int_0^\pi {\D^2p\over (2\pi)^2}
\, F(k; m_0a, p)\right. - a^2 \int_0^\infty{\D p^2 \over 4\pi}\,
{\rm T}_J^{\rm (IR)} F(k; m_0a, pa) \nonumber \\
&-&\,a^2 \left.\int_0^{M^2} {\D p^2 \over 4\pi}\, {\rm T}_J^{(0)}
F(k; m_0a, pa)\right] + \left[a^2 \int_0^\infty{\D p^2 \over 4\pi}\,
F(k; m_0a, pa)\right. \nonumber \\
&-&\,a^2 \int_0^\infty {\D p^2 \over 4\pi}\, {\rm T}_J^{\rm(UV)}
F(k; m_0a, pa) - a^2 \left.\int_{M^2}^\infty {\D p^2 \over 4\pi}\,
{\rm T}_J^{(0)} F(k; m_0a, pa)\right]. \nonumber \\
\label{I-SM-reg}
\end{eqnarray}
Eq.~(\ref{I-SM-reg}) is the starting point for the series expansion in
powers of $a$.

The first contribution in brackets is ultraviolet-regular.  The lattice
integral and the continuum integrals are separately infrared-singular, the
infrared divergences canceling out in the sum.  In analogy with the
standard notation for ultraviolet regularization (of.\
Eq.~(\ref{ren-gap-cont})), we always assume that such combinations of
(massless) separately infrared-divergent integrals stand for e.g.\
\begin{equation}
\int_0^\pi {\D^2p\over (2\pi)^2}\, \phi_{\,\rm latt}(p)
- \int_0^{\Lambda^2} {\D p^2 \over 4\pi}\, \phi_{\rm cont}(p) \equiv
\int_0^\pi {\D^2p\over (2\pi)^2} \left(\phi_{\,\rm latt}(p) -
\phi_{\rm cont}(p)\right)
- \int_{\scr R} {\D p^2 \over 4\pi}\, \phi_{\rm cont}(p),
\end{equation}
where ${\scr R}$ is the region comprised between the circle of radius
$\Lambda$ and the square of side $2\pi$ (assuming $\Lambda>\pi\sqrt2$).
The integral of the difference is infrared-regular by construction.

The second contribution in brackets is infrared-regular and ultraviolet
singularities are removed according to the prescriptions of the
sharp-momentum scheme, which means that the marginal (scaling) component of
$I(k;m_0a)$ receives a contribution exactly equal to the one obtained in
the corresponding continuum model: all lattice effects are included in the
first term.  Moreover, as we shall immediately show, by applying
Eq.~(\ref{I-SM-reg}) directly to the gap equation it is possible to fix the
value of $M^2$ in such a way that the sharp-momentum coupling constant can
be identified (for large $N$) with the lattice coupling.

\subsection{Expansion of the gap equation}

For our purposes it will be sufficient to truncate the expansion in
powers of $a$ to the second nontrivial order, i.e.\ $J=2$. The
regularized form of Eq.~(\ref{mass-gap}) is
\begin{eqnarray}
{1\over2f} &\approx& \int_0^\pi {\D^2p\over (2\pi)^2}\, {1\over
m_0^2a^2 + \sum_\mu4\sin^2\half p_\mu - c_2 \sum_\mu4\sin^4\half p_\mu}
\nonumber \\ &+&\, a^2 \int_0^\infty {\D p^2\over4\pi}\,
\left(1-{\rm T}_2\right)\,{1\over
m_0^2a^2 + \sum_\mu4\sin^2\half ap_\mu - c_2 \sum_\mu4\sin^4\half ap_\mu}
\end{eqnarray}
and, expanding to $O(a^2)$, we obtain
\begin{eqnarray}
{1\over2f} &\approx& \int_0^\pi {\D^2p\over (2\pi)^2}\,
   \left[{1\over\hat p^2-{1\over4}\, c_2\hat p^4} -
      {a^2m_0^2\over\left(\hat p^2
         - {1\over4}\, c_2\hat p^4\right)^2}\right]
\nonumber \\
&+&\,\int_0^\infty
   {\D p^2\over4\pi}\,\left[{1\over p^2+m_0^2} + {1\over4}\,
   {a^2\left(c_2+\third\right)\sum_\mu p_\mu^4 \over
      \left(p^2+m_0^2\right)^2}
   \right] \nonumber \\
&-&\,\int_0^\infty {\D p^2\over4\pi}\,\left[{1\over p^2}
- {m_0^2\over\left(p^2\right)^2} +
{1\over4}\,{a^2\left(c_2+\third\right)
   \sum_\mu p_\mu^4\over\left(p^2\right)^2}-
{1\over2}\,{a^2m_0^2\left(c_2+\third\right)
   \sum_\mu p_\mu^4\over\left(p^2\right)^3}
\right]. \nonumber \\
\end{eqnarray}
According to Eq.~(\ref{I-SM-reg}) we can write
\begin{eqnarray}
{1\over2f} &\approx& \int_0^\pi {\D^2p\over (2\pi)^2}\,
\left[{1\over\hat p^2-\fourth c_2\hat p^4} -
{a^2m_0^2\over\left(\hat p^2-\fourth c_2\hat p^4\right)^2}\right]
+ \int_0^\infty {\D p^2\over4\pi}\,{m_0^2\over \left(p^2\right)^2}
\nonumber \\
&-&\,\int_0^{M^2} {\D p^2\over4\pi}\,\left[{1\over p^2} -
{1\over2}\,{a^2 m_0^2\over p^2}\left(c_2+\third\right)
{\sum_\mu p_\mu^4\over\left(p^2\right)^2}\right] \nonumber \\
&+&\,\int_0^\infty {\D p^2\over4\pi}\,\left[{1\over p^2+m_0^2} +
{3\over16}\,{a^2\left(c_2+\third\right)\left(p^2\right)^2 \over
\left(p^2+m_0^2\right)^2} - {3\over16}\,a^2\left(c_2+\third\right)
\right] \nonumber \\
&-&\,\int_{M^2}^\infty {\D p^2\over4\pi}\,\left[{1\over p^2}
- {3\over8}\,{a^2m_0^2\over p^2}\left(c_2+\third\right)\right].
\end{eqnarray}
The continuum integrals can be explicitly evaluated and the final result
is
\begin{eqnarray}
{1\over2f} &\approx& \left[\int_0^\pi {\D^2p\over (2\pi)^2}\,
{1\over\hat p^2-\fourth c_2\hat p^4} -
\int_0^{M^2} {\D p^2\over4\pi}\,{1\over p^2}\right]
+ {1\over4\pi}\ln{M^2\over m_0^2}  \nonumber \\
&-&\, a^2m_0^2\left[\int_0^\pi {\D^2p\over (2\pi)^2}\,
{1\over\left(\hat p^2-\fourth c_2\hat p^4\right)^2}
- \int_0^\infty {\D p^2\over4\pi}\,{1\over \left(p^2\right)^2}
- \int_0^{M^2} {\D p^2\over4\pi}\,
{1\over2}\left(c_2+\third\right){\sum_\mu p_\mu^4\over\left(p^2\right)^3}
\right] \nonumber \\
&+&\,a^2m_0^2\,{3\over64\pi}\left(c_2+\third\right)
\left(1 - 2 \ln{M^2\over m_0^2}\right).
\label{f-messy}
\end{eqnarray}
Eq.~(\ref{f-messy}) deserves some comments:

A rescaling $p\to p/a$ in the terms originated by ${\rm T}^{\rm(IR)}_J$
must be performed {\it after\/} the expansion in powers of $a$ has been
accomplished. It may sound arbitrary, because we are working with
divergent quantities without an intrinsic scale, but it is mathematically
sound as one may recognize by going back to the proof of regularization.

The effect of a Symanzik improvement is made apparent by the cancellation
of the logarithmic dependence on $m_0$ in the first irrelevant
contribution when we choose the value $c_2=-\third$.

In order to identify sharp-momentum and lattice couplings we must
choose $M^2_L$ such that
\begin{equation}
\int_0^\pi {\D^2p\over (2\pi)^2}\,
{1\over\hat p^2-\fourth c_2\hat p^4} -
\int_0^{M^2_L} {\D p^2\over4\pi}\,{1\over p^2} = 0,
\label{ML}
\end{equation}
implying
\begin{equation}
{1\over2f}  =  {1\over4\pi}\ln{M^2_L\over m_0^2}  +  a^2 m_0^2
\left[\alpha_1(c_2) - {1+3c_2\over32\pi}\ln{M_L^2\over m_0^2}\right]
+ O(m_0^4).
\label{ML-f}
\end{equation}
The choice implied by Eq.~(\ref{ML}) will allow us to express our results
in a form immediately comparable to the lattice weak coupling expansion,
because the difference between the couplings is purely nonperturbative, as
shown by Eq.~(\ref{ML-f}). Any other choice would correspond to a finite
renormalization and would require a perturbative adjustment of $f$ in order
to recover standard perturbation theory. $M^2_L$ and $\alpha_1$ are plotted
as functions of $c_2$ in Fig.~\ref{M2L}. Let us notice the special values
\begin{mathletters}
\begin{eqnarray}
M^2_L\bigr|_{c_2=0} = 32,&&\qquad
M^2_L\bigr|_{c_2=-1/3} \cong 17.68967299, \\
\alpha_1\bigr|_{c_2=0} = {1\over32\pi}\,,&&\qquad
\alpha_1\bigr|_{c_2=-1/3} \cong -0.00479767.
\end{eqnarray}
\end{mathletters}

\subsection{Expansion of the propagators}

In practice, when willing to compute the expansion indicated in
Eqs.~(\ref{props-expansion}), much work can be saved by the following
observations:

The functions $B_n(k)$ are related in a simple way to the coefficients
of the $1/\varepsilon$ poles resulting from the continuum integration
in the representation (\ref{I-definition}). These coefficients are
related to essentially trivial one-loop continuum integrals that can
all be computed in closed form. Therefore the functions $B_n(k)$ can be
expressed in terms of elementary functions of $\hat k$ and do not
require an integral representation.

Moreover, let us assume that the integrand in the representation
(\ref{I-form}) of an inverse propagator $\Delta^{-1}$ be a function
$F_\Delta(k; m_0,p)$. By shifting the integration variable it is
possible to symmetrize $F$ to $F^{\rm sym}_\Delta(k; m_0,p)$, where
$F^{\rm sym}$ is an even function of $k_\mu$ with all singularities
located in $p_\mu = \pm\half k_\mu$.  Let us now consider the function
\begin{eqnarray}
&&\intp F^{\rm sym}_\Delta(k; m_0,p) -
\beta\sum_{n=0}^\infty B_n(k)\left(m_0^2\right)^n \nonumber \\
\equiv&&\intp\left[F^{\rm sym}_\Delta(k; m_0,p) - {1\over2}\left(
{1\over\overline{p{+}\half k}^2} +
{1\over\overline{p{-}\half k}^2}\right)
\sum_{n=0}^\infty B_n(k)\left(m_0^2\right)^n\right].
\end{eqnarray}
By construction, according to Eq.~(\ref{props-expansion}), this function
is analytic in $m_0$ around $m_0=0$. Therefore we immediately obtain the
following integral representation of $A_n(k)$:
\begin{eqnarray}
A_n(k) =&& {1\over n!} \intp {\D^n\over\D\left(m_0^2\right)^n}
\left[F^{\rm sym}_\Delta(k; m_0,p)
\vphantom{1\over\overline{p{-}\half k}^2 + m_0^2}
\right. \nonumber \\ &&\quad-\; \left.\left.
{1\over2}\left(
{1\over\overline{p{+}\half k}^2 + m_0^2} +
{1\over\overline{p{-}\half k}^2 + m_0^2}\right)
\sum_{l=0}^n B_l(k)\left(m_0^2\right)^l\right]\right|_{m_0^2=0}\,.
\end{eqnarray}
The symmetrization procedure insures us that all the singularities in
the integrand are {\it locally\/} canceled out. We are still left with
singularities regularized by counterterms of the form
\begin{eqnarray}
{\cos 4l\theta \over q^{2r}} \,,
\end{eqnarray}
where $q$ and $\theta$ are the polar coordinates in the $p$ plane with
center $\pm\half k_\mu$ and $l\ne0$; these counterterms are implicitly
introduced whenever needed, or equivalently polar integration in two small
circles around $p_\mu = \pm\half k_\mu$ is understood performed first.  In
practice, we shall only be interested in the functions $A_0(k)$ and
$A_1(k)$, whose integral representations we explicitly quote:
\begin{mathletters}
\label{A0+A1}
\begin{eqnarray}
A_0(k) =&& \intp\left[F^{\rm sym}_\Delta(k; 0,p) - {1\over2}\left(
{1\over\overline{p{+}\half k}^2} +
{1\over\overline{p{-}\half k}^2}\right) B_0(k)\right], \\
A_1(k) =&& \intp\left[\left.{\D F^{\rm sym}_\Delta(k; m_0,p) \over
    \D m_0^2}\right|_{m_0^2=0}
+ {1\over2}\left({1\over\left(\overline{p{+}\half k}^2\right)^2} +
{1\over\left(\overline{p{-}\half k}^2\right)^2}\right) B_0(k)\right.
\nonumber \\ &&\quad-\; \left.
{1\over2}\left({1\over\overline{p{+}\half k}^2} +
{1\over\overline{p{-}\half k}^2}\right) B_1(k)
\vphantom{1\over\left(\overline{p{-}\half k}^2\right)^2}\right].
\end{eqnarray}
\end{mathletters}
Hence for our present purposes we shall only have to work out the
functions $B_0(k)$ and $B_1(k)$. Let us illustrate the procedure by
evaluating $B_0^{(\alpha)}(k)$ and $B_1^{(\alpha)}(k)$. Our starting
point will be the expansion in powers of $a$ of the continuum integral
\begin{eqnarray}
2\int_0^\infty&&{\D^dp\over(2\pi)^d}\,a^d
\left[m_0^2a^2 + \sum_\mu 4\sin^2\left(\half ap_\mu\right)
    - c_2\sum_\mu 4\sin^4\left(\half ap_\mu\right)\right]^{-1}
\nonumber \\
&&\quad\times\;
  \left[m_0^2a^2 + \sum_\mu 4\sin^2\left(\half ap_\mu+k_\mu\right)
    - c_2\sum_\mu 4\sin^4\left(\half ap_\mu+k_\mu\right)\right]^{-1},
\label{I-continuum}
\end{eqnarray}
where the factor 2 has been included in order to take into account the
effects of the (excluded) expansion around the second singularity in
$p=-k$. Since we are interested in the expansion to second nontrivial
order we may replace Eq.~(\ref{I-continuum}) with
\begin{eqnarray}
2 a^\varepsilon\int_0^\infty {\D^dp\over(2\pi)^d} \,&&
{1\over\overline{pa}^2 + m_0^2a^2} \left\{{1\over{\bar k}^2} -
   {a^2\sum_\mu p^2_\mu\left(c_1\cos k_\mu + c_2\cos 2k_\mu\right)
    + m_0^2 \over \left({\bar k}^2\right)^2} \right.\nonumber \\
&&\quad+\;\left.
   {a^2\left[\sum_\mu p_\mu\left(2c_1\sin k_\mu + c_2\sin 2k_\mu\right)
    \right]^2 \over \left({\bar k}^2\right)^3} \right\}.
\label{I-varepsilon}
\end{eqnarray}
Within the desired approximation however
\begin{equation}
\int_0^\infty {\D^dp\over(2\pi)^d}\,{p_\mu p_\nu\over{\bar p}^2 + m_0^2}
\approx {1\over d}\int_0^\infty {\D^dp\over(2\pi)^d}\,
{{\bar p}^2\delta_{\mu\nu}\over{\bar p}^2 + m_0^2} =
-{m_0^2\over d}\,\delta_{\mu\nu}\int_0^\infty {\D^dp\over(2\pi)^d}\,
{1\over{\bar p}^2 + m_0^2}\,.
\end{equation}
Moreover, we are only interested in the pole part, and we can therefore
replace Eq.~(\ref{I-varepsilon}) with
\begin{eqnarray}
2 \int_0^\infty {\D^dp\over(2\pi)^d} \,&&
{1\over{\bar p}^2 + m_0^2} \left\{{1\over{\bar k}^2} -
   m_0^2 a^2 \left[
   {1-{1\over2}\sum_\mu\left(c_1\cos k_\mu + c_2\cos 2k_\mu\right)
    \over \left({\bar k}^2\right)^2} \right.\right.\nonumber \\
&&\quad-\;\left.\left.{1\over2}\,
   {\sum_\mu \left(2c_1\sin k_\mu + c_2\sin 2k_\mu\right)^2
    \over \left({\bar k}^2\right)^3} \right]\right\}
\label{pole-part}
\end{eqnarray}
However,
\[
\int_0^\infty {\D^dp\over(2\pi)^d}\,{1\over{\bar p}^2+m_0^2}
\]
is an exact representation of the pole part in the asymptotic
expansion of $\beta$, and therefore we can read $B_0^{(\alpha)}$ and
$B_1^{(\alpha)}$ directly off Eq.~(\ref{pole-part}) and find
\begin{mathletters}
\label{B0-1-alpha}
\begin{eqnarray}
B_0^{(\alpha)} &=& {2\over{\bar k}^2} \,, \\
B_1^{(\alpha)} &=&
-2\sum_\mu{\sin^2\half k_\mu \left(c_1+4c_2\cos^2\half k_\mu\right)^2
  \over\left({\bar k}^2\right)^2}
-4\sum_\mu{\sin^2 k_\mu \left(c_1+c_2\cos^2 k_\mu\right)^2
  \over\left({\bar k}^2\right)^3} \,. \nonumber \\
\end{eqnarray}
\end{mathletters}
An important representation of Eq.~(\ref{B0-1-alpha}b) is
\begin{equation}
B_1^{(\alpha)} = B_0^{(\alpha)} \left[\half\sum_\mu
    \partial_\mu\partial_\mu \ln{\bar k}^2
+ \sum_\mu{\sin^2\half k_\mu \left(c_1+4c_2\cos^2\half k_\mu\right)^2
    \over{\bar k}^2} - {2\over{\bar k}^2}\right].
\end{equation}

The computation of $B_0^{(\theta)}$ and $B_1^{(\theta)}$ is in no sense
conceptually more involved than the computation we have just presented.
However many more terms have to be taken into account and a few algebraic
tricks exploiting explicitly the fact that there are only two vector
components in two dimensions have to be employed in order to simplify the
result. Let us only quote the final result:
\begin{mathletters}
\begin{eqnarray}
B_0^{(\theta)} =&& 2\sum_\mu\left(c_1+c_2\cos^2\half k_\mu\right) -
2\sum_\mu{\hat k^2_\mu\left(c_1+c_2\cos^2\half k_\mu\right)^2
    \over{\bar k}^2} \nonumber \\
=&& 2\prod_\mu\left(c_1+c_2\cos^2\half k_\mu\right)
{\hat k^2\over{\bar k}^2} \, \\
B_1^{(\theta)} =&& B_0^{(\theta)}\left\{
  \sum_\mu\left[{c_1+4c_2\cos^2\half k_\mu\over{\bar k}^2}
    + {8\cos^2\half k_\mu\left(c_1+c_2\cos k_\mu\right)^2
       \over\left({\bar k}^2\right)^2}\right]\right. \nonumber \\
&&\quad\times\;\left.\left[{\sum_\nu\sin^2\half k_\nu
    \left(c_1+c_2\cos^2\half k_\nu\right)\over c_1+c_2\cos^2\half k_\mu}
  - \sin^2\half k_\mu\right]
  \vphantom{1\over\left({\bar k}^2\right)^2}
\vphantom{\left(c_1+c_2\cos^2\half k_\nu\right) \over 1}\right\}
\nonumber \\
=&& B_0^{(\theta)} \left\{
  \half\sum_\mu\partial_\mu\partial_\mu\ln{\bar k}^2
  + \sum_\mu{\sin^2\half k_\mu\left(c_1-4c_2\cos^2\half k_\mu\right)
    \over{\bar k}^2} + {2\over{\bar k}^2}\right. \nonumber \\
&&\quad+\;\left.\fourth\sum_\mu{1\over c_1+c_2 \cos^2\half k_\mu}
  \left(c_1+4c_2\cos^2\half k_\mu
    - {8 c_1\sin^2\half k_\mu\over{\bar k}^2}\right)\right\}.\qquad
\end{eqnarray}
\end{mathletters}

Finally we mention that it is possible to define an asymptotic
expansion also for the functions $\Delta^{-1}_{(\theta)\,\mu}(k)$ defined
in Sect.~\ref{lattice 1/N}.  The form of the expansion is
\begin{equation}
\beta\Delta^{-1}_{(\theta)\,\mu}(k) = \sum_{n=0}^\infty \left[
C_{n\mu} + \beta D_{n\mu}\right]\left(m_0^2\right)^n,
\label{CD-expansion}
\end{equation}
and it is easy to obtain
\begin{mathletters}
\begin{eqnarray}
D_{0\mu} =&& c_1+c_2\cos^2\half k_\mu , \\
C_{0\mu} =&& -\fourth\intp {c_1\hat p^2
    + 4c_2\sum_\nu\hat p^2_\nu\cos^2\half p_\nu
  \cos^2\half k_\mu \over{\bar p}^2} \, , \\
D_{1\mu} =&& \fourth\left(c_1+4c_2\cos^2\half k_\mu\right) , \\
C_{1\mu} =&& -\fourth\intp \left[
  {c_1 + 4c_2\cos^2\half k_\mu \over{\bar p}^2} -
  {c_1\hat p^2 + 4c_2\sum_\nu\hat p^2_\nu\cos^2\half p_\nu
    \cos^2\half k_\mu \over\left({\bar p}^2\right)^2}\right],
\nonumber \\
\end{eqnarray}
\end{mathletters}
All these results take a particularly simple form in the case $c_2=0$; in
this case we obtain
\begin{mathletters}
\begin{eqnarray}
B_0^{(\alpha)}(k) &=& {2\over\hat k^2} \, , \\
B_1^{(\alpha)}(k) &=& \left(-{1\over2} - {4\over\hat k^2} +
  {\hat k^4\over\bigl(\hat k^2\bigr)^2}\right){1\over\hat k^2}
= B_0^{(\alpha)}(k)\left({1\over2}\sum_\mu\partial_\mu\partial_\mu
    \ln\hat k^2 + {1\over4} - {2\over\hat k^2}\right) \, , \\
B_0^{(\theta)}(k) &=& 2 \, , \\
B_1^{(\theta)}(k) &=& -{1\over2} + {4\over\hat k^2} +
  {\hat k^4\over\bigl(\hat k^2\bigr)^2}
= B_0^{(\theta)}(k)\left({1\over2}\sum_\mu\partial_\mu\partial_\mu
    \ln\hat k^2 + {1\over4} + {2\over\hat k^2}\right) \, , \\
 D_{0\mu} &=& 1, \qquad\qquad C_{0\mu} = -\fourth,  \qquad\qquad
   D_{1\mu} = \fourth, \qquad\qquad C_{1\mu} = 0,
\end{eqnarray}
\end{mathletters}
and one can easily construct the representations of $A_0(k)$ and
$A_1(k)$ by a trivial application of Eqs.~(\ref{A0+A1}). We must
observe that these representations are well defined and they can be used
to compute numerically $A_0(k)$ and $A_1(k)$; however, the integrands
are plagued with a very singular behavior, especially for small $k$,
and it is very hard to perform accurate numerical integrations.

\section{Asymptotic expansions using integral representations}
\label{asymptotic integral representation}

The methods developed in the previous Section apply perfectly well to the
case $c_2=0$. However this is not the most convenient way of performing the
asymptotic expansion when we possess an integral representation like
Eqs.~(\ref{Delta-repr}). In this case we start from the observation that
\begin{equation}
I(k; m_0 a) = \int_0^1\D x\,G(\hat k_\mu, m_0 a, x).
\end{equation}
Moreover, if we perform an homogeneous expansion of $G$ in the powers of
$\hat k_\mu$ and $m_0$, the $x$-integration of the resulting coefficients
can be performed explicitly. As a consequence, we may write
\begin{equation}
I(k; m_0 a) = \int_0^1\D x\left[1-{\rm T}^{(a)}_J\right]
G(\hat k_\mu a, m_0 a, x)
+ \int_0^1\D x \, {\rm T}^{(a)}_J G(\hat k_\mu a, m_0 a, x),
\end{equation}
where the unconventional notation $\hat k_\mu a$ indicates that the
parameters of the Taylor expansion are $\hat k_\mu a$, not $k_\mu a$.  The
first integral is now regular up to $O(a^J)$, and the integrand can
therefore be expanded in powers of $m_0^2$. The second integral in turn can
be performed analytically and the result, which is a nonanalytic function
of $m_0$, can be expanded in an asymptotic series in the powers of $m_0$.
This procedure is rather cumbersome, but the payoffs are very high, as we
shall see in the following.

Let us present the essential ingredients of the computation and sketch
the derivation of the asymptotic expansion in the case of
$\Delta^{-1}_{(\alpha)}$.  Let us define
\begin{equation}
q_\mu^2 = \hat k_\mu^2\,x(1-x), \qquad q^2 = \sum_\mu q_\mu^2,
\qquad q^4 = \sum_\mu q_\mu^4,
\end{equation}
and expand all relevant quantities in powers of $m_0^2$ and $\hat k_\mu^2$:
\begin{mathletters}
\begin{eqnarray}
a_\mu &\approx& 2 - q_\mu^2, \\
\zeta^2 &\approx& 1 - \half\left(m_0^2+q^2\right), \\
E(\zeta) &\approx& 1 - {m_0^2 + q^2 \over8}
\left(\ln {m_0^2 + q^2 \over32} + 1\right) , \\
K(\zeta) &\approx& - {1\over2}\ln {m_0^2 + q^2 \over32} -
{m_0^2 + q^2 \over16}\ln {m_0^2 + q^2 \over32} + {m_0^2-q^2\over16}
-{1\over8}\,{q^4\over q^2+m_0^2}\,.
\end{eqnarray}
\end{mathletters}
The relevant integrals are:
\begin{mathletters}
\begin{eqnarray}
&&\int_0^1 {\D x\over q^2+m_0^2} =
{2\over\hat k^2\hat\xi}\ln{\hat\xi+1\over\hat\xi-1}\,, \\
&&\int_0^1\D x \ln{q^2+m_0^2\over32} =
\hat\xi\ln{\hat\xi+1\over\hat\xi-1} - 2 - \ln{32\over m_0^2} \,, \\
&&\int_0^1\D x\,{q^4\over\left(q^2+m_0^2\right)^2} =
{\hat k^4\over\bigl(\hat k^2\bigr)^2}\left[
    {1\over\hat\xi}\ln{\hat\xi+1\over\hat\xi-1}\left(
    {1\over4\hat\xi^2} + {1\over2} - {3\over4}\,\hat\xi^2\right)
    - {1\over2\hat\xi^2} + {3\over2}\right] ,
\end{eqnarray}
\end{mathletters}
where
\begin{eqnarray}
\hat\xi \equiv \sqrt{1 + {4 m_0^2\over \hat k^2}} \,.
\end{eqnarray}
Then, starting from Eq.~(\ref{Delta-repr}a), we obtain
\begin{eqnarray}
\Delta^{-1}_{(\alpha)} &\approx&
 {1\over4\pi}\int_0^1\D x\,{1\over m_0^2+q^2}
\left[1-{m_0^2+q^2\over8}\ln{q^2+m_0^2\over32} - {m_0^2\over4} -
{q^4\over4(m_0^2+q^2)}\right] \nonumber \\
&=& {1\over2\pi}\left[{1\over\hat k^2\hat\xi}
      \ln{\hat\xi+1\over\hat\xi-1}
    - {1\over16}\,{2\hat\xi^2-1\over\hat\xi}
      \ln{\hat\xi+1\over\hat\xi-1}
    + {1\over8} + {1\over16}\ln{32\over m_0^2}
\vphantom{\hat k^4\over\bigl(\hat k^2\bigr)^2}\right. \nonumber \\
&&\qquad-\;\left. {1\over8}\,{\hat k^4\over\bigl(\hat k^2\bigr)^2}
      \left({(1-\hat\xi^2)(1 + 3\hat\xi^2)\over4\hat\xi^3}
        \ln{\hat\xi+1\over\hat\xi-1}
	+ {3\hat\xi^2-1\over2\hat\xi^2}\right)\right].
\label{I-alpha}
\end{eqnarray}
Therefore by expanding both sides of Eq.~(\ref{I-alpha}) we can introduce
the regulator in the form
\begin{eqnarray}
&-&\,{1\over2\pi}\int_0^1\D x\,{1\over2q^2} +
{1\over2\pi\hat k^2}\ln{\hat k^2\over32} + 2\beta\,{1\over\hat k^2}
  + {m_0^2\over2\pi}\int_0^1\D x \left[2\,{1\over\left(q^2\right)^2}
  + {3\over16}\,{1\over q^2}
  - {1\over4}\,{\hat k^4\over\bigl(\hat k^2\bigr)^2}\,
  {1\over q^2}\right]
\nonumber \\
&+&\,{m_0^2\over2\pi\hat k^2}
   \left[\left(-2\,{1\over\hat k^2} - {3\over8} +
   {1\over2}\,{\hat k^4\over\bigl(\hat k^2\bigr)^2}\right)
   \ln{\hat k^2\over32} + 2\,{1\over\hat k^2} - {1\over4}
   - {1\over4}\,{\hat k^4\over\bigl(\hat k^2\bigr)^2}\right]
\nonumber \\
&+&\,{m_0^2\over\hat k^2}\left(-{4\over\hat k^2} - {1\over2}
+ {\hat k^4\over\bigl(\hat k^2\bigr)^2}\right)\beta.
\end{eqnarray}
As a consequence, the known values of $B_0^{(\alpha)}$ and
$B_1^{(\alpha)}$ are reproduced, and we find the representations
\begin{mathletters}
\begin{eqnarray}
A_0^{(\alpha)} &=& {1\over\pi}\int_0^1\D x\left\{
{\zeta_0^3\over(a_1a_2)^{3/2}}
\,{E(\zeta_0)\over1-\zeta_0^2} - {1\over4q^2}\right\} +
{1\over2\pi\hat k^2}\ln{\hat k^2\over32}, \nonumber \\ \\
A_1^{(\alpha)} &=& {1\over4\pi}\int_0^1{\D x}
\,{\zeta_0^3\over(a_1a_2)^{3/2}}
\left\{{E(\zeta_0)\over1-\zeta_0^2} +
{4\zeta_0^2\over a_1a_2}
\left[{K(\zeta_0) - 2 E(\zeta_0)\over1-\zeta_0^2}
   - {2 E(\zeta_0)\over\left(1-\zeta_0^2\right)^2}\right]\right\}
\nonumber \\
&+&\, {1\over4\pi}\int_0^1{\D x} \left[{1\over\left(q^2\right)^2}
   + {3\over8}\,{1\over q^2}
   - {1\over2}\,{q^4\over\left(q^2\right)^3}\right] \nonumber \\
&+&\, {1\over2\pi}\left[
{2\over\bigl(\hat k^2\bigr)^2} - {1\over4\hat k^2}
-{1\over4}\,{\hat k^4\over\bigl(\hat k^2\bigr)^3} +
{1\over\hat k^2}\ln{\hat k^2\over32}\,\left(-{2\over\hat k^2} - {3\over8}
  + {1\over2}\,{\hat k^4\over\bigl(\hat k^2\bigr)^2}\right)\right],
\end{eqnarray}
\end{mathletters}
where
\begin{equation}
\zeta_0 \equiv \sqrt{{4 a_1 a_2 \over 16 - \left(a_1-a_2\right)^2}} \,.
\end{equation}

By repeating the analysis for $\Delta^{-1}_{(\theta)}$ we find that
\begin{eqnarray}
\Delta^{-1}_{(\theta)} &\approx&
 {\hat k^2\over4\pi}\int_0^1\D x\,{(1-2x)^2\over m_0^2+q^2}
\left[1+{3\over4}\,m_0^2 + {3\over2}\,q^2 + {3\over8}(q^2+m_0^2)
\ln{q^2+m_0^2\over32} - {1\over4}\,{q^4\over m_0^2+q^2}\right]
\nonumber \\
&=& {1\over2\pi} \left[\hat\xi\ln{\hat\xi+1\over\hat\xi-1} - 2
+ {1\over16}\,\hat k^2 (3-2\hat\xi^2)
  \left(\hat\xi\ln{\hat\xi+1\over\hat\xi-1} - 2\right)
+ {5\over24}\,\hat k^2 - {1\over16}\,\hat k^2\ln{32\over m_0^2}\right.
\nonumber \\
&&\qquad\left.+\;{1\over8}\,{\hat k^4\over\hat k^2}\left(
{(1-\xi^2)(1-5\hat\xi^2)\over4\hat\xi}\ln{\hat\xi+1\over\hat\xi-1}
+ {13\over6} - {5\over2}\,\hat\xi^2\right)\right],
\label{D-theta-expa}
\end{eqnarray}
and expanding both sides of Eq.~(\ref{D-theta-expa}) we can introduce the
regulator in the form
\begin{eqnarray}
&-&\,{1\over2\pi}\int_0^1\D x\,{\hat k^2\over2q^2} +
{1\over2\pi}\ln{\hat k^2\over32} + 2\beta
+ {m_0^2\over2\pi}\int_0^1\D x \left[{\hat k^2\over2\left(q^2\right)^2}
   - {2\over q^2} + {3\over16}\,{\hat k^2\over q^2}
  - {1\over4}\,{\hat k^4\over\hat k^2 q^2}\right]
\nonumber \\
&+&\,{m_0^2\over2\pi}
   \left[\left(2\,{1\over\hat k^2} - {3\over8} +
   {1\over2}\,{\hat k^4\over\bigl(\hat k^2\bigr)^2}\right)
   \ln{\hat k^2\over32} + 2\,{1\over\hat k^2} + {1\over4}
   - {1\over4}\,{\hat k^4\over\bigl(\hat k^2\bigr)^2}\right]
\nonumber \\
&+&\,m_0^2\left[4\,{1\over\hat k^2} - {1\over2}
+ {\hat k^4\over\bigl(\hat k^2\bigr)^2}\right]\beta.
\end{eqnarray}
The values of $B_0^{(\theta)}$ and $B_1^{(\theta)}$ are reproduced and we
obtain the representations
\begin{mathletters}
\begin{eqnarray}
A_0^{(\theta)} &=& {4\hat k^2\over\pi}\int_0^1\D x \,
{(1-2x)^2\over(a_1a_2)^{5/2}}
\,\left[{\zeta_0^3 E(\zeta_0)\over1-\zeta_0^2}
 + 2\zeta_0\bigl(E(\zeta_0) - K(\zeta_0)\bigr)\right]
\nonumber \\
&-&\,{1\over2\pi}\int_0^1\D x\,{\hat k^2\over2q^2}
+{1\over2\pi}\ln{\hat k^2\over32} \,,
\\
A_1^{(\theta)} &=& {\hat k^2\over\pi}\int_0^1\D x \,
{(1-2x)^2\over(a_1a_2)^{5/2}}\,\zeta_0\,
\left[{E(\zeta_0)\over1-\zeta_0^2} + E(\zeta_0) - 2 K(\zeta_0)\right]
\nonumber \\
&-&\,{4\hat k^2\over\pi}\int_0^1\D x \,
{(1-2x)^2\over(a_1a_2)^{7/2}}\,\zeta_0^3\, \left[
2\,{1-\zeta_0^2+\zeta_0^4\over\left(1-\zeta_0^2\right)^2}\,E(\zeta_0)
- {2-\zeta_0^2\over1-\zeta_0^2}\,K(\zeta_0) \right] \nonumber \\
&+&\,{\hat k^2\over2\pi}\int_0^1\D x
\left[{1\over2}\,{1\over\left(q^2\right)^2}
   - 2\,{1\over q^2\hat k^2} + {3\over16}\,{1\over q^2}
   - {1\over4}\,{\hat k^4\over\left(\hat k^2\right)^2 q^2}\right]
\nonumber \\
&+&\,{1\over2\pi} \left[\left(2\,{1\over\hat k^2} - {3\over8} +
   {1\over2}\,{\hat k^4\over\bigl(\hat k^2\bigr)^2}\right)
   \ln{\hat k^2\over32} + 2\,{1\over\hat k^2} + {1\over4}
   - {1\over4}\,{\hat k^4\over\bigl(\hat k^2\bigr)^2}\right].
\end{eqnarray}
\end{mathletters}
The contour plots of $A_0^{(\alpha)}$ and $A_0^{(\theta)}$ in the $k$-plane
are drawn in Fig.~\ref{A0-fig}.

We finally present the results of the evaluation of the inverse propagators
along the principal diagonal of the momentum lattice, i.e.\ when $k_1 = k_2
\equiv l$. In this case direct manipulations of Eqs.~(\ref{Delta-diagonal})
lead to
\begin{mathletters}
\begin{eqnarray}
A_0^{(\alpha)}(l,l) &=& {1\over16\pi\sin^2\half l}\,
\cos\half l\ln{1 - \cos\half l\over1 + \cos\half l}\,, \\
A_1^{(\alpha)}(l,l) &=& {1\over64\pi\sin^4\half l}
\left[\cos^2\half l - (2-\cos^2\half l) \cos\half l\,
\ln{1 - \cos\half l\over1 + \cos\half l}\right], \\
A_0^{(\theta)}(l,l) &=& {1\over2\pi}\left[{1\over\cos\half l}\,
\ln{1 - \cos\half l\over1 + \cos\half l} - 2\right]\,, \\
A_1^{(\theta)}(l,l) &=& {1\over8\pi\sin^2\half l}
\left[2-\cos^2\half l + \cos\half l\,
\ln{1 - \cos\half l\over1 + \cos\half l}\right].
\end{eqnarray}
\end{mathletters}

\section{Evaluation of physical quantities in the scaling region}
\label{physical quantities}

The regularization (\ref{I-SM-reg}) of lattice integrals in the scaling
region and the asymptotic expansion (\ref{props-expansion}) of the
effective propagators are the essential ingredients for the evaluation of
the scaling contributions to any physical quantity to $O(1/N)$, involving
one-loop integrals over the effective fields propagators.

Actually there is no real need of evaluating separately every physical
quantity: whenever two operators have the same (possibly anomalous)
dimension, their ratio is a pure number that is scheme-independent, and
therefore it can be computed in the simplest available scheme; only one
of the two operators, possibly the easier to compute, must be evaluated
on the lattice in the scaling region. These statements obviously rest on
the existence of a renormalization-group behavior in the scaling regime
of the model involved, but they can be explicitly verified in the $1/N$
expansion, as we shall see in the following.

\subsection{Lattice correlation length}

A prototype dimensionful quantity is the inverse correlation length (mass
gap), that sets the scale for all quantities having canonical dimension.
For sufficiently large values of $\kappa$, as discussed in Sect.~\ref{spin
models}, it is meaningful to extract the mass gap from the asymptotic
behavior of the two-point correlation function of the fundamental fields.
This asymptotic behavior is determined by the location of the pole of the
two-point function, as in the continuum case.  There is however a slight
complication due to the absence of rotation invariance outside the scaling
region. For sake of completeness, let us discuss this point, that was
solved in the ${\rm O}(N)$ case by M\"uller, Raddatz and Ruhl
\cite{Muller-Raddatz-Ruhl} by applying saddle-point techniques to estimate
the large-distance behavior of $G(x)$ (see also Cristofano {\it et al.}\
\cite{Cristofano-Musto-Nicodemi-Pettorino-Pezzella}).

For sake of definiteness and simplicity we shall now focus on the case
$c_2 = 0$, where the inverse two-point function is
\begin{equation}
G^{-1}(p) \equiv \hat p^2 + m_0^2 + {1\over N}\,\Sigma(p).
\end{equation}
Let us introduce the function $\mu(\theta)$, indicating the coefficient
of the large-distance exponential decay, which is dependent on the polar
angle $\theta$ in the $(x,y)$ plane, and define
\begin{equation}
\mu(\theta) = \mu_x(\theta) \cos\theta + \mu_y(\theta) \sin\theta,
\end{equation}
where $\mu_x$ and $\mu_y$ are functions of $\theta$ such that
\begin{equation}
G^{-1}\left(p_i{=}i\mu_i\right) = 0.
\label{mu-G}
\end{equation}
Eq.~(\ref{mu-G}) is not sufficient to determine $\mu_i(\theta)$, but,
guided by the saddle-point analysis of Ref.~\cite{Muller-Raddatz-Ruhl}, we
make the large-$N$ Ansatz
\begin{mathletters}
\label{mu-Ansatz}
\begin{eqnarray}
\mu_i &=& \mu_{0i} + {1\over N}\mu_{1i} + O\!\left(1\over N^2\right), \\
\sinh\mu_{0x} &=& \nu_0 \cos\theta, \qquad
\sinh\mu_{0y} = \nu_0 \sin\theta.
\end{eqnarray}
\end{mathletters}
The generalization to $c_2\ne0$ would consist in the replacement
\begin{equation}
\sinh\mu_{0i} \to c_1\sinh\mu_{0i} + \half c_2\sinh2\mu_{0i}\,.
\end{equation}
We can now solve the equation
\begin{equation}
\hat p^2 + m_0^2 = 0
\end{equation}
for $p_i = i \mu_{0i}$ with the Ansatz (\ref{mu-Ansatz}) and obtain
\begin{equation}
\nu_0 = m_0\sqrt{2+\fourth m_0^2}
    \left[1+\sqrt{1 - m_0^2 \left(8+m_0^2\right)
	\left(\cos2\theta\over4+m_0^2\right)^2}\;\right]^{-1/2},
\end{equation}
hence
\begin{equation}
\mu_0(\theta) = \cos\theta\arsinh(\nu_0\cos\theta) +
   \sin\theta\arsinh(\nu_0\sin\theta) = m_0 + O(m_0^3) \,.
\label{mu0-scaling}
\end{equation}
Now replacing Eq.~(\ref{mu-Ansatz}a) in the condition (\ref{mu-G}) we
immediately get
\begin{equation}
-2\left[\nu_0\mu_{1x}\cos\theta + \nu_0\mu_{1y}\sin\theta\right]
+ \Sigma_1(i\mu_{0i}) = 0,
\end{equation}
implying
\begin{equation}
\mu_1(\theta) \equiv \mu_{1x}\cos\theta + \mu_{1y}\sin\theta
= {\Sigma_1(i\mu_{0i})\over2\nu_0}
\end{equation}
and
\begin{equation}
\mu^2(\theta) = \mu^2_0(\theta) + {1\over N}\,{\mu_0(\theta)\over
    \nu_0(\theta)}\,\Sigma_1(i\mu_{0i}) + O\!\left(1\over N^2\right).
\label{mu-expansion}
\end{equation}

These results are quite general and may be useful in detailed studies of
lattice models outside the scaling region. One may check that in the
appropriate (scaling) limit the dependence on $\theta$ disappears in
$\mu_1(\theta)$ as well as in $\mu_0$.  However, in practice we may limit
our attention to only two special values of $\theta$, corresponding to the
two extrema $\theta=0$ (side correlation) and $\theta=\fourth\pi$ (diagonal
correlation). In both cases no special Ansatz is needed in order to find
the expression for the mass-gap: in the first case the lattice symmetry
implies $\mu_y(0) = 0$ and therefore $\mu(0) = \mu_x(0)$; in the second
case we have $\mu_x(\fourth\pi) = \mu_y(\fourth\pi) \equiv \mu_d$ and
therefore $\mu(\fourth\pi) = \sqrt{2}\,\mu_d$. Eq.~(\ref{mu-G}) reduces in
both cases to a single-variable equation.  It is easy to check that, when
$c_2=0$,
\begin{mathletters}
\begin{eqnarray}
\mu_0(0) &=& 2\arsinh \half m_0, \\
\mu_0(\fourth\pi) &=& 2\sqrt2\,\arsinh{m_0\over2\sqrt2}\,;
\end{eqnarray}
\end{mathletters}
$\mu_0^2(0)$ and $\mu_0^2(\fourth\pi)$ are plotted as functions of $f$ in
Fig.~\ref{large-N-fig}.
It is also pleasant to check that, in these two specific cases, the
resulting definition of the mass gap coincides with that obtained from the
so-called wall-wall correlations
\begin{mathletters}
\begin{eqnarray}
G_s(y-x) &=& {1\over L}\sum_{x_1,y_1} G(x_1,x; y_1,y), \\
G_d\left(y-x\over\sqrt2\right) &=&
{\sqrt2\over L}\sum_{x_1,y_1} G(x_1,x-x_1; y_1,y-y_1),
\end{eqnarray}
\end{mathletters}
whose effective one-dimensional propagators in momentum space are
\begin{mathletters}
\begin{eqnarray}
G_s^{-1} &=& G^{-1}(p_x{=}0, p_y), \\
G_d^{-1} &=& {1\over\sqrt2}\,G^{-1}(p_x{=}p_d, p_y{=}p_d).
\end{eqnarray}
\end{mathletters}
Since side and diagonal correlations are the two extremal cases on the
lattice, rotation invariance may be verified simply by checking that
\begin{equation}
G_s(x) = G_d(x)
\end{equation}
for all values of $x$.

As long as we are interested only in the scaling behavior, in order to
minimize the effort we can focus on the case $\theta=0$. We can be fairly
general, because in the scaling region
\begin{mathletters}
\label{scaling-mu-nu}
\begin{eqnarray}
\mu_0(\theta) &=& m_0 + O(m_0^3), \\
\nu_0(\theta) &=& m_0 + O(m_0^3),
\end{eqnarray}
\end{mathletters}
independently of the choice of the lattice action.
Therefore
\begin{equation}
\mu^2(\theta) = m_0^2 + {1\over N}\,\Sigma_1(im_0)
+ O(m_0^4) + O\!\left(1\over N^2\right),
\label{mu-scaling}
\end{equation}
and we need only to compute the lattice $O(1/N)$ contribution to the
self-energy.

\subsection{Scaling behavior of the free energy}

Before attacking the above-mentioned problem, let us however notice that,
if we believe standard renormalization-group arguments, we may extract the
scaling behavior on the lattice directly and with lesser effort from the
expression of the lattice free energy.  The subtle point in the evaluation
of the lattice free energy has to do with the existence of a perturbative
tail that obscures the scaling behavior of this physical quantity. However
this turns out to be a tractable problem, because the asymptotic expansion
introduced in Sect.~\ref{asymptotic expansions} allows us to isolate and
remove unambiguously the perturbative tail of the free energy.  In
practice, without belaboring in this point (cf.\
Ref.~\cite{Campostrini-Rossi-condensates} for more details), we must simply
translate Eq.~(\ref{F-cont}) into the lattice language and write, in
analogy with the continuum notation,
\begin{eqnarray}
F^{({\rm scal})} &=& N\Tr\ln \left({\bar p}^2 + m_0^2\right)
    - N\Tr\ln {\bar p}^2 - N \beta m_0^2 \nonumber \\
&+&\, \half\Tr\ln\Delta_{(\alpha)}^{-1}(p) -
	\half\Tr\ln\Delta_{(\alpha)}^{-1}(p)\bigr|_{m_0^2=0} \nonumber \\
&+&\, \half\Tr\ln\Delta_{(\theta)\,\mu\nu}^{-1}(p) -
	\half\Tr\ln\Delta_{(\theta)\,\mu\nu}^{-1}(p)\bigr|_{m_0^2=0},
\label{F-scal}
\end{eqnarray}
where we need to recall that
\begin{equation}
\Delta_{(\theta)\,\mu\nu}^{-1}(p) =
\kappa\Delta^{-1}_{(\theta)\,\mu}(p)\,\delta_{\mu\nu} +
\Delta^{-1}_{(\theta)}(p)\,\delta_{\mu\nu}^t(p)
\end{equation}
and
\begin{equation}
\Tr\ln\Delta_{(\theta)\,\mu\nu}^{-1}(p) =
\intp \ln\det\Delta_{(\theta)\,\mu\nu}^{-1}(p).
\end{equation}
Hence we only need to evaluate
\begin{equation}
\det\Delta_{(\theta)\,\mu\nu}^{-1}(p) =
\kappa\left[\sum_\mu\Delta^{-1}_{(\theta)\,\mu}
   \,{\hat p_\mu^2\over\hat p^2}\,\Delta^{-1}_{(\theta)}(p)
+ \kappa \prod_\mu\Delta^{-1}_{(\theta)\,\mu}(p)\right].
\end{equation}
The $m_0\to0$ limits of the relevant expressions are obtained by the
replacements
\begin{mathletters}
\begin{eqnarray}
\Delta^{-1}_{(\alpha)}(p) &\to& A_0^{(\alpha)}
   + \beta B_0^{(\alpha)}, \\
\Delta^{-1}_{(\theta)}(p) &\to&
    A_0^{(\theta)} + \beta B_0^{(\theta)}, \\
\beta\Delta^{-1}_{(\theta)\,\mu}(p) &\to&
    C_{0\mu} + \beta D_{0\mu}\,.
\end{eqnarray}
\end{mathletters}

Let us now notice that, according to the rules of the asymptotic
expansions, we may evaluate
\begin{eqnarray}
&&N\Tr\ln\left({\bar p}^2 + m_0^2\right)
- N\Tr\ln{\bar p}^2 - N\beta m_0^2 \nonumber \\
=&& N\intp{m_0^2\over{\bar p}^2} - N\beta m_0^2 +
N\intp\left[\ln\!\left(1 + {m_0^2\over p^2}\right)
   - {m_0^2\over p^2}\right]
+ O(m_0^4) \qquad  \nonumber \\
=&& {N\over4\pi}\,m_0^2 + O(m_0^4).
\end{eqnarray}
Moreover, if we treat $m_0^2$ and $\beta$ as formally independent variables
in Eqs.~(\ref{props-expansion}) and (\ref{CD-expansion}), we can introduce
partial derivatives via the relationship
\begin{equation}
{\D\over\D m_0^2} \equiv {\partial\over\partial m_0^2} +
{\D\beta\over\D m_0^2}\,{\partial\over\partial\beta}
= {\partial\over\partial m_0^2} -
\Delta^{-1}_{(\alpha)}(0)\,{\partial\over\partial\beta}\,;
\end{equation}
this allows us to reformulate Eq.~(\ref{F-scal}), obtaining the following
representation of the scaling lattice free energy to $O(1/N)$:
\begin{eqnarray}
F^{({\rm scal})} &=& {N\over4\pi}\,m_0^2
+ {1\over2}\,m_0^2\,{\partial\over\partial m_0^2}
\left[\intp\ln\Delta^{-1}_{(\alpha)}(p) +
\intp\ln\det\Delta_{(\theta)\,\mu\nu}^{-1}(p)
\right]\left.\vphantom{\sum_{\hat p}}\right|_{m_0^2=0}
\nonumber \\
&+&\,\hbox{continuum counterterms}.
\label{F-scal-cont}
\end{eqnarray}
The continuum conterterms in Eq.~(\ref{F-scal-cont}) are to be introduced
according to the usual rules and will generate a contribution akin to the
counterterms used in the sharp-momentum cutoff evaluation of the free
energy. We remind that the continuum conterparts of the relevant
quantities appearing in Eq.~(\ref{F-scal-cont}) are
\begin{mathletters}
\label{continuum-conterparts}
\begin{eqnarray}
\Delta^{-1}_{(\alpha)}(p) &\to&
{1\over2\pi p^2\xi}\ln{\xi+1\over\xi-1} + O(a^2), \\
\Delta^{-1}_{(\theta)}(p) &\to&
{\xi\over2\pi}\ln{\xi+1\over\xi-1} - {1\over\pi} + O(a^2), \\
\Delta^{-1}_{(\theta)\,\mu}(p) &\to& 1 - {1\over4\beta}\intq
{c_1\hat q^2 + 4 c_2\sum_\nu\hat q^2_\nu\cos^2\half q_\nu \over
 c_1\hat q^2 + c_2\sum_\nu\hat q^2_\nu\cos^2\half q_\nu}
\equiv 1 + {C_0\over\beta}\,, \\
C_0 &=& C_{0\mu}\bigr|_{p=0}. \nonumber
\end{eqnarray}
\end{mathletters}
The form of the continuum counterterms is therefore
\begin{equation}
{1\over2}\intp\left[1-{\rm T}_2^{(m_0)}\right]
\left\{\ln\left({1\over2\pi p^2\xi}\ln{\xi+1\over\xi-1}\right)
+ \ln\left[{\xi\over2\pi}\ln{\xi+1\over\xi-1} - {1\over\pi}
+ \kappa\left(1 + {C_0\over\beta}\right)\right]\right\}\!.
\label{F-counterterms}
\end{equation}
We confirm the observation that, in order to get a better agreement with
the corresponding continuum result, it is convenient to redefine $\kappa$
by
\begin{equation}
\kappa = {\kappatilde\over1+C_0/\beta},
\label{kappa-redef}
\end{equation}
where $\kappatilde$ is to be kept constant along the renormalization-group
trajectory. The redefinition (\ref{kappa-redef}) is not going to affect
Eq.~(\ref{F-scal-cont}), because $\kappa$ turns out to be a perturbative
function of $\beta$ and is therefore not affected by the partial derivative
with respect to $m_0^2$. Nonperturbative redefinitions of $\kappa$ are not
allowed; as we shall see, they would spoil the scaling properties of the
physical quantities.

Eq.~(\ref{F-counterterms}) can be analyzed in the light of the results
presented in Sect.~\ref{continuum 1/N}, and we recognize that it can be
rephrased in the form
\begin{eqnarray}
&&{m_0^2\over4\pi}\left[\ln4\pi\beta + (3-2\pi\kappatilde)
\ln(4\pi\beta + 2\pi\kappatilde-2) + c_F(\kappatilde)\right]
\nonumber \\
-&&\,\int_0^{M_L^2} {\D p^2\over4\pi}\,{4\pi\over p^2}
\left[{1\over\ln(p^2/m_0^2)}
+ {3-2\pi\kappatilde\over\ln(p^2/m_0^2)
+ 2\pi\kappatilde - 2}\right],
\label{F-counter-new}
\end{eqnarray}
implying
\begin{equation}
F^{({\rm scal})} = F^{({\rm SM})} + \delta F,
\label{F-delta}
\end{equation}
where
\begin{eqnarray}
\delta F &=& {1\over2}\,m_0^2\,{\partial\over\partial m_0^2}
\left.\left[\intp\ln\Delta^{-1}_{(\alpha)}(p) +
\intp\ln\det\Delta_{(\theta)\,\mu\nu}^{-1}(p)
\right]\right|_{m_0^2=0} \nonumber \\
&-&\,\int_0^{M_L^2} {\D p^2\over4\pi}\,{4\pi\over p^2}
\left[{1\over\ln(p^2/m_0^2)}
+ {3-2\pi\kappatilde\over\ln(p^2/m_0^2)
+ 2\pi\kappatilde - 2}\right]
\end{eqnarray}
and $F^{({\rm SM})}$ is given by Eq.~(\ref{F-ren-cont}) (with $\kappa$
replaced by $\kappatilde$).

\subsection{Scaling behavior of the self-energy}

The $O(1/N)$ contribution to the self-energy is obtained by computing the
Feynman diagrams drawn in Fig.~\ref{Feynman-diagrams-fig}.
Evaluating the diagrams, according to the rules presented in
Fig.~\ref{Feynman-rules-latt}, leads to
\begin{eqnarray}
\Sigma_1(p) &=& \intk
{\Delta_{(\alpha)}(k)\over\overline{p{+}k}^2 + m_0^2}
+ {1\over2}\,\Delta_{(\alpha)}(0) \intk\Delta_{(\alpha)}(k)\,
{\D\over\D m_0^2}\Delta^{-1}_{(\alpha)}(k) \nonumber \\
&+&\,\intk \sum_{\mu\nu}\Delta_{(\theta)\,\mu\nu}(k)
\biggl\{(1+\kappa f)\,\delta_{\mu\nu}V_{4\mu}(p,k)
\vphantom{1\over\left(\overline{p{+}k}^2 + m_0^2\right)}
\nonumber \\
&&\qquad-\;
\exp\left[\half i (k_\mu-k_\nu)\right]
{V_{3\mu}(p{+}\half k,k)\,V_{3\nu}(p{+}\half k,k)\over
   \overline{p{+}k}^2 + m_0^2}
\biggr\} \nonumber \\
&+&\, {1\over2}\Delta_{(\alpha)}(0) \intk \sum_{\mu\nu}
   \Delta_{(\theta)\,\mu\nu}(k)  \nonumber \\ &&\qquad\times\;
{\D\over\D m_0^2}\left[
 \Delta^{-1}_{(\theta)\,\mu\nu}(k) - 2(1+\kappa f)\,
  \delta_{\mu\nu}\intq{V_{4\mu}(q,k)\over\bar q^2 + m_0^2}\right]
\nonumber \\
&+&\, (1+\kappa f)\Delta_{(\alpha)}(0) \intk \sum_{\mu\nu}
 \Delta_{(\theta)\,\mu\nu}(k)\,\delta_{\mu\nu}
  {\D\over\D m_0^2}
  \intq{V_{4\mu}(q,k)\over\bar q^2 + m_0^2} \,.
\label{Sigma1}
\end{eqnarray}
Let us now apply the decomposition of the total derivative with respect
to the mass into partial derivatives, perform some simplifications, and
rearrange the terms into the form
\begin{eqnarray}
\Sigma_1(p) &=& \intk\Delta_{(\alpha)}(k)
\left[{1\over\overline{p{+}k}^2 + m_0^2} - {1\over2}
   \,{\partial\over\partial\beta}\Delta^{-1}_{(\alpha)}(k)\right]
\nonumber \\
&+&\, {1\over2}\,\Delta_{(\alpha)}(0) \intk {\partial\over\partial m_0^2}
  \ln\Delta^{-1}_{(\alpha)}(k) \nonumber \\
&+&\,\intk \sum_{\mu\nu}\Delta_{(\theta)\,\mu\nu}(k)
\biggl\{\delta_{\mu\nu}V_{4\mu}(p,k)
\nonumber \\
&&\qquad-\;
\exp\left[\half i (k_\mu-k_\nu)\right]
{V_{3\mu}(p{+}\half k,k)\,V_{3\nu}(p{+}\half k,k)\over
   \overline{p{+}k}^2 + m_0^2}
- {1\over2}\,{\partial\over\partial\beta}
 \Delta^{-1}_{(\theta)}(k)\,\delta_{\mu\nu}^t(k)\biggr\} \nonumber \\
&+&\, {1\over2}\Delta_{(\alpha)}(0) \intk \left[
  {\partial\over\partial m_0^2}\ln\det\Delta^{-1}_{(\theta)\,\mu\nu}(k)
  - \sum_{\mu\nu}
   \Delta_{(\theta)\,\mu\nu}(k)\,\delta_{\mu\nu}\Delta^{-1}_{\mu}(k)
    {\partial\kappa\over\partial m_0^2}\right] \nonumber \\
&+&\, \kappa f \intk \sum_{\mu\nu}
 \Delta_{(\theta)\,\mu\nu}(k)\,\delta_{\mu\nu}
  \left[V_{4\mu}(p,k) - {\partial\over\partial\beta}
  \left(\beta\Delta^{-1}_{\mu}(k)\right)\right].
\label{Sigma-interm}
\end{eqnarray}
Eqs.~(\ref{gauge-lattice}) imply the following relationship:
\begin{eqnarray}
&&\sum_\mu \exp\left(-\half i k_\mu\right) \hat k_\mu
\left\{\delta_{\mu\nu}V_{4\mu}(p,k) -
\exp\left[\half i (k_\mu-k_\nu)\right]
{V_{3\mu}(p{+}\half k,k) V_{3\nu}(p{+}\half k,k)\over
\overline{p{+}k}^2 + m_0^2}\right\} \nonumber \\
=&&\, \exp\left(-\half i k_\nu\right)\left[
{\bar p^2 + m_0^2 \over \overline{p{+}k}^2 + m_0^2}\,
V_{3\nu}(p{+}\half k,k) - {1\over2}\left(
V_{3\nu}(p{+}\half k,k) + V_{3\nu}(p{-}\half k,k)\right)\right].
\nonumber \\
\end{eqnarray}
Eliminating terms vanishing under symmetric $k$-integration, we can
perform the replacement
\begin{eqnarray}
&&\delta_{\mu\nu}V_{4\mu}(p,k) -
\exp\left[\half i (k_\mu-k_\nu)\right]
{V_{3\mu}(p{+}\half k,k) V_{3\nu}(p{+}\half k,k)\over
\overline{p{+}k}^2 + m_0^2} \nonumber \\
\to&&\, W(p,k)\,\delta_{\mu\nu}^t(k) +
{\bar p^2 + m_0^2 \over \overline{p{+}k}^2 + m_0^2} \left[
{V_{3\mu}(p{+}\half k,k)\over\hat k_\mu}\,\delta_{\mu\nu}
- \sum_\rho{V_{3\rho}(p{+}\half k,k)\over\hat k_\rho}
   \,\delta_{\mu\nu}^t(k)
\right], \nonumber \\
\end{eqnarray}
where
\begin{equation}
W(p,k) = \sum_\mu V_{4\mu}(p,k) -
\sum_\mu{\left[V_{3\mu}(p{+}\half k,k)\right]^2\over
\overline{p{+}k}^2 + m_0^2} \,.
\end{equation}

In order to evaluate the function $\Sigma_1(p)$ in the scaling region, we
must keep in mind that the formalism introduced in Sect.~\ref{asymptotic
expansions} allows for a direct expansion of the lattice integrals in terms
of the ``external'' variables $m_0$ and $p$, since all the nonanaliticity
is accumulated in the continuum integrals, which in turn have already been
computed in the sharp-momentum regularization scheme.  We now realize that,
as far as the lattice contribution is concerned, we can make explicit use
of the following relationship, holding for any regular (analytic) function
$f$:
\begin{equation}
f(k;p,m_0) \cong f(k;0,0) + p^2\,{\partial\over\partial p^2}f(k;0,0)
+ m_0^2\,{\partial\over\partial m_0^2}f(k;0,0).
\label{f-Taylor}
\end{equation}
In particular, when imposing the on-shell condition $p^2+m_0^2=0$, we
find
\begin{equation}
f(k;im_0,m_0) \cong f(k;0,0) - m_0^2\,{\partial\over\partial p^2}f(k;0,0)
+ m_0^2\,{\partial\over\partial m_0^2}f(k;0,0),
\label{f-onshell}
\end{equation}
and, as a consequence,
\begin{equation}
f(k;p,m_0) \cong f(k;im_0,m_0) + (p^2+m_0^2)\,
{\partial\over\partial p^2}f(k;0,0).
\label{f-off-shell}
\end{equation}
Applying Eq.~(\ref{f-off-shell}) to $\Sigma_1(p)$, we come to the
conclusion that
\begin{equation}
\Sigma_1(p,m_0) \cong \Sigma_1^{({\rm SM})} + \delta m_1^2
- (p^2+m_0^2)\,\delta Z_1 \,,
\label{Sigma-off-shell}
\end{equation}
where $\Sigma_1^{({\rm SM})}$ is the value of the self-energy in the
sharp-momentum cutoff regularization scheme, and $\delta m_1^2$, $\delta
Z_1$ are the $O(1/N)$ lattice contributions to mass and wavefunction
renormalization, and are amenable to finite lattice integrals whose
infrared regularization is provided by the appropriate continuum
counterterms.

\subsection{Lattice contribution to mass renormalization}

The $O(1/N)$ contribution to the mass, (for arbitrary values of the
coupling constant), can be obtained from Eqs.~(\ref{mu-expansion}) and
(\ref{Sigma-interm}):
\begin{eqnarray}
\mu_1^2 &\equiv& {\mu_0\over\nu_0}\,\Sigma_1(i\mu_0) =
{\mu_0\over\nu_0} \Biggl\{ \intk\Delta_{(\alpha)}(k) \left[
{1\over\overline{k{+}i\mu_0}^2 + m_0^2} + {1\over2}\,\Delta_{(\alpha)}(0)\,
{\D\over\D m_0^2}\Delta^{-1}_{(\alpha)}(k)\right] \nonumber \\
&&\quad+\;\intk \sum_{\mu\nu}
  \Delta_{(\theta)\,\mu\nu}(k)\,\delta_{\mu\nu}^t(k) \left[
W(i\mu_0,k) + {1\over2}\,\Delta_{(\alpha)}(0)\,
{\D\over\D m_0^2}\Delta^{-1}_{(\theta)}(k)\right] \nonumber \\
&&\quad+\;\kappa f\intk \sum_{\mu\nu}
   \Delta_{(\theta)\,\mu\nu}(k)\,\delta_{\mu\nu}
\left[V_{4\mu}(i\mu_0,k) +\Delta_{(\alpha)}(0)\,{\D\over\D m_0^2}
\left(\beta\Delta^{-1}_{(\theta)\,\mu}(k)\right)\right] \Biggr\}
\nonumber \\
&=&{\mu_0\over\nu_0} \Biggl\{ \intk\Delta_{(\alpha)}(k) \left[
{1\over\overline{k{+}i\mu_0}^2 + m_0^2} - {1\over2}\,
{\partial\over\partial\beta}\Delta^{-1}_{(\alpha)}(k)\right] \nonumber \\
&&\quad+\;\intk \sum_{\mu\nu}
  \Delta_{(\theta)\,\mu\nu}(k)\,\delta_{\mu\nu}^t(k) \left[
W(i\mu_0,k) - {1\over2}\, {\partial\over\partial\beta}
\Delta^{-1}_{(\theta)}(k)\right] \nonumber \\
&&\quad+\;\kappa f\intk \sum_{\mu\nu}
   \Delta_{(\theta)\,\mu\nu}(k)\,\delta_{\mu\nu}
\left[V_{4\mu}(i\mu_0,k) - {\partial\over\partial\beta}
\left(\beta\Delta^{-1}_{(\theta)\,\mu}(k)\right)\right] \nonumber \\
&&\quad+\;{1\over2}\,\Delta_{(\alpha)}(0)\intk
{\partial\over\partial m_0^2}\left[\ln\Delta^{-1}_{(\alpha)}(k)
+ \ln\det\Delta^{-1}_{(\theta)\,\mu\nu}(k)\right] \nonumber \\
&&\quad-\;{1\over2}\,\Delta_{(\alpha)}(0)\intk \sum_{\mu\nu}
\Delta_{(\theta)\,\mu\nu}(k)\,\delta_{\mu\nu}
\Delta^{-1}_{(\theta)\,\mu}(k)\,
{\partial\kappa\over\partial m_0^2} \Biggr\} .
\label{m1-latt}
\end{eqnarray}

In order to evaluate $\delta m_1^2$, let us first complete the lattice
computation of $m_1^2$ in the scaling region from Eq.~(\ref{mu-scaling}).
{}From the definition of the asymptotic expansion presented in
Sect.~\ref{asymptotic expansions} it is easy to check that, under symmetric
$k$-integration, the following replacements are allowed:
\begin{mathletters}
\begin{eqnarray}
{1\over\overline{k{+}i\mu_0}^2 + m_0^2} &\to& {1\over2}\left[
B_0^{(\alpha)}(k) + m_0^2 B_1^{(\alpha)}(k)\right] + O(m_0^4)
= {1\over2}\,{\partial\over\partial\beta}\Delta^{-1}_{(\alpha)}(k)
+ O(m_0^4), \nonumber \\ \\
W(i \mu_0,k) &\to& {1\over2}\,{\partial\over\partial\beta}
\Delta^{-1}_{(\theta)}(k) + O(m_0^4), \\
V_{4\mu}(i \mu_0,k) &\to& {\partial\over\partial\beta}
\left(\beta\Delta^{-1}_{(\theta)\,\mu}(k)\right) + O(m_0^4).
\end{eqnarray}
\end{mathletters}
We make now the crucial assumption
\begin{equation}
{\partial\kappa\over\partial m_0^2} = 0,
\label{RG-lattice}
\end{equation}
which corresponds to our choice of renormalization-group trajectories and
defines the class of theories we are going to generate in the continuum
limit. The renormalization-group flow trajectories defined by
Eq.~(\ref{kappa-redef}), consistently with Eq.~(\ref{RG-lattice}), are
plotted in Fig.~\ref{RG-lattice-fig} (for $c_2=0$).  Under such assumption
and making use of the identity
\begin{equation}
\Delta_{(\alpha)}(0) = 4\pi m_0^2 + O(m_0^4),
\end{equation}
we can conclude that
\begin{eqnarray}
\Sigma_1(i m_0) &=& 4\pi m_0^2\times
{1\over2}\left.\intk{\partial\over\partial m_0^2}\left[
\ln\Delta^{-1}_{(\alpha)}(k) + \ln\det\Delta^{-1}_{(\theta)\,\mu\nu}(k)
\right]\right|_{m_0^2=0} \nonumber \\
&&\qquad+\;O(m_0^4) + \hbox{continuum counterterms}.
\end{eqnarray}
In order to write down explicitly the continuum counterterms appropriate
to $\Sigma_1(im_0)$, it is convenient to go back to Eq.~(\ref{Sigma1})
and perform the replacements indicated in
Eqs.~(\ref{continuum-conterparts}), supplemented by the substitutions
\begin{mathletters}
\label{substitutions}
\begin{eqnarray}
{1\over\overline{k{+}im_0}^2 + m_0^2} &\to& {1\over k^2\xi}\,, \\
W(im_0,k) &\to& \xi, \\
V_{4\mu}(im_0,k) &\to& 1.
\end{eqnarray}
\end{mathletters}
The resulting expression for the continuum conterterms is then
\begin{equation}
\intk\left[1-{\rm T}_2^{(m_0)}\right]\left\{
{2\pi\over\displaystyle\ln{\xi+1\over\xi-1}}\left(1-{1\over\xi}\right) +
{2\pi\over\displaystyle\xi\ln{\xi+1\over\xi-1}-2+2\pi\kappatilde}
\left[\xi - {1\over\xi^2}
   + \pi\kappatilde\left({1\over\xi^2}-1\right) \right]\right\}.
\label{Sigma1-counterterms}
\end{equation}
Let us notice that Eq.~(\ref{Sigma1-counterterms}) was derived from
Eq.~(\ref{Sigma1}) without making any assumption on the dependence of
$\kappa$ on $\beta$.

In the light of the results of Sect.~\ref{continuum 1/N},
Eq.~(\ref{Sigma1-counterterms}) turns into
\begin{eqnarray}
&&{m_0^2\over4\pi}\left[\ln4\pi\beta + (3-2\pi\kappatilde)
\ln(4\pi\beta + 2\pi\kappatilde-2) + c_m(\kappatilde)\right]
\nonumber \\
-&&\,\int_0^{M_L^2} {\D k^2\over4\pi}\,{4\pi\over k^2}
\left[{1\over\ln(k^2/m_0^2)} + {3-2\pi\kappatilde\over
   \ln(k^2/m_0^2) + 2\pi\kappatilde - 2}\right],
\label{m-counter-new}
\end{eqnarray}
implying
\begin{equation}
m_1^2 = m_1^{2\,({\rm SM})} + \delta m_1^2,
\label{m-delta}
\end{equation}
where
\begin{equation}
\delta m_1^2 = 4\pi\,\delta F
\label{delta-F-m}
\end{equation}
and $m_1^{({\rm SM})}$ is given by Eq.~(\ref{m-Z-cont}a) (with $\kappa$
replaced by $\kappatilde$).

In conclusion, we can summarize the results contained in
Eqs.~(\ref{F-delta}) and (\ref{delta-F-m}) into the relationship
\begin{equation}
F^{({\rm scal})} = \left[N - c_m(\kappatilde)
   + c_F(\kappatilde)\right]
{m^2\over4\pi} + O\!\left(1\over N\right) + O(m_0^4).
\label{F/m-latt}
\end{equation}
Eq.~(\ref{F/m-latt}) is one of the main results of the present Section.
It shows that, performing a proper {\it perturbative\/} redefinition of
the parameter $\kappa$, it is possible to map the lattice theory in the
scaling region into the corresponding continuum theory in such a way that
the fundamental scaling relationships are preserved. Eq.~(\ref{F/m-latt})
is identical in form to the corresponding continuum relationship
(\ref{F/m-cont}). Let us notice once more that nonperturbative
redefinitions of $\kappa$ would spoil the scaling relationship, because
they would change $F^{({\rm scal})}$ without affecting $\Sigma_1$ (the
partial derivative $\partial\kappa/\partial m_0^2$ cancels in
Eq.~(\ref{m1-latt}) with the corresponding term coming from
$\partial\Delta^{-1}_{(\theta)\,\mu\nu}/\partial m_0^2$).

Recalling the explicit form of the asymptotic expansions presented in
Sect.~\ref{asymptotic expansions}, we can now construct the following
representation of the $O(1/N)$ contribution to the physical mass gap in
the scaling region:
\begin{eqnarray}
\delta m_1^2 &=& 2\pi m_0^2
  \left\{\intk {A_1^{(\alpha)} + \beta B_1^{(\alpha)}
  \over A_0^{(\alpha)} + \beta B_0^{(\alpha)}} \right. \nonumber \\
&&\quad+\; \intk \left[
  \sum_\mu\left(C_{0\mu} + \beta D_{0\mu}\right)
  {\hat k^2_\mu\over\hat k^2}
  \left(A_1^{(\theta)} + \beta B_1^{(\theta)}\right) \right. \nonumber \\
&&\quad\qquad+\;\left.\sum_\mu\left(C_{1\mu} + \beta D_{1\mu}\right)
  {\hat k^2_\mu\over\hat k^2}
  \left(A_0^{(\theta)} + \beta B_0^{(\theta)}\right)\right. \nonumber \\
&&\quad\qquad+\;\left.\Bigl[
  (C_{01} + \beta D_{01}) (C_{12} + \beta D_{12}) +
  (C_{02} + \beta D_{02}) (C_{11} + \beta D_{11})\Bigr]
  {\kappatilde\over C_0+\beta}
  \vphantom{\hat k^2_\mu\over\hat k^2}\right]\nonumber \\
&&\qquad\times\left[\sum_\mu\left(C_{0\mu} + \beta D_{0\mu}\right)
  {\hat k^2_\mu\over\hat k^2}
  \left(A_0^{(\theta)} + \beta B_0^{(\theta)}\right)
  + {\kappatilde\over C_0 + \beta}
  \prod_\mu\left(C_{0\mu} + \beta D_{0\mu}\right) \right]^{-1}
\nonumber \\
&&\quad-\; \left.\int_0^{M^2_L}{\D k^2\over4\pi}\,{2\over k^2}
  \left[{1\over\ln\left(k^2/m_0^2\right)} +
  {3-2\pi\kappatilde\over
     \ln\left(k^2/m_0^2\right)+2\pi\kappatilde-2}
  \right]\right\} .
\label{m1-scaling}
\end{eqnarray}
The integral representation is improper, since the denominators vanish on
certain curves; the generalized principal-part prescription giving a
precise meaning to Eq.~(\ref{m1-scaling}) will be discussed in
Subsect.~\ref{lattice integrals} (cf.\
Ref.~\cite{Campostrini-Rossi-condensates}).

A rather straightforward consequence of Eqs.~(\ref{f-Taylor}) and
(\ref{f-onshell}) is the relationship
\begin{equation}
f(k;im_0,m_0) \cong f(k;0,m_0) -
m_0^2\,{\partial\over\partial p^2}f(k;0,m_0) + O(m_0^4),
\end{equation}
implying that the lattice contribution to $\Sigma_1(im_0)$ and to
$\Sigma_1(0) - m_0^2\Sigma_1'(0)$ have the same scaling limit. As a
consequence we should have obtained the same expression for $\delta m_1^2$
if we had computed the lattice counterpart of $m_R^2$ in the scaling
region.

We are now ready to apply our knowledge of $\delta m_1^2$ to the problem of
finding an explicit expression for any physical quantity with known scaling
properties, when we analyze it on the lattice in the scaling regime.  Let
us assume $Q$ to be a physical quantity with canonical scaling dimension
$2\delta_Q$ in mass units (i.e.\ no anomalous dimension), and let us
compute its large $N$ limit and $O(1/N)$ corrections:
\begin{equation}
Q = \left(m_0^2\right)^{\delta_Q}\left[q_0 + {1\over N}q_1(\beta)
+ O\!\left(1\over N^2\right)\right],
\end{equation}
evaluating the $O(1/N)$ corrections in the SM regularization scheme.
Since $q_0$ is universal and $\beta$-independent, we can immediately
predict the lattice scaling behavior of $Q$ to be
\begin{equation}
Q^{(L)} = \left(m_0^2\right)^{\delta_Q}\left[q_0 + {1\over N}q_1(\beta)
 + {1\over N}\delta_Q q_0\,{\delta m_1^2(\beta)\over m_0^2}
+ O\!\left(1\over N^2\right)\right]
+ O\!\left(\left(m_0^2\right)^{\delta_Q+2}\right).
\label{Q-scaling}
\end{equation}
Eq.~(\ref{Q-scaling}) realizes the promise of evaluating every physical
quantity in the scaling regime by means of a single lattice computation.

\subsection{Lattice contribution to wavefunction renormalization}

The discussion of the wavefunction renormalization goes along the same
lines. We recall from Eq.~(\ref{f-off-shell}) and (\ref{Sigma-off-shell})
that
\begin{equation}
\delta Z_1 = -{\partial\over\partial p^2}\Sigma_1(p,m_0)
\Bigr|_{\stackrel{\scriptstyle p=0}{\scriptstyle m_0=0}}
\end{equation}
{}From Eq.~(\ref{Sigma-interm}) we obtain
\begin{eqnarray}
-\delta Z_1 &=& \intk\Delta_{(\alpha)}(k)\,{\partial\over\partial p^2}
\,\left.{1\over\overline{p{+}k}^2}\right|_{p=0} +
\intk\Delta_{(\theta)\,\mu\nu}(k)\,\delta_{\mu\nu}^t(k)
\,{\partial\over\partial p^2}W(p,k)\Bigr|_{p=0} \nonumber \\
&+&\,\intk\Delta_{(\theta)\,\mu\nu}(k)\,{1\over\bar k^2}
\left[D_{0\mu}(k)\,\delta_{\mu\nu} -
\sum_\rho D_{0\rho}(k)\,\delta_{\mu\nu}^t(k) \right] \nonumber \\
&-&\,\kappa f \intk\Delta_{(\theta)\,\mu\nu}(k)\,\delta_{\mu\nu}\,
D_{1\mu}(k) + \hbox{continuum conterterms},
\end{eqnarray}
where we made use of the relationships
\begin{mathletters}
\label{V4-p0}
\begin{eqnarray}
&&V_{4\mu}(0,k) = {V_{3\mu}(\half k,k)\over\hat k_\mu}
= D_{0\mu}(k)\,, \\
&&{\partial\over\partial p^2}V_{4\mu}(p,k)\Bigr|_{p=0}
= -D_{1\mu}(k)\,.
\end{eqnarray}
\end{mathletters}
Let us further notice that
\begin{mathletters}
\begin{eqnarray}
-{\partial\over\partial p^2}
   \left.{1\over\overline{p{+}k}^2}\right|_{p=0}
&=& {1\over2}B_1^{(\alpha)}(k) + {1\over\bigl(\bar k^2\bigr)^2}\,, \\
-{\partial\over\partial p^2}W(p,k)\Bigr|_{p=0}
&=& {1\over2}B_1^{(\theta)}(k) - {1\over\bigl(\bar k^2\bigr)^2}
\sum_\mu\hat k_\mu^2[D_{0\mu}(k)]^2.
\end{eqnarray}
\end{mathletters}
We are finally ready to write down the complete explicit expression for
$\delta Z_1$:
\begin{eqnarray}
\delta Z_1 &=& \intk {1\over A_0^{(\alpha)} + \beta B_0^{(\alpha)}}
\left[{1\over2}B_1^{(\alpha)} + {1\over\bigl(\bar k^2\bigr)^2}\right]
\nonumber \\ &+&\, \intk
\left[\sum_\mu(C_{0\mu}+\beta D_{0\mu})
{\hat k_\mu^2 \over \hat k^2}(A_0^{(\theta)} + \beta B_0^{(\theta)}) +
{\kappatilde\over\beta+C_0}
\prod_\mu(C_{0\mu}+\beta D_{0\mu})\right]^{-1}
\nonumber \\ &&\qquad\times\left\{\sum_\mu(C_{0\mu}+\beta D_{0\mu})
{\hat k_\mu^2 \over \hat k^2} \left[{1\over2} B_1^{(\theta)} -
{1\over\bigl(\bar k^2\bigr)^2}\sum_\mu\hat k_\mu^2[D_{0\mu}(k)]^2\right]
\right.\nonumber \\ &&\qquad\qquad -\;{1\over\bar k^2}\left[
(C_{01} + \beta D_{01})D_{02} + (C_{02} + \beta D_{02})D_{01}
- \sum_{\mu\nu}(C_{0\mu} + \beta D_{0\mu})
{\hat k_\mu^2 \over \hat k^2} D_{0\nu}\right] \nonumber \\
&&\qquad\qquad -\;{\beta+C_0\over\kappatilde}\,{1\over\bar k^2}
\sum_\mu D_{0\mu}{\hat k_\mu^2 \over \hat k^2}
(A_0^{(\theta)} + \beta B_0^{(\theta)}) +
{1\over2} (A_0^{(\theta)} + \beta B_0^{(\theta)})
\sum_\mu D_{1\mu}{\hat k_\mu^2 \over \hat k^2} \nonumber \\
&&\qquad\qquad +\;\left.{1\over2}\,{\kappatilde\over\beta+C_0}
\left[(C_{01} + \beta D_{01})D_{12} + (C_{02} + \beta D_{02})D_{11}\right]
\vphantom{\sum_\mu}\right\}
\nonumber \\&+&\, \int_0^{M^2_L}{\D k^2\over4\pi}
  \left[{2\pi\over\ln\left(k^2/m_0^2\right)} -
  {4\pi\over\ln\left(k^2/m_0^2\right)+2\pi\kappatilde-2}
  + {1\over\kappatilde}\right]{1\over k^2} \,,
\label{Z1-scaling}
\end{eqnarray}
and we may verify that
\begin{equation}
\lim_{\beta\to\infty}\delta Z_1 = 0.
\end{equation}

\subsection{$\pbbox{\CPN}$ and $\pbbox{{\rm O}(2N)}$ models}

At the borders of the parameter space, $\CPN$ ($\beta_{\rm v}=0$) and ${\rm
O}(2N)$ ($\beta_{\rm g}=0$) models require a separate discussion, because
of some subtleties related to the order of the limiting procedures in the
lattice formulation.

We know that, because of gauge invariance and confinement,
\begin{eqnarray}
\lim_{\kappatilde\to0} m_1^2 = \infty.
\end{eqnarray}
However, the quantity $\delta m_1^2$ stays finite, and this is important
because of its renormalization-group interpretation: it allows us to
define a renormalization-group-invariant, $1/N$-expandable scale in
lattice $\CPN$ models. When computing $\delta m_1^2$ for $\CPN$ models,
it is necessary to keep in mind the transversality of the vector
propagator, due to gauge invariance. As a consequence, one may extract
directly from Eq.~(\ref{m1-latt}) the correct expression
\begin{eqnarray}
\delta m_1^2\bigr|_{\beta_{\rm v}=0} &=&
2\pi m_0^2 \left\{\intk \left[
   {A_1^{(\alpha)} + \beta B_1^{(\alpha)} \over
    A_0^{(\alpha)} + \beta B_0^{(\alpha)}} +
   {A_1^{(\theta)} + \beta B_1^{(\theta)} \over
    A_0^{(\theta)} + \beta B_0^{(\theta)}}
\right]\right. \nonumber \\
&&\qquad-\;  \left.\int_0^{M^2_L}{\D k^2\over4\pi}\,{2\over k^2}
  \left[{1\over\ln\left(k^2/m_0^2\right)} +
  {3\over\ln\left(k^2/m_0^2\right)-2}\right]
   \vphantom{B_1^{(\alpha)}\over B_0^{(\alpha)}}\right\}.
\label{m1-scaling-CPN}
\end{eqnarray}
Eq.~(\ref{m1-scaling-CPN}) differs from the $\kappatilde\to0$ limit
of Eq.~(\ref{m1-scaling}):
\begin{eqnarray}
\lim_{\kappatilde\to0} \delta m_1^2 &=& 2\pi m_0^2 \Biggl\{\intk
\left[{A_1^{(\alpha)} + \beta B_1^{(\alpha)}\over
      A_0^{(\alpha)} + \beta B_0^{(\alpha)}} +
     {A_1^{(\theta)} + \beta B_1^{(\theta)}\over
      A_0^{(\theta)} + \beta B_0^{(\theta)}} +
    {\sum_\mu (C_{1\mu}+\beta D_{1\mu})\hat k_\mu^2 \over
     \sum_\mu (C_{0\mu}+\beta D_{0\mu})\hat k_\mu^2}\right] \nonumber \\
&&\qquad-\,\int_0^{M^2_L}{\D k^2\over4\pi}\,
  {2\over k^2} \left[{1\over\ln\left(k^2/m_0^2\right)} +
    {3\over\ln\left(k^2/m_0^2\right)-2}\right]\Biggr\}.
\label{m1-scaling-kappa0}
\end{eqnarray}

On the other side, when we consider ${\rm O}(2N)$ models we find from
Eq.~(\ref{m1-latt})
\begin{mathletters}
\label{m-Z-scaling-ON}
\begin{equation}
\delta m_1^2\bigr|_{\beta_{\rm g}=0} = 2\pi m_0^2  \left\{\intk
{A_1^{(\alpha)} + \beta B_1^{(\alpha)}\over
 A_0^{(\alpha)} + \beta B_0^{(\alpha)}}
 - \int_0^{M_L^2} {\D k^2\over4\pi}\,
{2\over k^2}\left[{1\over\ln\left(k^2/m_0^2\right)} - 1\right]\right\},
\end{equation}
and
\begin{equation}
\delta Z_1\bigr|_{\beta_{\rm g}=0} =
\intk{1\over A_0^{(\alpha)} + \beta B_0^{(\alpha)}}
\left[{1\over2}\,B_1^{(\alpha)} + {1\over\left(\bar k^2\right)^2}\right]
+ \int_0^{M_L^2} {\D k^2\over4\pi}\,
{1\over k^2}\,{2\pi\over\ln\left(k^2/m_0^2\right)}\,.
\end{equation}
\end{mathletters}
Eqs.~(\ref{m-Z-scaling-ON}) ought to be compared with the
$\kappatilde\to\infty$ limit of Eqs.~(\ref{m1-scaling}) and
(\ref{Z1-scaling}):
\begin{mathletters}
\label{m-Z-scaling-kappainf}
\begin{eqnarray}
\lim_{\kappatilde\to\infty} \delta m_1^2 &=&
2\pi m_0^2\Biggl\{\intk{A_1^{(\alpha)} + \beta B_1^{(\alpha)}\over
 A_0^{(\alpha)} + \beta B_0^{(\alpha)}}
  - \int_0^{M_L^2} {\D k^2\over4\pi}\,
{2\over k^2}\left[{1\over\ln\left(k^2/m_0^2\right)} - 1\right]
\nonumber \\ &&\qquad+\, \intk\sum_\mu
{C_{1\mu} + \beta D_{1\mu} \over C_{0\mu} + \beta D_{0\mu}}\Biggr\}
\end{eqnarray}
and
\begin{eqnarray}
\lim_{\kappatilde\to\infty} \delta Z_1 &=&
\intk{1\over A_0^{(\alpha)} + \beta B_0^{(\alpha)}}
\left[{1\over2} B_1^{(\alpha)} + {1\over\bigl(\bar k^2\bigr)^2}\right]
+ \int_0^{M_L^2} {\D k^2\over4\pi}\,
{1\over k^2}\,{2\pi\over\ln\left(k^2/m_0^2\right)}\nonumber \\
&+&\,{1\over2} \intk\sum_\mu
{D_{1\mu} \over C_{0\mu} + \beta D_{0\mu}}\,.
\end{eqnarray}
\end{mathletters}
We may observe that Eqs.~(\ref{m1-scaling-CPN}) and (\ref{m-Z-scaling-ON})
differ from Eqs.~(\ref{m1-scaling-kappa0}) and
(\ref{m-Z-scaling-kappainf}) respectively because of a term depending on
$C_{1\mu}$ and $D_{1\mu}$.  Therefore this difference would be completely
absent in the SM and in all other continuum regularization schemes.  On the
lattice these contributions are nonvanishing; however they are perturbative
in $1/\beta$, and in particular when $\kappatilde\to\infty$ they do
not affect the scaling relationship (\ref{F/m-latt}), where
\begin{equation}
\lim_{\kappatilde\to\infty}
\left[c_m(\kappatilde) -  c_F(\kappatilde)\right] = 1.
\end{equation}
Therefore in the scaling region they are amenable to a perturbative
redefinition of the coupling $f$, i.e.\ to a different regularization
scheme of the same theory.  We shall come back to this phenomenon in
Sect.~\ref{Lambda parameter}, where we shall explicitly discuss the
$\beta\to\infty$ limit of the above expressions.

\subsection{Evaluation of lattice integrals}
\label{lattice integrals}

The numerical evaluation of Eqs.~(\ref{m1-scaling}) and (\ref{Z1-scaling})
is a fairly nontrivial task.  The general structure of $\delta m_1^2$ and
$\delta Z_1$ is that of a difference between an infrared-singular lattice
integral and an ultraviolet-cutoffed continuum integral, with the same
singular infrared behavior as the lattice integral.

By replacing $\ln(k^2/m_0^2)$ with $4\pi\beta + \ln(k^2/M_L^2)$, it is
possible to perform the expansion of $\delta m_1^2$ in a power series in
the powers of $f = 1/(2\beta)$. The coefficients of this weak-coupling
expansion are individually infrared-regular combinations of lattice and
continuum integrals.  It is easy to generate the numerical values of the
coefficients up to high orders, simply by expanding the integrands and
integrating term by term.  However the full expressions themselves are not
proper integrals, reflecting the fact that the series are not
Borel-summable.  As a consequence, the approximation by series expansion
will fail for sufficiently large values of $f$.

In practice, we found that numerical predictions from series evaluation
become unstable (with respect to different truncations) at the border of
the scaling region.  We don't mean that perturbative evaluations be
intrinsically meaningless; we only stress that extrapolation to the
intermediate coupling may be dangerous and need to be carefully tested
against instabilities.  In particular, disagreement between Monte Carlo
results and low orders of perturbation theory cannot be automatically
interpreted as absence of scaling and/or failure of field theoretical
predictions.

In any case, in the models we are discussing an explicit way out of these
difficulties can be found, since a generalized principal-part prescription
allows an unambiguous evaluation and resummation of (the scaling part
of) the series (cf.\ Ref.~\cite{Campostrini-Rossi-condensates}).  Our
prescription is expressed as follows:
\begin{enumerate}
\item
represent the integrals as sums of individual terms of the form
\begin{equation}
\intk {M(k)\over\left[\beta+N(k)\right]^n}\,;
\end{equation}
\item
define complex variable functions
\begin{equation}
C(z) = \int_0^\infty{\D z'\over z-z'}\intk M(k)\,\delta(z'+N(k));
\end{equation}
\item
identify
\begin{equation}
\intk {M(k)\over\left[\beta+N(k)\right]^n} \equiv
{(-1)^{n-1}\over(n-1)!}\,
{\D^{n-1}\over\D \beta^{n-1}}\lim_{\varepsilon\to0}
\left[C(\beta+i\varepsilon)+C(\beta-i\varepsilon)\over2\right].
\label{princ-prescription}
\end{equation}
\end{enumerate}

{}From a numerical point of view, a direct principal-part integration is
possible but very
unstable, since the line of vanishing denominator of the integrands has to
be found numerically, and moreover it shrinks rapidly towards zero in the
limit $f\to0$.  The solution is to adopt the method described in Ref.\
\cite{Campostrini-Rossi-condensates}: the integral is split into the two
regions $k^2<\rho^2$ and $k^2>\rho^2$, choosing $\rho$ such that all the
singularities are included into the first region. The integral over the
second region is perfectly regular and it is easily computed numerically.
In the first region, we expand the integrand in a power series in $k$ and
integrate analytically term by term, resulting in the exponential-integral
and related functions.  The typical analytic integration involved in the
abovementioned procedure has the form
\begin{equation}
\princint_0^y\D x\,{x^n\over(\ln x)^{s+1}} =
{(n+1)^s\over s!}\,y^{n+1}\,{\scr E}_s\left((n+1)\ln y\right),
\end{equation}
where
\begin{equation}
{\scr E}_s(z) = e^{-z}\Ei(z) - \sum_{t=0}^{s-1} {t!\over z^{t+1}} \sim
\sum_{t=s}^\infty {t!\over z^{t+1}} \,,
\end{equation}
and $\Ei$ is the standard exponential-integral function.  Tuning $\rho$ as
a function of $f$, the $n$th order of this expansion approaches the exact
result with an error decreasing as fast as $\exp[-4\pi(n+1)\beta]$.  The
series expansion is shown numerically to approximate extremely well the
integrand in the relevant region.

It is worth noticing that the resummation procedure we have sketched
turns out to be useful also in the evaluation of the so-called perturbative
tails of composite operators, i.e.\ those non-scaling contributions to
vacuum expectation values resulting from spin waves that must be explicitly
subtracted in order to derive normal-ordered quantum expectation values
\cite{Campostrini-Rossi-condensates}.

We may notice that the ambiguity in $\lim_{z\to\beta}C(z)$ is related to
the Borel ambiguity in the resummation of the perturbative series observed
by David \cite{David-ambiguity,David-OPE}.  However the coefficients of the
asymptotic expansion in powers of $m_0^2$ of a physical $1/N$-expandable
quantity like the mass gap are real functions, whose expansion in powers of
the coupling is the standard perturbative series.

\section{Evaluation of physical quantities for nearest-neighbor
interactions.}
\label{physical standard}

As already discussed at the end of Sect.~\ref{integral representations},
the task of performing accurate numerical evaluations of the integrals
entering the $O(1/N)$ contributions to physical quantities is greatly
simplified in the case of nearest-neighbor interactions $c_2=0$, where we
possess integral representations of the lattice propagators.

Notable simplifications occur in Eq.~(\ref{m1-latt}) in this special case:
the exact $O(1/N)$ contribution to the mass is
\begin{eqnarray}
\mu_1^2 = {\mu_0\over\nu_0} \Biggl\{ &&\intk
\Delta_{(\alpha)}(k)\left[{1\over\widehat{k{+}i\mu_0}^2 + m_0^2} +
  {1\over2}\Delta_{(\alpha)}(0)\,
{\D\Delta^{-1}_{(\alpha)}(k)\over\D m_0^2}\right]
+ {1\over8}\,{\Delta_{(\alpha)}(0)\over\Delta^{-1}_\mu}
\nonumber \\  +\;&&\intk \left
[\kappa\Delta^{-1}_\mu +\Delta^{-1}_{(\theta)}(k)\right]^{-1}
\left[{1\over2}(4+m_0^2)
- {\widehat{k{+}2i\mu_0}^2\over\widehat{k{+}i\mu_0}^2 + m_0^2}\right.
\nonumber \\ &&\qquad\qquad \left.
+\;{1\over2}\Delta_{(\alpha)}(0)\left(
{\D\Delta^{-1}_{(\theta)}(k)\over\D m_0^2}
+ {\kappa\over4}\right)\right] \Biggr\},
\label{mu1-c0}
\end{eqnarray}
where
\begin{mathletters}
\begin{eqnarray}
\Delta^{-1}_\mu &=& 1 - {1\over4\beta} + {m_0^2\over4}\,, \\
\kappa &=& \kappatilde\,{4\beta\over4\beta-1}\,,
\end{eqnarray}
\end{mathletters}
and $\Delta_{(\alpha)}$, $\Delta_{(\theta)}$ can be computed from
Eqs.~(\ref{Delta-repr}).

Eq.~(\ref{mu1-c0}) can be tested, for sufficiently small $\beta$, against
existing strong-coupling results. Strong-coupling expansions for the mass
gap in ${\rm O}(N)$ and $\CPN$ models were pioneered in
Refs.~\cite{Rabinovici-Samuel,DiVecchia-Holtkamp-Musto-Nicodemi-Pettorino,%
Affleck-Levine,Musto-Nicodemi-Pettorino-Clarizia}, and in particular it was
recognized that, in the $\CPN$ case, strong-coupling and large-$N$ limits
cannot trivially commute: indeed the ${\rm U}(1)$ gauge invariance is
spontaneously broken at $N=\infty$ (states in the fundamental
representation are deconfined free particles); on the other hand, it cannot
be broken in strong coupling for any finite value of $N$. A signal for this
phenomenon is the nonexistence of a regular strong-coupling expansion of
Eq.~(\ref{mu1-c0}) at $\kappa=0$.  In the ${\rm O}(N)$ case, extended
strong-coupling series for the mass gap (up to $O(\beta^{11})$) were
obtained by Butera and coworkers in
Ref.~\cite{Butera-Comi-Marchesini-sigma} and analyzed in the large-$N$
limit in Ref.~\cite{Butera-Comi-Marchesini-Onofri}. Results up to
$O(\beta^{14})$ for the magnetic susceptibility and related quantities were
obtained in Refs.~\cite{Bonnier-Hontebeyrie,Butera-Comi-Marchesini-series}.
Strong-coupling expansion of ${\rm O}(N)$ models with improved action was
discussed in Ref.~\cite{Clarizia-Cristofano-Musto-Nicodemi-Pettorino}.

Eq.~(\ref{mu1-c0}) can be further simplified in the
large-$\kappatilde$ regime, where one can perform a
$1/\kappatilde$ expansion of the integral involving the vector field
propagator, leading to an expression which can be evaluated exactly to
$O(1/\kappatilde)$.  We obtain the result
\begin{eqnarray}
\mu_1^2 = {\mu_0\over\nu_0} \Biggl\{ &&\intk
\Delta_{(\alpha)}(k)\left[{1\over\widehat{k{+}i\mu_0}^2 + m_0^2} +
  {1\over2}\Delta_{(\alpha)}(0)\,
{\D\Delta^{-1}_{(\alpha)}(k)\over\D m_0^2}\right]
+ {1\over4}\,{\Delta_{(\alpha)}(0)\over\Delta^{-1}_\mu}
   \nonumber \\  +\;&& {1\over\kappatilde}\,{\beta^2\over2}
   \left(1-{1\over4\beta}\right){\Delta_{(\alpha)}(0)\over\Delta^{-1}_\mu}
   \left(\Delta^{-1}_\mu + {1\over\Delta^{-1}_\mu}\right)
+ O\!\left(1\over\kappatilde^2\right) \Biggr\}.
\label{mu1-c0-1ok}
\end{eqnarray}
Eq.~(\ref{mu1-c0-1ok}) is a rather good approximation of Eq.~(\ref{mu1-c0})
even for quite small values of $\kappatilde$, as long as we consider not
too large values of $\beta$. Eq.~(\ref{mu1-c0-1ok}) should be compared with
the exact $O(1/N)$ contribution to the mass of the ${\rm O}(2N)$ model
\begin{equation}
\mu_1^2 = {\mu_0\over\nu_0} \left\{\intk
\Delta_{(\alpha)}(k)\left[{1\over\widehat{k{+}i\mu_0}^2 + m_0^2} +
  {1\over2}\Delta_{(\alpha)}(0)\,
{\D\Delta^{-1}_{(\alpha)}(k)\over\D m_0^2}\right]\right\}.
\label{mu1-c0-ON}
\end{equation}

When $c_2=0$ significant simplifications occur also in the scaling region
contributions obtained in Eqs.~(\ref{m1-scaling}) and (\ref{Z1-scaling}).
The explicit forms of $\delta m_1^2$ and $\delta Z_1$ are respectively
\begin{eqnarray}
\delta m_1^2 &=& 2\pi m_0^2 \left\{\intk
\left[{A_1^{(\alpha)} + \beta B_1^{(\alpha)}\over
 A_0^{(\alpha)} + \beta B_0^{(\alpha)}} +
{A_1^{(\theta)} + \beta B_1^{(\theta)}
 + \beta\kappatilde/(4\beta-1)\over
 A_0^{(\theta)} + \beta B_0^{(\theta)} + \kappatilde}\right]
+ {\beta\over4\beta-1} \right.\nonumber \\
&&\qquad-\,\left.\int_0^{32} {\D k^2\over4\pi}\,
{2\over k^2}\left[{1\over\ln\left(k^2/m_0^2\right)} +
	{3-2\pi\kappatilde\over
	\ln\left(k^2/m_0^2\right)+2\pi\kappatilde-2}
\right]\right\}
\label{mu10-c0}
\end{eqnarray}
and
\begin{eqnarray}
\delta Z_1 &=& \intk {\half B_1^{(\alpha)} + 1/\bigl(\hat k^2\bigr)^2
	\over A_0^{(\alpha)} + \beta B_0^{(\alpha)}}
 + {1\over2} \intk {B_1^{(\theta)} + \kappatilde/(4\beta-1)
	\over A_0^{(\theta)} + \beta B_0^{(\theta)} + \kappatilde}
 \nonumber \\
&+&\,{1\over2}\,{1\over4\beta-1} +
\int_0^{32} {\D k^2\over4\pi}\,{1\over k^2}
\left[{2\pi\over\ln\left(k^2/m_0^2\right)} -
{4\pi\over\ln\left(k^2/m_0^2\right)+2\pi\kappatilde-2}\right].
\label{Z10-c0}
\end{eqnarray}
It is easy to compute the first nontrivial orders in the weak coupling
expansion of Eqs.~(\ref{mu10-c0}) and (\ref{Z10-c0}), obtaining
\begin{equation}
{\delta m_1^2\over m_0^2} = \left({3\pi\over2} - 2\right) +
\left(c_1^{(\alpha)} + c_1^{(\theta)} + \kappatilde\right) f +
O\left(f^2\right),
\label{m1-pert}
\end{equation}
and
\begin{equation}
\delta Z_1 = {f\over2\pi}\left[\left({3\pi\over2} - 2\right) +
\left(c_1^{(\alpha)} + c_1^{(\theta)} + \kappatilde
	+ {1\over\pi}\right) f + O\left(f^2\right)\right],
\label{Z1-pert}
\end{equation}
where
\begin{mathletters}
\begin{eqnarray}
c_1^{(\alpha)} &=& 4\pi\intk\left[{A_1^{(\alpha)}\over B_0^{(\alpha)}}
- {A_0^{(\alpha)}\over B_0^{(\alpha)}}\,{B_1^{(\alpha)}\over B_0^{(\alpha)}}
- {1\over2\pi\hat k^2}\right], \\
c_1^{(\theta)} &=& 4\pi\intk\left[{A_1^{(\theta)}\over B_0^{(\theta)}}
- {A_0^{(\theta)}\over B_0^{(\theta)}}\,{B_1^{(\theta)}\over B_0^{(\theta)}}
+ {1\over16} - {3\over2\pi\hat k^2}\right].
\end{eqnarray}
\end{mathletters}
We can now make use of a number of identities holding for regulated lattice
integrals, discussed in Appendix~\ref{lattice integr}, to obtain
\begin{mathletters}
\begin{eqnarray}
c_1^{(\alpha)} = -{\pi\over8} - {1\over2\pi}+ 4\pi G_1^{(\alpha)}
	&\cong& 0.0282552, \\
c_1^{(\theta)} = -{\pi\over8} + {3\over2\pi}+ 4\pi G_1^{(\theta)}
	&\cong& 1.7120726
\end{eqnarray}
\end{mathletters}
(cf.\ Eq.~(\ref{G1-at})).

Eqs.~(\ref{m1-pert}) and (\ref{Z1-pert}) have a straightforward
relationship with the three-loop computation of the lattice $\beta$ and
$\gamma$ functions.  In particular the ${\rm O}(2N)$ results
\begin{mathletters}
\begin{eqnarray}
{\delta m_1^2\over m_0^2} &=& \left({\pi\over2} - 1\right)
+ c_1^{(\alpha)} f + O\left(f^2\right), \\
\delta Z_1 &=& {f\over2\pi}\left[\left({\pi\over2} - 1\right) +
\left(c_1^{(\alpha)} + {1\over2\pi}\right) f + O\left(f^2\right)\right],
\end{eqnarray}
\end{mathletters}
reproduce the original calculation by Falcioni and Treves
\cite{Falcioni-Treves}, already confirmed in
Refs.~\cite{Biscari-Campostrini-Rossi,Luscher-Weisz-Wolff}.

Higher-order coefficients of the weak coupling series can be evaluated
numerically with high precision, and different truncations of the series
can be compared, checking for stability.

We can also compute numerically to high precision the difference between
the exact lattice representation (\ref{mu1-c0}) of $\mu_1^2$ and its SM
continuum counterpart $m^2_{1\,\rm SM}$ as expressed by
Eq.~(\ref{m1-cont}), in the region $\beta\lesssim1.5$.

Finally we may evaluate the representation (\ref{mu10-c0}) of the scaling
contribution to the mass gap $\delta m_1^2$, as well as the representation
(\ref{Z10-c0}) of the renormalization function $\delta Z_1$.  We evaluated
the integrals in Eqs.~(\ref{mu10-c0}) and (\ref{Z10-c0}) using the
expansion in exponential-integral functions described in
Sect.~\ref{physical quantities} and
Ref.~\cite{Campostrini-Rossi-condensates}.

These different evaluations should all agree with each other in the very
weak-coupling domain, where truncated perturbative series are accurate.
Moreover the difference between the last two determinations of $\delta
m_1^2$ is entirely due to scaling violations; therefore it can be compared
with independent determinations of the scaling region, such as the study of
rotation invariance properties of the mass gap.

All the relevant numerical results are presented in Figs.~\ref{delta-m} and
\ref{delta-Z}, where $\delta m_1^2$ and $\delta Z_1$ respectively are
plotted as functions of $f$ for different values of $\kappatilde$.

In the case $c_2=0$ we can also perform a $1/\kappatilde$ expansion of
Eq.~(\ref{mu10-c0}).  The result is simply
\begin{eqnarray}
\delta m_1^2 &=& 2\pi m_0^2  \Biggl\{\intk
{A_1^{(\alpha)} + \beta B_1^{(\alpha)}\over
 A_0^{(\alpha)} + \beta B_0^{(\alpha)}}
 - \int_0^{32} {\D k^2\over4\pi}\,
{2\over k^2}\left[{1\over\ln\left(k^2/m_0^2\right)} - 1\right]
\nonumber \\
&&\qquad+\,{2\beta\over4\beta-1} + {1\over\kappatilde}\left[
-{\beta\over\pi} + {\beta\over4}\,{1\over4\beta-1} - {1\over4\pi^2}\right]
+ O\!\left(1\over\kappatilde^2\right) \Biggr\}.
\label{mu10-c0-1ok}
\end{eqnarray}
It is possible to check directly the consistency of Eq.~(\ref{mu10-c0-1ok})
with the asymptotic expansion of Eq.~(\ref{mu1-c0-1ok}).  As a byproduct,
one may also verify that the asymptotic behavior of $c_m(\kappatilde)$ is
correctly represented by Eq.~(\ref{cm-1/k}).  The corresponding result for
the ${\rm O}(2N)$ models is expressed by Eq.~(\ref{m-Z-scaling-ON}a) and is
plotted in Fig.~\ref{delta-m-ON}.

In the opposite limit, the result for $\CPN$ models as expressed by
Eq.~(\ref{m1-scaling-CPN}) is plotted in Fig.~\ref{delta-m-CPN}.

\section{Topological operators}
\label{lattice topology}

We already mentioned at the end of Sect.~\ref{continuum 1/N} the special
r\^ole played by topological properties in the limit $\kappa=0$,
corresponding to pure $\CPN$ models and ${\rm U}(1)$ gauge invariance. The
problem of defining a sensible lattice counterpart of the topological
charge density (\ref{qt}) has long been debated in the literature.

The geometrical definition originally proposed by Berg and L\"uscher
\cite{Berg-Luscher-O3} amounts to defining
\begin{equation}
q_n^{\rm g} = {1\over2\pi}\,\Im\left\{\ln\tr(P_{n+\mu+\nu}P_{n+\mu}P_n)
+ \ln\tr(P_{n+\nu}P_{n+\mu+\nu}P_n)\right\}, \qquad \mu\ne\nu,
\label{qt-Luscher}
\end{equation}
where
\begin{equation}
P_{n,ij} = \bar z_{n,i} z_{n,j}\,.
\end{equation}
$q_n^{\rm g}$ has the advantage of generating integer values of the
topological charge for any given field configuration, and one can prove the
absence of a perturbative tail in the $1/N$ expansion
\cite{DiVecchia-Musto-Nicodemi-Pettorino-Rossi}.
In formulations involving an explicit ${\rm U}(1)$ gauge field
$\lambda_{n,\mu} = \exp i\theta_{n,\mu}$, an alternative geometrical
definition is obtained by defining
\begin{equation}
q_n = {1\over4\pi}\sum_{\mu\nu}\varepsilon_{\mu\nu}
(\theta_{n,\mu} + \theta_{n+\mu,\nu} - \theta_{n+\nu,\mu} - \theta_{n,\nu}).
\label{qt-lambda}
\end{equation}
$q_n$ enjoys the same properties of $q_n^{\rm g}$ and has the same $1/N$
expansion.

Unfortunately for finite $N$ geometrical definitions are plagued by the
so-called ``dislocations'', and therefore one cannot extract the correct
scaling behavior from numerical data
\cite{Berg-letter,Luscher-lattice-chi}.  It is however possible to express
the topological charge in terms of a local operator constructed from the
gauge fields. Let us define the plaquette operator (elementary Wilson
loop):
\begin{mathletters}
\begin{eqnarray}
u_{n,\mu\nu} &=& \lambda_{n,\mu} \lambda_{n+\mu,\nu}
\bar\lambda_{n+\nu,\mu} \bar\lambda_{n,\nu}, \qquad\mu\ne\nu, \\
u_n &=& u_{n,12} = \bar u_{n,21} = \exp\{2\pi i q_n\}.
\end{eqnarray}
\end{mathletters}
Taking proper combinations of higher powers of the plaquette operator, it
is possible to construct an infinite sequence of local operators
\begin{equation}
2\pi q_n^{(k)} = \sum_{l=1}^k {(-1)^{l+1}\over l}
\pmatrix{ 2k \cr k-l \cr} {2k \over \pmatrix{ 2k \cr k \cr}}
\,\Im\left\{(u_n)^l\right\},
\end{equation}
whose formal $k\to\infty$ limit is exactly Eq.~(\ref{qt-lambda}).
One can then construct the sequence of topological susceptibilities
\begin{equation}
\chi_t^{(k)} = \left<\sum_n q_n^{(k)} q_0^{(k)}\right>.
\end{equation}

The perturbative evaluation of these quantities is obtained by considering
that $q_n$ is linear in the effective Lagrangian field $\theta_{n,\mu}$,
and expanding $q_n^{(k)}$ in a power series in $q_n$. One may easily show
that
\begin{equation}
2\pi q_n^{(k)} = 2\pi q_n - {(k!)^2\over(2k+1)!}
(2\pi q_n)^{2k+1} + O(q_n^{2k+3}).
\end{equation}
In standard perturbation theory, using lowest-order momentum space
propagators
\begin{equation}
\left<q(p)\,q(-p)\right>_0 = {1\over(2\pi)^2}\,{f\over N}\,\hat p^2,
\end{equation}
one can prove the relationship
\begin{equation}
\left<q^{(k)}(p)\,q^{(k)}(-p)\right> \cong \left<q(p)\,q(-p)\right>
\left[1 - 2 k! \left(2f\over N\right)^k
  + O\!\left(\left(f\over N\right)^{k+1}\right)\right].
\end{equation}
A more refined analysis, based on the relationship
\begin{equation}
\chi_t^{(k)} = \lim_{p^2\to0} \left<q^{(k)}(p)\,q^{(k)}(-p)\right>
\end{equation}
and the observation that
\begin{equation}
\lim_{p^2\to0} \left<q(p)\,q(-p)\right>_0 = 0,
\end{equation}
allows us to prove that, in the leading order,
\begin{equation}
\chi_t^{(k)} \simeq \chi_t + c_k\left(2f\over N\right)^{2k+1},
\label{chi-k}
\end{equation}
where
\begin{equation}
c_k \cong {(k!)^4 (4k+2)! \over ((2k+1)!)^3} \sim (2k)!
\end{equation}
for large $k$. Eq.~(\ref{chi-k}) shows that the perturbative tail of the
topological susceptibility involves for high $k$ only very high powers of
$f$. The same property might easily be shown to hold also for other mixing
coefficients. However, the corresponding numerical weights are growing so
fast with $k$ that the convergence to the geometrical definition $\chi_t$
cannot be uniform, i.e.\ the limit $k\to\infty$ does not commute with the
continuum limit ($f\to0$). This phenomenon leads to a perturbative
explanation of the observed discrepancy between geometrical and
local-operator definitions of the topological susceptibility, and shows
that, for fixed $f/N$, an optimal value of $k$ should exist such that the
mixing is minimized.

{}From the point of view of the $1/N$ expansion, the situation is however
quite different: for fixed $f$ the perturbative tail is depressed by a
factor $(1/N)^{2k}$, and therefore the absence of a perturbative tail of
the geometrical definition in the $1/N$ expansion is confirmed; the
difference $\chi_t^{(k)}-\chi_t$ is calculable order by order in $1/N$ by
generalizing the techniques described in the previous Sections, while
$\chi_t$ itself is simply obtained from the lattice counterpart of
Eq.~(\ref{chidef}):
\begin{equation}
\chi_t = \lim_{p^2\to0}\,{1\over(2\pi)^2}
\,\hat p^2\tilde\Delta_{(\theta)}(p),
\label{lattice-chi}
\end{equation}
where $\tilde\Delta_{(\theta)}(p)$ is the full lattice propagator of the
field $\theta_\mu$. As a matter of illustration, let us consider the first
contribution to the difference $\chi_t^{(1)}-\chi_t$, drawn in
Fig.~\ref{chi-diff}. A factor of $\hat p^2\Delta_{(\theta)}(p)$ is
associated with each wavy line; as a consequence, the infrared behavior is
regular.

\section{Ratio of $\pbbox{\Lambda}$ parameters and
renormalization-group functions}
\label{Lambda parameter}

It is possible to analyze the results presented in Sect.~\ref{physical
quantities}, and especially Eq.~(\ref{m1-scaling}), from the point of view
of the perturbative renormalization group. Let us focus on the
contributions to $m_1^2$ that depend on the specific lattice model adopted,
i.e.\ the quantity $\delta m_1^2$ defined by Eq.~(\ref{m-delta}).

\subsection{Ratio of $\pbbox{\Lambda}$ parameters}

Let us consider the $\beta\to\infty$ limit of Eq.~(\ref{m1-scaling}) and
notice that even in this limit a ($\kappatilde$-independent)
contribution to $\delta m_1^2$ survives:
\begin{equation}
\lim_{\beta\to\infty} \delta m_1^2 = 2\pi m_0^2 \intk \left[
   {B_1^{(\alpha)}\over B_0^{(\alpha)}}
 + {B_1^{(\theta)}\over B_0^{(\theta)}}
 + {\sum_\mu D_{1\mu}\,\hat k_\mu^2 \over
    \sum_\mu D_{0\mu}\,\hat k_\mu^2} \right].
\label{m1-beta-infinity}
\end{equation}
Because of the noncommutativity of the limits, we must consider separately
the borders of the parameter space. Indeed at $\kappatilde=0$ we find
\begin{equation}
\lim_{\beta\to\infty} \delta m_1^2\bigr|_{\beta_{\rm v}=0} =
  2\pi m_0^2 \intk \left[
   {B_1^{(\alpha)}\over B_0^{(\alpha)}}
 + {B_1^{(\theta)}\over B_0^{(\theta)}}\right].
\label{m1-beta-infinity-kappa0}
\end{equation}
When $\kappatilde\to\infty$ we must face an even more involved
situation, because the limits $\kappatilde\to\infty$ and
$\beta\to\infty$ do not commute, as one may easily check directly from the
standard perturbative expansion.  From Eq.~(\ref{m-Z-scaling-kappainf}a)
we obtain
\begin{equation}
\lim_{\beta\to\infty}\lim_{\kappatilde\to\infty} \delta m_1^2 =
2\pi m_0^2 \intk \left[{B_1^{(\alpha)}\over B_0^{(\alpha)}} +
  \sum_\mu {D_{1\mu} \over  D_{0\mu}} + {2\over\bar k^2} \right].
\label{m1-beta-infinity-kappa-infinity}
\end{equation}
Finally when $\kappatilde=\infty$ we obtain from Eq.~(\ref{m-Z-scaling-ON}a)
\begin{equation}
\lim_{\beta\to\infty} \delta m_1^2\bigr|_{\beta_{\rm g}=0} =
  2\pi m_0^2 \intk \left[
   {B_1^{(\alpha)}\over B_0^{(\alpha)}} + {2\over\bar k^2}\right].
\label{m1-beta-infinity-ON}
\end{equation}
We observe that the difference between different regularizations of the
same physical models is amenable in this limit to the condition
$D_{1\mu} \ne 0$.

Let us now come to the physical interpretation of
Eqs.~(\ref{m1-beta-infinity}), (\ref{m1-beta-infinity-kappa0}),
(\ref{m1-beta-infinity-kappa-infinity}), and (\ref{m1-beta-infinity-ON}).
These quantities are obviously related to the ratio of the so-called
lattice $\Lambda$ parameter to the continuum (SM) $\Lambda$ parameter in
the models at hand \cite{Parisi-coupling,Berg-Lambda}.  To be more precise,
and recalling Eqs.~(\ref{ML}) and (\ref{ML-f}), we may write the
relationship
\begin{equation}
{\Lambda_{\rm SM}\over\Lambda_L} \approx M_L
\left(1 + {1\over2N}\lim_{\beta\to\infty}
   {\delta m_1^2\over m_0^2}\right).
\label{Lambda-ratio-approx}
\end{equation}
Actually we can do better than Eq.~(\ref{Lambda-ratio-approx}); we can
exploit the fact that the ratio of the $\Lambda$ parameters is essentially
a one-loop phenomenon and our knowledge of the first coefficient of the
renormalization-group $\beta$ function ($N$ for all
$\kappatilde\ne\infty$, $N-1$ when $\kappatilde\to\infty$) to
exponentiate Eq.~(\ref{Lambda-ratio-approx}) and obtain the exact
relationships
\begin{equation}
{\Lambda_{\rm SM}\over\Lambda_L} = M_L
\,\exp\left\{{\pi\over N}\intk\left[{B_1^{(\alpha)}\over B_0^{(\alpha)}}
+ {B_1^{(\theta)}\over B_0^{(\theta)}}
\right]\right\}
\label{Lambda-ratio-kappa-0}
\end{equation}
when $\kappatilde=0$ ($\CPN$ models),
\begin{equation}
{\Lambda_{\rm SM}\over\Lambda_L} = M_L
\,\exp\left\{{\pi\over N}\intk\left[{B_1^{(\alpha)}\over B_0^{(\alpha)}}
+ {B_1^{(\theta)}\over B_0^{(\theta)}}
+ {\sum_\mu D_{1\mu}\,\hat k_\mu^2 \over \sum_\mu D_{0\mu}\,\hat k_\mu^2}
\right]\right\}
\label{Lambda-ratio-kappa-generic}
\end{equation}
when $\kappatilde\ne0$ and $\kappatilde\ne\infty$, and finally
\begin{equation}
{\Lambda_{\rm SM}\over\Lambda_L} = M_L \,\exp\left\{
   {\pi\over N-1}\intk\left[{B_1^{(\alpha)}\over B_0^{(\alpha)}}
   + {2\over\bar k^2} + \sum_\mu {D_{1\mu}\over D_{0\mu}} \right]\right\}
\label{Lambda-ratio-kappa-infty}
\end{equation}
when $\kappatilde\to\infty$.
Eq.~(\ref{Lambda-ratio-kappa-infty}) is to be compared with
\begin{equation}
{\Lambda_{\rm SM}\over\Lambda_L} = M_L \,\exp\left\{
   {\pi\over N-1}\intk\left[{B_1^{(\alpha)}\over B_0^{(\alpha)}}
   + {2\over\bar k^2} \right]\right\},
\label{Lambda-ratio-O2N}
\end{equation}
which we would obtain by setting $\beta_{\rm g}=0$ from the very beginning
(standard ${\rm O}(2N)$ models).

We can now obtain explicit representations of the quantities entering
Eqs.~(\ref{Lambda-ratio-kappa-0}), (\ref{Lambda-ratio-kappa-generic}),
(\ref{Lambda-ratio-kappa-infty}), and (\ref{Lambda-ratio-O2N}). First we
notice that
\begin{mathletters}
\begin{eqnarray}
{\sum_\mu D_{1\mu}\,\hat k_\mu^2 \over \sum_\mu D_{0\mu}\,\hat k_\mu^2}
&=& {1\over\bar k^2}\sum_\mu\sin^2\half k_\mu
\left(c_1 + 4 c_2 \cos^2\half k_\mu\right), \\
\sum_\mu {D_{1\mu}\over D_{0\mu}} &=& {1\over4}\sum_\mu
{c_1 + 4 c_2 \cos^2\half k_\mu \over c_1 + c_2 \cos^2\half k_\mu} \,.
\end{eqnarray}
\end{mathletters}
Furthermore, using the results presented in Sect.~\ref{asymptotic
expansions} we can show that
\begin{mathletters}
\begin{eqnarray}
{B_1^{(\alpha)}\over B_0^{(\alpha)}} +
{B_1^{(\theta)}\over B_0^{(\theta)}} &=&
\sum_\mu \partial_\mu \partial_\mu \ln\bar k^2
- 2{c_1 c_2\over\bar k^2}\sum_\mu
{\sin^4\half k_\mu\over c_1 + c_2 \cos^2\half k_\mu}
+ \sum_\mu {D_{1\mu}\over D_{0\mu}}\,, \\
{B_1^{(\alpha)}\over B_0^{(\alpha)}} + {2\over\bar k^2} &=&
{1\over2}\sum_\mu \partial_\mu \partial_\mu \ln\bar k^2 +
 {\sum_\mu D_{1\mu}\,\hat k_\mu^2 \over
  \sum_\mu D_{0\mu}\,\hat k_\mu^2}\,.
\end{eqnarray}
\end{mathletters}
We notice that terms proportional to $\sum_\mu \partial_\mu \partial_\mu
\ln\bar k^2$ are total derivatives that can be integrated exactly for any
physically acceptable form of $\bar k^2$:
\begin{eqnarray}
\intk\sum_\mu \partial_\mu \partial_\mu \ln\bar k^2 = -{1\over\pi}\,.
\end{eqnarray}
These terms are not really lattice artifacts: they are related to the
ratio $\Lambda_{\rm SM}/\Lambda_{\overline{\rm MS}}$, where
$\Lambda_{\overline{\rm MS}}$ is the $\Lambda$ parameter defined in the
dimensional regularization scheme with minimal subtraction (notice that
in dimensional regularization the integral of a total derivative vanishes
exactly). Therefore we obtain
\begin{mathletters}
\label{Lambda-change}
\begin{eqnarray}
&\displaystyle{\Lambda_{\rm SM}\over\Lambda_{\overline{\rm MS}}} =
\exp\left[-{1\over N}\right]\qquad\qquad&(\kappatilde\ne\infty),\\
&\displaystyle{\Lambda_{\rm SM}\over\Lambda_{\overline{\rm MS}}} =
\exp\left[-{1\over 2(N-1)}\right]\qquad\qquad&
(\kappatilde\to\infty).
\end{eqnarray}
\end{mathletters}
Eqs.~(\ref{Lambda-change}) are crucial in finding the variable change from
SM to $\overline{\rm MS}$ scheme and verifying the perturbative consistency
of the continuum results.  The ratio of $\Lambda$ parameters can now be
expressed in the more conventional form
\begin{equation}
{\Lambda_{\overline{\rm MS}}\over\Lambda_L} = M_L
\,\exp\left\{{\pi\over N}\intk\left[-{2c_1c_2\over\bar k^2}
\sum_\mu{\sin^4\half k_\mu\over c_1 + c_2 \cos^2\half k_\mu}
+ \sum_\mu{D_{1\mu}\over D_{0\mu}}
\right]\right\}
\label{Lambda-conv1}
\end{equation}
when $\kappatilde=0$,
\begin{equation}
{\Lambda_{\overline{\rm MS}}\over\Lambda_L} = M_L
\,\exp\left\{{\pi\over N}\intk\left[-{2c_1c_2\over\bar k^2}
\sum_\mu{\sin^4\half k_\mu\over c_1 + c_2 \cos^2\half k_\mu}
+ \sum_\mu{D_{1\mu}\over D_{0\mu}}
+ {\sum_\mu D_{1\mu}\,\hat k^2_\mu \over\sum_\mu D_{0\mu}\,\hat k^2_\mu}
\right]\right\}
\label{Lambda-conv2}
\end{equation}
when $\kappatilde\ne0$ and $\kappatilde\ne\infty$,
\begin{equation}
{\Lambda_{\overline{\rm MS}}\over\Lambda_L} = M_L
\,\exp\left\{{\pi\over N-1}\intk\left[
\sum_\mu{D_{1\mu}\over D_{0\mu}}
+ {\sum_\mu D_{1\mu}\,\hat k^2_\mu \over\sum_\mu D_{0\mu}\,\hat k^2_\mu}
\right]\right\}
\label{Lambda-conv3}
\end{equation}
when $\kappatilde\to\infty$, and
\begin{equation}
{\Lambda_{\overline{\rm MS}}\over\Lambda_L} = M_L
\,\exp\left\{{\pi\over N-1}\intk
{\sum_\mu D_{1\mu}\,\hat k^2_\mu \over\sum_\mu D_{0\mu}\,\hat k^2_\mu}
\right\}
\label{Lambda-conv4}
\end{equation}
in the pure ${\rm O}(2N)$ case.

When $c_2=0$ all integrals can be computed in closed form; we obtain
\begin{eqnarray}
&\displaystyle{\Lambda_{\overline{\rm MS}}\over\Lambda_L} =
\sqrt{32}\,\exp\left[{\pi\over 2N}\right]
\qquad\qquad& (\CPN), \\
&\displaystyle{\Lambda_{\overline{\rm MS}}\over\Lambda_L} =
\sqrt{32}\,\exp\left[{3\pi\over 4N}\right]
\qquad\qquad& (\kappatilde\ne0), \\
&\displaystyle{\Lambda_{\overline{\rm MS}}\over\Lambda_L} =
\sqrt{32}\,\exp\left[{3\pi\over 4(N-1)}\right]
\qquad\qquad& (\kappatilde\to\infty), \\
&\displaystyle{\Lambda_{\overline{\rm MS}}\over\Lambda_L} =
\sqrt{32}\,\exp\left[{\pi\over 4(N-1)}\right]
\qquad\qquad& ({\rm O}(2N)).
\end{eqnarray}

Eqs.~(\ref{Lambda-conv1}), (\ref{Lambda-conv2}), (\ref{Lambda-conv3}),
and (\ref{Lambda-conv4}) have been all explicitly verified in standard
perturbation theory. In particular, the difference between
Eq.~(\ref{Lambda-conv1}) and Eq.~(\ref{Lambda-conv2}) can be traced to
the contribution of the field associated to the phase of the last
component of the field $z_N$, which cannot be eliminated by a gauge
transformation when $\kappatilde\ne0$.  In turn the difference between
Eq.~(\ref{Lambda-conv3}) and Eq.~(\ref{Lambda-conv4}) is originated by
the unsuppressed contribution of the tadpole graphs involving closed
loops of vector propagators, shown in Fig.~\ref{tadpole-graph}. Trivial
power counting arguments show that no inverse powers of $1+\kappa f$
appear in the perturbative evaluation of these diagrams.

The agreement between perturbative and $1/N$ evaluation of the ratios of
$\Lambda$ parameters is a strong confirmation of the commutativity of the
$1/\beta$ and $1/N$ expansion with respect to renormalization-group
properties of the models. Even the apparent singularity of the
$\kappatilde\to0$ and $\kappatilde\to\infty$ limits has no consequences on
the exchange of the $\beta\to\infty$ and $N\to\infty$ limits.  Further
confirmations can be obtained by the comparison of the perturbative
$\beta$-function coefficients with those obtained by expanding the resummed
$1/N$ lattice $\beta$ function, which can be easily obtained from $\delta
m_1^2$.

\subsection{Lattice $\pbbox{\beta}$ and  $\pbbox{\gamma}$ functions}

Let us come to the evaluation of the $O(1/N)$ lattice renormalization-group
$\beta$ function.  We may apply the homogeneous renormalization-group
equations to the expression of the mass gap
\begin{equation}
\left[M\,{\partial\over\partial M} +
\beta(f)\,{\partial\over\partial f}\right] m^2(M,f) = 0.
\label{homogeneous-RG}
\end{equation}
We can expand $\beta(f)$ in powers of $1/N$ in the form
\begin{equation}
\beta(f) = \beta_0(f) + {1\over N}\beta_1(f) + O\!\left(1\over N^2\right).
\end{equation}
Our choice of lattice action allows us to use the relationship
\begin{equation}
m_0^2 = M^2\,\exp\left(-{2\pi\over f}\right).
\end{equation}
in Eq.~(\ref{homogeneous-RG}) to obtain
\begin{equation}
\beta_0(f) = -{f^2\over\pi} \,.
\end{equation}
Further substitutions in Eq.~(\ref{homogeneous-RG}) lead to
\begin{equation}
\beta_1(f) = \left[\beta_0(f)\right]^2{1\over2}\,
{\partial\over\partial f}\,{m_1^2\over m_0^2}
\end{equation}
and, since we know $\beta_1$ in the SM scheme, we immediately obtain
\begin{equation}
\beta_1^{(L)}(f) = \beta_1^{(\rm SM)}(f) + {f^2\over\pi} \,
{f^2\over2\pi} \, {\partial\over\partial f}
\left(\delta m_1^2\over m_0^2\right).
\label{beta-latt}
\end{equation}
Eq.~(\ref{beta-latt}) admits a natural interpretation.  We must recognize
that a change in the regularization scheme corresponds to a
reparametrization of the model, i.e.\ $f'=f'(f)$.  Covariance of the
renormalization-group equations under reparametrization implies
\begin{equation}
\beta'(f') = \beta(f(f'))\left(\partial f\over\partial f'\right)^{-1}.
\end{equation}
As a consequence Eq.~(\ref{beta-latt}) implies
\begin{equation}
f^{({\rm SM})} = f - {1\over N}\,\beta_0(f)\,{1\over2}\,
{\delta m_1^2\over m_0^2} + O\!\left(1\over N^2\right).
\end{equation}

The improper integral obtained by substituting Eq.~(\ref{m1-scaling}) into
Eq.~(\ref{beta-latt}) is defined according to the prescription
(\ref{princ-prescription}). It is worth noticing that all the residues in
the complex integration vanish, in contrast with Eq.~(\ref{m1-scaling})
itself.  Taking the derivative with respect to $f\equiv1/(2\beta)$ in
Eq.~(\ref{m1-scaling}) is completely straightforward and we shall not write
down the result in the most general case. We shall however consider a few
interesting special cases.

For $\kappatilde=0$ ($\CPN$ models) we have
\begin{eqnarray}
\beta(f) &=& -{f^2\over\pi}\left\{1
  + {1\over N}\,{f\over2\pi}\left(1+{3\over1-f/\pi}\right)
  - {1\over N}\intk {1\over2}\,
    {A_1^{(\alpha)} B_0^{(\alpha)} - B_1^{(\alpha)} A_0^{(\alpha)} \over
     \left(A_0^{(\alpha)} + \beta B_0^{(\alpha)}\right)^2}
\right.\nonumber \\
&&\qquad-\;{1\over N}\intk {1\over2}\,
    {A_1^{(\theta)} B_0^{(\theta)} - B_1^{(\theta)} A_0^{(\theta)} \over
     \left(A_0^{(\theta)} + \beta B_0^{(\theta)}\right)^2} \nonumber \\
&&\qquad+\;\left.{1\over N}\int^{M_L^2} {\D k^2\over4\pi}\,
    {4\pi\over k^2}\left({1\over\ln^2(k^2/m_0^2)}
        + {3\over\bigl(\ln(k^2/m_0^2)-2\bigr)^2}\right)
\vphantom{A_1^{(\alpha)} B_0^{(\alpha)} - B_1^{(\alpha)} A_0^{(\alpha)}
\over \left(A_0^{(\alpha)} + \beta B_0^{(\alpha)}\right)^2} \right\}.
\label{sample-den2}
\end{eqnarray}

For $\kappatilde\ne0$ and $c_2=0$ we have
\begin{eqnarray}
\beta(f) &=& -{f^2\over\pi}\left\{1
  + {1\over N}\,{f\over2\pi}
  \left[1+{3 - 2\pi\kappatilde \over
           1 + f(\kappatilde - 1/\pi)}\right]
  - {1\over N}\intk {1\over2}\,
    {A_1^{(\alpha)} B_0^{(\alpha)} - B_1^{(\alpha)} A_0^{(\alpha)} \over
     \left(A_0^{(\alpha)} + \beta B_0^{(\alpha)}\right)^2}
\right.\nonumber \\
&&\qquad-\;{1\over N}\intk {1\over2}\,
  \left(A_0^{(\theta)} + \beta B_0^{(\theta)}
      + \kappatilde\right)^{-2}
\left[A_1^{(\theta)} B_0^{(\theta)} - B_1^{(\theta)} A_0^{(\theta)}
\vphantom{A_0^{(\theta)}\over(4\beta-1)^2}\right.\nonumber \\
&&\qquad\qquad\qquad\qquad+\;\left.
       \kappatilde\left({A_0^{(\theta)}\over(4\beta-1)^2}
       + {4\beta^2 B_0^{(\theta)}\over(4\beta-1)^2}
       - B_1^{(\theta)}\right) + {\kappatilde^2\over(4\beta-1)^2}
   \right] \nonumber \\
&&\qquad-\;{1\over2N}\,{1\over(4\beta-1)^2} \nonumber \\
&&\qquad+\;\left.{1\over N}\int^{32} {\D k^2\over4\pi}\,
    {4\pi\over k^2}\left({1\over\ln^2(k^2/m_0^2)}
        + {3 - 2\pi\kappatilde \over
           \bigl(\ln(k^2/m_0^2) + 2\pi\kappatilde - 2\bigr)^2}
    \right)
\vphantom{A_1^{(\alpha)} B_0^{(\alpha)} - B_1^{(\alpha)} A_0^{(\alpha)}
\over \left(A_0^{(\alpha)} + \beta B_0^{(\alpha)}\right)^2} \right\}.
\end{eqnarray}

For $\kappatilde\to\infty$ we have
\begin{eqnarray}
\beta(f) &=& -{f^2\over\pi}\left\{1 - {1\over N} + {1\over N}\,{f\over2\pi}
  - {1\over N}\intk {1\over2}\,
    {A_1^{(\alpha)} B_0^{(\alpha)} - B_1^{(\alpha)} A_0^{(\alpha)} \over
     \left(A_0^{(\alpha)} + \beta B_0^{(\alpha)}\right)^2}
\right.\nonumber \\
&&\qquad+\;\left.{1\over N}\intk {1\over2}\,
  \sum_\mu{D_{1\mu}C_{0\mu} - C_{1\mu}D_{0\mu} \over
    (C_{0\mu} + \beta D_{0\mu})^2}
  + {1\over N} \int^{M_L^2} {\D k^2\over4\pi}\,
    {4\pi\over k^2}\,{1\over\ln^2(k^2/m_0^2)}
\vphantom{A_1^{(\alpha)} B_0^{(\alpha)} - B_1^{(\alpha)} A_0^{(\alpha)}
\over \left(A_0^{(\alpha)} + \beta B_0^{(\alpha)}\right)^2} \right\}.
\nonumber \\
\end{eqnarray}

Finally, for ${\rm O}(2N)$ models we have
\begin{eqnarray}
\beta(f) &=& -{f^2\over\pi}
\left\{1 - {1\over N} + {1\over N}\,{f\over2\pi}
  - {1\over N}\intk {1\over2}\,
    {A_1^{(\alpha)} B_0^{(\alpha)} - B_1^{(\alpha)} A_0^{(\alpha)} \over
     \left(A_0^{(\alpha)} + \beta B_0^{(\alpha)}\right)^2}
\right.\nonumber \\
&&\qquad+\;\left.{1\over N} \int^{M_L^2} {\D k^2\over4\pi}\,
    {4\pi\over k^2}\,{1\over\ln^2(k^2/m_0^2)}
\vphantom{A_1^{(\alpha)} B_0^{(\alpha)} - B_1^{(\alpha)} A_0^{(\alpha)}
\over \left(A_0^{(\alpha)} + \beta B_0^{(\alpha)}\right)^2} \right\}.
\end{eqnarray}

The evaluation of the lattice renormalization-group function $\gamma$
follows essentially the same path.  From Eq.~(\ref{gamma-SM}) we obtain the
relationships
\begin{mathletters}
\begin{eqnarray}
\gamma_0(f) &=& -{\beta_0(f)\over f} = {f\over\pi}, \\
\gamma_1(f) &=& -\beta_0(f)\,{\partial\over\partial f}Z_1(f)
- {\beta_1(f)\over f}.
\end{eqnarray}
\end{mathletters}
Since we know $\gamma_1$ in the SM scheme, we obtain
\begin{eqnarray}
\gamma_1^{(L)}(f) &=& \gamma_1^{({\rm SM})}(f) - \beta_0(f)
\left[{\partial\over\partial f}\,\delta Z_1 + {\beta_0(f)\over f}\,
{1\over2}\,{\partial\over\partial f}\,{\delta m_1^2\over m_0^2}\right]
\nonumber \\
&=& \gamma_1^{({\rm SM})}(f) - {\partial\gamma_0(f)\over\partial f}
\,\beta_0(f)\,{1\over2}\,{\delta m_1^2\over m_0^2}
-\beta_0(f)\,{\partial\over\partial f}
\left(\delta Z_1 - \gamma_0(f)\,{1\over2}\,{\delta m_1^2\over m_0^2}\right).
\label{gamma-latt}
\end{eqnarray}
Eq.~(\ref{gamma-latt}) in turn is consistent with
\begin{equation}
\gamma^{(L)}(f) = \gamma^{({\rm SM})}\left(f^{({\rm SM})}(f)\right)
- \beta(f)\,{\partial\over\partial f}\ln\zeta(f),
\end{equation}
where
\begin{equation}
\zeta(f) = 1 + {1\over N}
\left(\delta Z_1 - \gamma_0(f)\,{1\over2}\,{\delta m_1^2\over m_0^2}\right)
 + O\!\left(1\over N^2\right)
\end{equation}
is the additional finite field-amplitude renormalization due to the change
of regularization scheme.  We may appreciate the fact that
$\zeta_1(f) = O(f^2)$, as expected.

\section{Finite size scaling}
\label{finite size scaling}

The study of finite-size effects is quite important, both from a purely
theoretical point of view and in the context of controlling systematic
deviations from the infinite-volume limit in numerical simulations.
Finite size scaling in the large-$N$ limit has been widely studied for
different geometries and space dimensionalities by Brezin and collaborators
\cite{Brezin-FSS,Brezin-Korutcheva-Jolicoeur-ZinnJustin}. A finite-volume
approach was applied to the problem of evaluating the low-lying spectrum in
two-dimensional spin models by L\"uscher \cite{Luscher-low-lying-states}
and extended by Floratos and coworkers
\cite{Floratos-Petcher-I,Floratos-Petcher-II,Floratos-Vlachos-CPN}.  The
systematic analysis of $1/N$ finite-size effects is however rather recent
\cite{Campostrini-Rossi-fss}.  Let us review the main results of this
analysis.

Any coordinate-independent physical quantity $Q$ defined in the context of
the $1/N$-expandable finite-lattice model will in general depend on four
different parameters:
\begin{equation}
Q = Q(f,a,L,N),
\end{equation}
where $L^d$ is the physical volume in $d$ dimensions and $a\sim 1/M_L$ is
the lattice spacing. In the infinite-volume limit and in the scaling region
(according to the discussion presented in Sect.~\ref{Lambda parameter}) all
separate dependence on $f$ and $a$ can be made disappear by parametrizing
everything in terms of the physical mass gap $m^2(a,f,N)$, solution of
Eq.~(\ref{homogeneous-RG}). The finite-size-scaling relation stems from the
observation that one can reach the infinite-volume limit ($L/a\to\infty$)
while simultaneously keeping a constant finite value of $mL$. As a
consequence
\begin{equation}
{Q(f,a,L,N)\over Q(f,a,\infty,N)}
\goto_{\stackrel{\scriptstyle f\to0}{\scriptstyle mL = {\rm const}}}
f^{(Q)}(mL,N)
\label{fss-def}
\end{equation}
The $1/N$ expandability in turn implies that, assuming
\begin{equation}
Q(f,a,L,N) =
Q_0(f,a,L) + {1\over N}\,Q_1(f,a,L) + O\!\left(1\over N^2\right)
\label{fss-Q}
\end{equation}
and
\begin{equation}
m(f,N) = m_0(f) + {1\over N}\,m_1(f) + O\!\left(1\over N^2\right),
\label{fss-m}
\end{equation}
we may expand the finite-size functions $f^{(Q)}$ in the form
\begin{equation}
f^{(Q)}(mL,N) = f^{(Q)}_0(mL) + {1\over N}\,f^{(Q)}_1(mL)
 + O\!\left(1\over N^2\right).
\label{fss-f}
\end{equation}
Substituting Eqs.~(\ref{fss-Q}), (\ref{fss-m}), and (\ref{fss-f}) into
Eq.~(\ref{fss-def}), we obtain
\begin{mathletters}
\label{fss-1/N}
\begin{eqnarray}
f^{(Q)}_0(m_0L) &=& {Q_0(f,a,L)\over Q_0(f,a,\infty)}\,, \\
{f^{(Q)}_1(m_0L)\over f^{(Q)}_0(m_0L)} &=&
{Q_1(f,a,L)\over Q_0(f,a,L)} - {Q_1(f,a,\infty)\over Q_0(f,a,\infty)}
- (m_0L) \left(m_1\over m_0\right)
{f^{\prime(Q)}_0(m_0L)\over f^{(Q)}_0(m_0L)}\,.
\end{eqnarray}
\end{mathletters}

Eq.~(\ref{fss-1/N}) is the most general form of the $1/N$-expanded
finite-size-scaling relation. In order to gain further insight, one must
consider the specific properties of the quantity under investigation. In
any case, a basic tool in the analysis is the knowledge of the
finite-size-scaling properties of the finite-size mass parameter $m_L$
defined by the gap equation:
\begin{equation}
{1\over L^2}\sum_q{1\over\bar q^2 + m_L^2} = {1\over2f} =
\intq{1\over\bar q^2 + m_0^2} \,,
\end{equation}
where the sum runs over the momentum lattice modes, i.e.\ $q_\mu=0, 2\pi/L,
..., 2\pi(L{-}1)/L$. In the case $c_2=0$ and in the scaling region,
defining $z_L=m_LL$ and $z_0=m_0L$, we could establish the relationship
\begin{equation}
z_0 = z_c\exp\left\{-\half\omega(z_L)\right\},
\end{equation}
where $z_c \cong 4.163948$ and in the region $z_L\le2\pi$ the function
$\omega$ may be defined by
\begin{mathletters}
\begin{eqnarray}
\omega(z_L) &=& {4\pi\over z_L^2} + 4\pi\sum_{n=1}^\infty
(-1)^nz_L^{2n}d_{n+1}\,, \\
d_n &=& {1\over(2\pi)^{2n}}
\sum_{\stackrel{\scriptstyle n_1,n_2=-\infty}{\scriptstyle
(n_1,n_2)\ne(0,0)}}^\infty
{1\over(n_1^2+n_2^2)^n}\,,\qquad n>1.
\end{eqnarray}
\end{mathletters}
The function $z_0(z_L)$ is monotonic and invertible. Therefore all
subsequent calculations can be performed making use of the auxiliary
variable $z_L$, which simplifies many computations.

Without giving further technical details, we mention that
finite-size-scaling functions were computed in the $1/N$ expansion of
masses and magnetic susceptibilities, both in the pure ${\rm O}(N)$ case
\cite{Campostrini-Rossi-fss} and in the pure $\CPN$ models
\cite{Rossi-Vicari}. In ${\rm O}(N)$ models the $1/N$ expansion of
finite-size functions is an accurate description of finite-size effects in
all possible regimes: small volume ($mL\ll2\pi$), where the results can
also be compared, by asymptotic freedom, with those obtained from
finite-volume weak-coupling perturbation theory
\cite{Hasenfratz-zero-modes,Brihaye-Spindel,Flyvbjerg-DS}; large volume
($mL\gg2\pi$) where, due to the existence of a physical mass gap, one
expects exponentially fast convergence to the infinite-volume limit
\cite{Luscher-volume-dependence}; and intermediate volume.

In $\CPN$ models at intermediate volume and for $N$ not too large
($N<100$), new phenomena occur: not every physical quantity is expandable
in a $1/N$ series; moreover, even if we limit ourselves to $1/N$-expandable
objects, we must observe that the scale of finite-size effects is not set
by the mass gap, but instead, since we are in presence of a ``weak''
confining potential, it depends on the (semiclassical) radius of the bound
states; the radius in turn grows like $N^{1/3}$ and therefore does not have
an analytic dependence on $1/N$, a result not unexpected in the light of
the semiclassical results on the bound state spectrum presented in
Appendix~\ref{Schrodinger equation}.

Another subtle point in the study of finite-size effects in $\CPN$ models
is related to the properties of the Abelian Wilson loop on finite lattices.
We only mention here that, defining the Polyakov ratio, corresponding to
the derivative of the static potential introduced in
Sect.~\ref{continuum 1/N},
\begin{equation}
\chi_P(R) = \ln{W(R{-}1,L)\over W(R,L)}\,,
\end{equation}
in the infinite-volume limit and in the scaling region one should find the
Abelian string tension
\begin{equation}
\chi_P(R) \goto_{R\to\infty} \sigma \goto_{N\to\infty} \sigma_0
\equiv {6\pi m_0^2\over N}\,.
\end{equation}
However, on finite lattices, even in the scaling region,
\begin{equation}
\lim_{m_0\to0}{\chi_P(R,L,a,f,N)\over\sigma} =
f^{(P)}\left(N,m_0L,R/L\right) \goto_{N\to\infty}
f_0^{(P)}\left(m_0L,R/L\right) ,
\end{equation}
and one may show that
\begin{equation}
\lim_{m_0L\to\infty} f_0^{(P)}\left(m_0L,R/L\right)
= 1 - {2R\over L}\,.
\end{equation}
We may define a function measuring the deviations from the infinite
(periodic) volume limit:
\begin{equation}
g_0^{(P)} = {f_0^{(P)}\left(m_0L,R/L\right)\over1 - 2R/L}\,;
\end{equation}
$g_0^{(P)}$ in turn can be shown numerically to enjoy a factorization
property: for large $m_0L$ and $m_0R$
\begin{equation}
g_0^{(P)}\left(m_0L,R/L\right) \approx 1 + \phi(m_0L)\,\psi(R/L),
\end{equation}
which can be understood in terms of an effective Yukawa interaction
replacing the Coulomb potential on finite lattices.

If we compare Eq.~(\ref{loop}) with the definition (\ref{lattice-chi}) of
the topological susceptibility, we recognize that the latter quantity is
strictly related in $\CPN$ models to the Abelian string tension, i.e.\
\begin{equation}
{\sigma\over2\pi^2\chi_t} \cong 1 - {64\pi^2\over5N^2}
\end{equation}
(cf.\ Eq.~(\ref{W2-zero})).  Therefore, as a side effect of the above
analysis, we are led to the (not unexpected) result that $\chi_t$ should
vanish on any finite lattice.  However, since the infinite-volume limit is
reached smoothly, it should always be possible to devise an appropriate
limiting procedure to extract infinite-volume information from
finite-volume, finite-$a$ measurements.  Effects of topology in finite
volumes were also studied, for different geometries, in
Ref.~\cite{Schultka-MullerPreussker}.

In closing the present Section, it is relevant to observe that finite size
scaling in ${\rm O}(N)$ models has been the subject of studies concerning
the three-dimensional case, where a second-order phase transition occurs at
a finite value $\beta_c$ of the coupling. In Ref.~\cite{Ruhl-critical}
finite three-dimensional lattices were studied in the context of the $1/N$
expansion. The method for treating near-critical behaviors in three
dimensions, originally developed in Ref.~\cite{Muller-Ruhl} for the
infinite-volume limit, is very reminiscent of the asymptotic expansion
techniques employed in the present work.

Another approach to finite-size effects in three-dimensional ${\rm O}(N)$
models near criticality was developed by Hasenfratz and Leutwyler employing
the techniques of chiral perturbation theory \cite{Hasenfratz-Leutwyler}.

\section{Higher orders of the $\pbbox{1/N}$ expansion on the
lattice}
\label{higher lattice 1/N}

We have not seriously addressed the problem of evaluating $O(1/N^2)$
contributions in the scaling region.  A finite-$\beta$, finite-lattice
calculation can certainly be performed with no conceptual problems, but
with some technical troubles in evaluating accurately two-loop lattice
integrals involving dressed propagators.

This approach has been put forward in recent years by Flyvbjerg and
collaborators
\cite{Flyvbjerg-1/N,Flyvbjerg-Varsted,Flyvbjerg-Larsen-Kristjansen}, who
explicitly studied the case of ${\rm O}(N)$ models. They evaluate the
two-point function of the ${\rm O}(N)$ nonlinear $\sigma$ models up to
$O(1/N^2)$, on finite square lattices and for fixed values of $N$,
typically $N=3,4$, in order to compare with existing Monte Carlo
simulations. From the two-point function they can extract the numerical
value of such physical quantities as the mass gap and the magnetic
susceptibility. The comparison with high-precision Monte Carlo results
allows an estimate of the systematic errors involved in the series
truncation to zeroth, first, and second order. These errors appear to be
uniform and smaller than the expected magnitude of the neglected terms.
Further insight is obtained by the use of Fourier-accelerated numerical
evaluation of Feynman diagrams and extrapolations of finite-volume results
to infinite volume by phenomenological finite size scaling
\cite{Flyvbjerg-Larsen-support,Flyvbjerg-Larsen-susceptibility}.

These improved results lead to agreement with Monte Carlo data, within the
expected errors, for $N\ge3$, and give for $N=3,4$ extrapolated mass gap --
$\Lambda$-parameter ratios consistent with the exact continuum results of
Ref.~\cite{Hasenfratz-Maggiore-Niedermayer}, reported in
Eq.~(\ref{Lambda-ON}).

However a fully analytic approach to higher orders in the $1/N$ expansion
would require extracting the scaling contributions along the lines defined
in principle in Sect.~\ref{physical quantities}.  This extraction in turn
would involve a proper treatment of the regularization problem, which may
not be straightforward in presence of higher loops, if we want to stick to
our favorite SM scheme: we might run into technical problems similar to
those involved in generalizing the BPHZ scheme to massless theories. In any
case, based on the proofs of renormalizability of the $1/N$ expansion, we
believe there should be no general obstruction to such a calculation.

\section{A different approach: Schwinger--Dyson equations}
\label{Schwinger-Dyson}

The approach to large $N$ based on the effective action and effective
Feynman rules is by no means the only way of generating an expansion that
is naturally organized in powers of $1/N$. Writing down Schwinger--Dyson
equations for ${\rm U}(N)$ (${\rm O}(2N)$) invariant correlation functions,
it is possible to recognize that $N$ occurs only polynomially in the
coefficients of the equations themselves. It is therefore possible to
truncate the (a priori infinite) set of Schwinger--Dyson equations by
keeping only terms and equations down to a chosen power of $N$. When the
resulting finite set of equations is solved, the solution depends on $1/N$
through all powers, and it is equal to the sum of an infinite subseries of
the exact $1/N$ series. It is therefore at least as accurate as the
corresponding truncated $1/N$ series; in practice one can get sensibly
higher accuracy, as shown in the original papers by Drouffe and Flyvbjerg,
explicitly concerned with ${\rm O}(N)$ models
\cite{Flyvbjerg-DS,Drouffe-Flyvbjerg-letter,Drouffe-Flyvbjerg}.

The derivation of the Schwinger--Dyson equations is essentially
straightforward in the generating functional formalism, and we refer to
\cite{Drouffe-Flyvbjerg} for all details. We just present in
Fig.~\ref{SD-graph} the graphical form the equations involving the
fundamental field and the $\alpha$-field propagators in the ${\rm O}(2N)$
model, truncated to $O(1/N^2)$.
There is an ambiguity in the location of the bare and dressed vertices; we
fixed it by the prescription that each $\alpha$-field propagator should
connect a bare and a dressed vertex.

We tested our prescription by a very simple and illuminating example, the
exactly-solvable continuum one-dimensional ${\rm O}(2N)$ model.  Since the
$O(1/N)$ truncation amounts to identifying the bare and dressed vertex, we
may try to solve the resulting equations by the Ansatz
\begin{equation}
G^{-1}(p) = A(p^2+m^2).
\end{equation}
Simple integrations (performed for convenience in dimensional
regularization) lead to an explicit solution in the form
\begin{mathletters}
\label{one-dim}
\begin{eqnarray}
m &=& {2N-1\over4N\beta}\,, \\
G^{-1}(p) &=& {2N\over2N-1}\,(p^2+m^2), \\
\Delta^{-1}(p) &=&
\left(2N-1\over2N\right)^2{N\over m}\,{1\over p^2+4m^2}\,.
\end{eqnarray}
\end{mathletters}
Actually Eqs.~(\ref{one-dim}a) and (\ref{one-dim}b) correspond to the
{\it exact\/} solution of the one-dimensional ${\rm O}(2N)$ model
\cite{Stanley}, thus showing the power of the approach: albeit truncated to
$O(1/N)$, the equations enable us to resum the whole $1/N$ series.

We must now verify that the $O(1/N^2)$ truncation does not spoil the
result. However, because of our prescription, this is simply achieved by
the Ansatz
\begin{mathletters}
\begin{eqnarray}
\Delta^{-1}(p) &=& {N\over m}
\left(2N-1\over2N\right)^2{1\over p^2+4m^2}\,V(p), \\
V_{(\alpha)}(p,p') &=& V(p-p'),
\end{eqnarray}
\end{mathletters}
insuring that the value of $G^{-1}(p)$ is unchanged, while by consistency
one can find
\begin{equation}
V(p) = {1\over\displaystyle1 - {1\over2N}
\left(1 + {8m^2\over p^2+4m^2}\right)}.
\label{one-V}
\end{equation}
Eq.~(\ref{one-V}) can be verified in the context of the standard $1/N$
expansion.

Although rather promising, the Schwinger--Dyson approach has till now only
be applied to the $O(1/N)$ truncation of the two-dimensional ${\rm O}(N)$
models \cite{Drouffe-Flyvbjerg}; by comparing their results with
high-precision Monte Carlo data for the ${\rm O}(3)$ and ${\rm O}(4)$
models, the authors find an apparent uniform systematic error of
$O(1/N^3)$. Higher orders and more general models involve, besides the
abovementioned problem of locating the dressed vertices, technical
difficulties related to performing the higher-loop integration with the
needed accuracy.

\section{Fermionic models}
\label{fermionic models}

The next obvious extension of the $1/N$ approach in the study of model
field theories is considering fermionic degrees of freedom. The natural
counterparts of the bosonic models we have discussed are theories with
four-Fermi interactions and $U(N)$ symmetry, power-counting renormalizable
in two dimensions and $1/N$ expandable.  These theories are ultraviolet
renormalizable order by order in the $1/N$ expansion in less than four
dimensions, as shown by Rosenstein and coworkers
\cite{Rosenstein-Warr-Park-renorm,Gat-Kovner-Rosenstein-Warr,%
Rosenstein-Warr-Park-breaking} and by Kikukawa and Yamawaki
\cite{Kikukawa-Yamawaki}, essentially because their ultraviolet limit has
the same relevant operator content as the infrared limit of the
superrenormalizable Yukawa model.  These arguments were further developed
in Refs.~\cite{He-Kuang-Wang-Yi,ZinnJustin,Hands-Kocic-Kogut-I}, and the
whole subject is carefully reviewed in the introduction of
Ref.~\cite{Hands-Kocic-Kogut-II}.  Critical indices were computed by Gracey
to $O(1/N^2)$ \cite{Gracey-GN-eta}.

A sufficiently general ${\rm U}(N)$-invariant two-dimensional continuum
Euclidean Lagrangian depends on three couplings
\cite{Bondi-Curci-Paffuti-Rossi-central-charge}:
\begin{equation}
{\scr L} = \bar\psi\slashed\partial\psi
- \half g_{\rm s}\left(\bar\psi\psi\right)^2
- \half g_{\rm p}\left(\bar\psi\gamma_5\psi\right)^2
- \half g_{\rm v}\left(\bar\psi\gamma_\mu\psi\right)^2,
\label{L-fermion-cont}
\end{equation}
where $\psi$ is an $N$-plet of Dirac fermions.  This Lagrangian
interpolates between the ${\rm O}(2N)$-symmetric Gross--Neveu model
($g_{\rm p} = g_{\rm v} = 0$) and the ${\rm SU}(N)$-symmetric Gross--Neveu
model  ($g_{\rm s} = g_{\rm p} = N g_{\rm v}$), enjoying a global axial
${\rm U}(1)$ invariance.

Gross--Neveu and chiral Gross--Neveu models possess factorized $S$-matrices
in two dimensions, and other exact results can be obtained, in full analogy
with their bosonic partners. The related literature has been steadily
growing in the last twenty years, and we shall not even try to give
references for the continuum formulation, addressing the interested reader
to the (partial) bibliography appearing in
Ref.~\cite{Abdalla-Abdalla-Rothe}.  However, as already observed for
bosonic models, only a minor effort was done in going beyond the leading
large-$N$ approximation, both in the continuum and in the lattice versions
of the models.

Lattice formulations, as it is well known, are plagued by the supplementary
problem of fermion doubling. The problem is solved in principle in any
dimension by the introduction of the Wilson term \cite{Wilson-lecture},
which however leads to an unavoidable complication in both analytical and
numerical computations. In the present context we only want to summarize
the available results concerning the large-$N$ limit and the $1/N$
expansion of lattice Gross--Neveu and the chiral Gross--Neveu models, and
include a new result coming as a rather natural extension of our previous
analysis.

The first lattice formulations of the Gross--Neveu models admitting the
correct continuum limit were presented in
Ref.~\cite{Cohen-Elitzur-Rabinovici}, and involved staggered fermions
\cite{Susskind}; therefore they described models with ${\rm O}(4N)$
symmetry, $N$ being the number of ``na\"\i ve'' fermionic components (cf.\
also Ref.~\cite{Jolicoeur-Morel-Petersson}).  A lattice formulation with
Wilson fermions of the chiral Gross--Neveu model was introduced and
discussed in the large-$N$ limit in Ref.~\cite{Eguchi-Nakayama} (see also
Ref.~\cite{Aoki-Higashijima}).  The Gross--Neveu model in the Wilson
formulation was studied to $O(1/N)$ in the seminal paper by David and
Hamber \cite{David-Hamber}, where the notion of an asymptotic expansion of
the effective propagator was introduced. In
Refs.~\cite{Wetzel-impr,Ma-Wetzel} the Symanzik improvement program was
applied to the large-$N$ Gross--Neveu model with Wilson fermions. A
systematic analysis of the $1/N$ contributions to the lattice Gross--Neveu
model was finally performed in the staggered version in
Ref.~\cite{Belanger-Lacaze-Morel-Attig-Petersson-Wolff}, and in the
Wilson--Symanzik version in Ref.~\cite{Campostrini-Curci-Rossi}.  For
completeness we must mention that a lattice formulation of $\CPN$ models
coupled to fermions was discussed in Ref.~\cite{Ichinose}, and that the
problem of formulating lattice versions of two-dimensional $1/N$-expandable
supersymmetric models was addressed by a few authors
\cite{DiVecchia-Musto-Nicodemi-Pettorino-Rossi,Abdalla-Abdalla-Kawamoto},
but never systematically investigated.

Our new result concerns the possibility of obtaining an integral
representation of the effective propagator $\Delta_{(\sigma)}^{-1}$ in a
staggered version of the Gross--Neveu model, defined by the action
\begin{equation}
S = \sum_{x,y} \bar\chi_x D_{xy} \chi_y + \sum_x \bar\chi_x\chi_x\Sigma_x
+ N \sum_x {\sigma^2_x\over2f}\,,
\end{equation}
where $\chi_x$ is a $N$-plet of (one-component) fermionic fields, $D_{xy}$
is the Susskind ``differential'' operator
\begin{equation}
D_{xy} = \half[\delta_{x,y{+}\hat1} - \delta_{x,y{-}\hat1}] +
\half(-1)^{x_1}[\delta_{x,y{+}\hat2} - \delta_{x,y{-}\hat2}],
\end{equation}
and the fermions are coupled to the Lagrange-multiplier field $\sigma$ by
\begin{equation}
\Sigma_x = \fourth(\sigma_x + \sigma_{x{-}\hat1}
 + \sigma_{x{-}\hat2} + \sigma_{x{-}\hat1{-}\hat2}).
\end{equation}
In this case, we can obtain an integral representation of the fermionic
integral
\begin{eqnarray}
\Delta_{(\sigma)}^{-1} &=& \intp \tr\left\{
{1\over i\gamma_\mu\,\overline{p{+}\half k}_\mu + m_0}\,
{1\over i\gamma_\mu\,\overline{p{-}\half k}_\mu + m_0}\right\} \nonumber \\
&=& 2\intp {m_0^2 - \sum_\mu\overline{p{+}\half k}_\mu\,
	\overline{p{-}\half k}_\mu \over
	\left(\overline{p{+}\half k}^2 + m_0^2\right)
	\left(\overline{p{-}\half k}^2 + m_0^2\right)}\,,
\end{eqnarray}
where $\bar p_\mu = \sin p_\mu$, and the mass parameter $m_0$ is
momentum-independent.  By repeating the arguments of Sect.~\ref{integral
representations}, we find that $\Delta_{(\sigma)}^{-1}$ can be computed in
closed form along the principal diagonal of the momentum lattice:
\begin{eqnarray}
\Delta^{-1}_{(\sigma)}(l,l) &=& {4\over\pi}\,{1\over1+m_0^2}
\left({m_0^2-\cos l\over m_0^2+1} + {1\over\cos l}\right)
\Pi\!\left({\cos^2 l\over\left(1+m_0^2\right)^2},{1\over1+m_0^2}\right)
\nonumber \\
&-&\, {4\over\pi}\,{1\over1+m_0^2}\,{1\over\cos l}\,
K\!\left({1\over1+m_0^2}\right).
\end{eqnarray}
More generally, using the standard Feynman parameter representation we
obtain
\begin{equation}
\Delta^{-1}_{(\sigma)} = 2 \int_0^1\D x
\intq {\bar c + \sum_\mu \bar b_\mu\cos q_\mu\over
\left[1+m_0^2 - \sum_\mu \bar a_\mu\cos q_\mu\right]^2},
\end{equation}
where
\begin{mathletters}
\begin{eqnarray}
\bar a_\mu &=& \half\sqrt{1-x(1-x)\hat k_\mu^2}\,, \\
\bar b_\mu &=& {\cos k_\mu\over4\bar a_\mu}\,, \\
\bar c &=& m_0^2 - \half \sum_\mu\cos k_\mu\,.
\end{eqnarray}
\end{mathletters}
By essentially trivial algebraic manipulation we are led to the final
result
\begin{eqnarray}
\Delta^{-1}_{(\sigma)}(k) &=&
{m_0^2 + \bar c(1+2m_0^2)\over4\pi}\int_0^1\D x \,
{1\over(\bar a_1\bar a_2)^{3/2}}\,{\bar\zeta^3 E(\bar\zeta)
	\over 1-\bar\zeta^2}
\nonumber \\ &+&\, {1\over2\pi}\int_0^1\D x \,
{\bar\zeta\over(\bar a_1\bar a_2)^{1/2}}\sum_\mu{\bar b_\mu\over\bar a_\mu}
\left[{E(\bar\zeta) \over 1-\bar\zeta^2} + E(\bar\zeta)
	- 2 K(\bar\zeta)\right] \,,
\end{eqnarray}
where
\begin{equation}
\bar\zeta = \sqrt{4\bar a_1\bar a_2 \over
	(1+m_0^2)^2 - (\bar a_1-\bar a_2)^2}\,.
\end{equation}

\section{Conclusions and outlook}
\label{conclusions}

In our opinion the most important conclusions that can be drawn from our
results are summarized by the following statements.
\begin{enumerate}
\item
Nontrivial asymptotically-free two-dimensional Euclidean field theories can
be constructed, in the context of the $1/N$ expansion, starting from a
lattice formulation and exhibiting explicitly the existence of a scaling
region.  The accuracy of our construction is $O(1/N)$, but there is no
obstruction to higher-order extensions.
\item
In the scaling region, results that are expressible in terms of
adimensional ratios of physical quantities are {\it universal}, i.e.\
they do not depend on the specific lattice model adopted as long as the
physical parameters are kept fixed. Moreover these results are unaffected
by the pathologies of standard perturbation theory and can be unambiguously
predicted.
\item
The {\it width\/} of the scaling region however necessarily depends on the
choice of a lattice action.  In turn, it is widely independent of the $1/N$
corrections, since these depend on the effective propagators and vertices,
whose scaling properties are fixed by the (large-$N$) effective action and
are modeled upon the scaling properties of the large-$N$ lattice mass gap
(cf.\ Eq.~(\ref{mu0-scaling}) and Fig.~\ref{large-N-fig}).  For standard
nearest-neighbor interactions, scaling within $10^{-3}$ is achieved
starting from $2f\simeq1.25$ ($\beta\simeq0.8$), corresponding (for
not too small $N$) to a correlation length $1/(ma)\simeq27$.
\item
``There ain't nothing like asymptotic scaling'' in the real world
(excluding very large $N$).  The asymptotic scaling region is in our
language the small-$f$ region, where the behavior of $m_1^2/m_0^2$ is well
approximated by the two-loop perturbative renormalization group, i.e.\ its
finite part (in the notation of Sect.~\ref{continuum 1/N}) is very close to
its value at $f=0$.  As an example, for the ${\rm O}(N)$ models (the ``best
case'') at $c_2=0$, with $N$ as large as 20, the mass gap is approximated
by the (two-loop) asymptotic formula within $10^{-3}$ for
$\beta\gtrsim3.3$, i.e.\ $1/(ma)>10^8$.
\item
Perturbation theory may however be a good guide to the physics of the
models, in that it commutes order by order with the $1/N$ expansion and it
leads to the same renormalization-group functions and asymptotic behaviors.
Moreover, by summing over a sufficient number of perturbative terms one may
reproduce the correct lattice $\Lambda$ parameter, renormalization
constants and perturbative tails throughout the whole scaling region (cf.\
Figs.~\ref{delta-m}--\ref{delta-m-CPN}).
\end{enumerate}

As we have shown, the $1/N$ approach may be successfully extended in many
different directions.  The major limitation we could not however bypass is
the restriction to models where the fields belong to the fundamental
(vector) representation of the symmetry group. The problem of extension to
fields in the adjoint (matrix) representation, like principal chiral models
and gauge theories, has been the stumbling block of the $1/N$ expansion in
the last decade. A breakthrough in this domain could turn the $1/N$
expansion from a toy in the theoretical playground into a major tool in the
analysis of realistic physical models of the fundamental interactions.

\subsection*{Acknowledgments}

We thank M. Maggiore, A. Pelissetto and E. Vicari for critical reading of
the manuscript.

\appendix
\section{Perturbative results}
\label{perturbative results}

In the literature on perturbative calculations, it is usual to report the
results in terms of a rescaled renormalized coupling $t$, following the
notation first adopted in Ref.~\cite{Brezin-ZinnJustin-breakdown}.  All
four-loop-order $\beta$ functions of nonlinear $\sigma$ models on symmetric
spaces can be found in Ref.~\cite{Wegner}.  Here we are only interested in
two special cases:
\begin{enumerate}
\item
${\rm O}(N)/{\rm O}(N{-}1)$ spaces
\begin{equation}
\beta(t) \cong \varepsilon t - (N-2)t^2\left[1 + t + \fourth(N+2)t^2 +
\left(-\case1/{12} N^2 + \case{11}/6 N - \case{17}/6 +
\case3/2(N-3)\zeta(3)\right) t^3\right]\!.
\end{equation}
The result for ${\rm O}(N)$ models is obtained by setting
$t=1/(2\pi N\beta_{\rm v})$.
\item
${\rm U}(N)/({\rm U}(N{-}1){\times}{\rm U}(1))$ spaces
\begin{equation}
\beta(t) \cong \varepsilon t - N t^2\left[1 + 2 t +
\left(\case3/2 N + 2\right)t^2 +
\left(\third N^2 + \case{13}/2 N + 1\right) t^3\right].
\end{equation}
The result for $\CPN$ models is obtained by setting
$t=1/(2\pi N\beta_{\rm g})$.
\end{enumerate}

The local nonderivative scaling operators can be expressed in terms of
orthogonal polynomials in the variable $\sigma^2 = \bar z_1 z_1$
\cite{Hof-Wegner} (see also \cite{Rossi-Brihaye}).  Their anomalous
dimensions were computed to four-loop order:
\begin{enumerate}
\item
${\rm O}(N)$ models
\item[ ]
The scaling operators are the Gegenbauer polynomials
$C_l^{({N\over2}-1)}(\sigma)$ and
\begin{eqnarray}
\gamma_l(t) &\cong& l(N+l-2)\bigl\{t + \case3/4(N-2)t^3 \nonumber \\
&&\quad+\; (N-2)
\left[-\third N + \case5/3 + \half \zeta(3)\left(1-\half l(N+l-2)\right)
\right]t^4\bigr\}.
\end{eqnarray}
\item
$\CPN$ models
\item[ ]
The scaling operators are the Jacobi polynomials
$P_k^{[N-2,0]}(2\sigma^2{-}1)$ and
\begin{equation}
\gamma_k(t) \cong 2 k(N+k-1)\left\{t + \case3/2 N t^3 + (N+6)
\left[\third N + \fourth\zeta(3)\left(N-k(N+k-1)\right) \right] t^4
\right\}.
\end{equation}
\end{enumerate}

The $\varepsilon$-expansion of the critical exponents may be extracted
from the above results by finding the critical point $t^*$ defined by
$\beta(t^*)=0$ and applying the relationships
\begin{mathletters}
\begin{eqnarray}
\eta &=& -\varepsilon + \gamma_1(t^*), \\
\nu &=& - {1\over\beta'(t^*)}\,,
\end{eqnarray}
\end{mathletters}
holding for ${\rm O}(N)$ models.

\section{Effective propagators in $\pbbox{\lowercase{d}}$ dimensions}
\label{d propagators}

The mass-gap equation takes the form
\begin{equation}
\beta = {\Gamma\left(1{-}\half d\right)\over(4\pi)^{d/2}}\,
(m_0^2)^{d/2 - 1}.
\end{equation}

The inverse propagator of the $\alpha$ field is
\begin{eqnarray}
\Delta^{-1}_{(\alpha)}(p) &=& \int {\D^d p \over (2\pi)^d}
\,{1\over q^2+m_0^2}\,{1\over (p+q)^2+m_0^2} \nonumber \\
&=& {\Gamma\left(2{-}\half d\right)\over(4\pi)^{d/2}}
\int_0^1{\D x \over \left[p^2x(1-x) + m_0^2\right]^{2-d/2}} \nonumber \\
&=& {\Gamma\left(2{-}\half d\right)\over(4\pi)^{d/2}}
\left(\fourth p^2+m_0^2\right)^{d/2-2}
F\!\left(2{-}{1\over2}d,{1\over2};{3\over2};{1\over\xi^2}\right),
\end{eqnarray}
where $F$ is the hypergeometric function.  For $p=0$, the inverse
propagator assumes the value
\begin{equation}
\Delta^{-1}_{(\alpha)}(0) =
{\Gamma\left(2{-}\half d\right)\over(4\pi)^{d/2}}
\left(m_0^2\right)^{d/2-2}.
\end{equation}

Special values for the lowest integer dimensions are
\begin{mathletters}
\begin{eqnarray}
d=0: &\quad& \Delta^{-1}_{(\alpha)}(p)
 = {8\over\left(p^2\right)^2}\,{1\over\xi^2}\left[{1\over\xi^2-1}
+{1\over2\xi}\ln{\xi+1\over\xi-1}\right]
= {2\beta\over p^2+4m_0^2} + {4\xi\over\left(p^2+4m_0^2\right)^2}
\ln{\xi+1\over\xi-1}\,, \nonumber \\ \\
d=1: &\quad& \Delta^{-1}_{(\alpha)}(p) = {1\over m_0}\,{1\over p^2+4m_0^2}
 = {2\beta\over p^2+4m_0^2}\,, \\
d=2: &\quad& \Delta^{-1}_{(\alpha)}(p)
 = {1\over2\pi p^2}\,{1\over\xi}\ln{\xi+1\over\xi-1}
 = {1\over p^2+4m_0^2}\,{\xi\over2\pi}\ln{\xi+1\over\xi-1}\,, \\
d=3: &\quad& \Delta^{-1}_{(\alpha)}(p)
 = {1\over4\pi p}\,\arctan{p\over2m_0}\,, \\
d=4: &\quad& \Delta^{-1}_{(\alpha)}(p)
 = -{1\over(4\pi)^2}\left[\xi\ln{\xi+1\over\xi-1} - 1\right]
   - {\beta\over m_0^2}\,.
\end{eqnarray}
\end{mathletters}

The inverse propagator of the $\theta$ field is
\begin{equation}
\Delta^{-1}_{(\theta)\,\mu\nu}(p) =
\left(\delta_{\mu\nu}-{p_\mu p_\nu\over p^2}\right)
\Delta^{-1}_{(\theta)}(p),
\end{equation}
where
\begin{eqnarray}
\Delta^{-1}_{(\theta)}(p) &=&
\int {\D^d p \over (2\pi)^d}\,{2\over q^2+m_0^2} - {2\over d-1}
\int {\D^d p \over (2\pi)^d}\,{q\cdot(p+q) - m_0^2
  \over \left[q^2+m_0^2\right] \left[(p+q)^2+m_0^2\right]} \nonumber \\
&=& 2\,{\Gamma\left(1{-}\half d\right)\over
	(4\pi)^{d/2}} \left[(m_0^2)^{d/2-1}
- \int_0^1{\D x \over \left[p^2x(1-x) + m_0^2\right]^{1-d/2}}\right]
 \nonumber \\
&=& 2\,{\Gamma\left(1{-}\half d\right)\over
	(4\pi)^{d/2}} \left[ (m_0^2)^{d/2-1}
- \left(\fourth p^2+m_0^2\right)^{d/2-1}
F\!\left(1{-}{1\over2}d,{1\over2};{3\over2};{1\over\xi^2}\right)\right].
\end{eqnarray}
For $p=0$, we have
\begin{equation}
\Delta^{-1}_{(\theta)}(0) = 0.
\end{equation}

\section{Continuum integrals}
\label{continuum integrals}

A number of continuum integrals occurring in the evaluation of the
constants appearing in $O(1/N)$ results can be computed analytically.  A
typical dimensionless SM-regulated one-loop integral in the $1/N$
expansion, after angular integration has been performed, takes the form
\begin{equation}
I^{\rm reg} = m_0^{-\Delta-2}\left[
\int_0^\infty {\D p^2\over4\pi}\,F(p^2,m_0^2) -
\int_0^\infty {\D p^2\over4\pi}\,{\rm T}^{\rm(UV)}F(p^2,m_0^2) -
\int_{M^2}^\infty {\D p^2\over4\pi}\,{\rm T}^0 F(p^2,m_0^2) \right],
\end{equation}
where ${\rm T}^{\rm(UV)}F$ are the terms in the series expansion of $F$ in
powers of $m_0^2$ that are only ultraviolet-divergent (``perturbative
tails''), while ${\rm T}^0 F$ are the terms that are both ultraviolet and
infrared divergent; $\Delta$ is the canonical dimension of $F$.
$F(p^2,m_0^2)$ may be represented in the form
\begin{equation}
m_0^\Delta\hat F\left(p^2\over m_0^2\right) = m_0^\Delta f(\xi),
\end{equation}
where $\xi = \sqrt{1+4 m_0^2/p^2}$.

It is now convenient to perform the following changes of integration
variable: in the first integral we set
\begin{equation}
t = {\xi+1\over\xi-1}, \qquad \D p^2 = m_0^2\left(1-{1\over t^2}\right)\D t;
\end{equation}
in the last two integrals we set
\begin{equation}
u = {p^2\over m_0^2} = {(t-1)^2\over t}, \qquad \D p^2 = m_0^2\,\D u.
\end{equation}
By observing that $t = u + 2 + O(1/u)$, we obtain the representation
\begin{equation}
4\pi I^{\rm reg} = \lim_{\Lambda\to\infty}\left\{
\int_1^{\Lambda+2}\D t\left(1-{1\over t^2}\right)f\left(t+1\over t-1\right)
- \int_0^\Lambda\D u\,{\rm T}^{\rm(UV)}\hat F(u)
- \int_{M^2/m_0^2}^\Lambda\D u\,{\rm T}^0\hat F(u)\right\}.
\end{equation}

Now, in order to parametrize explicitly the regulator-dependent part of
$I^{\rm  reg}$ we can split the last integral into the two regions
$M^2/m_0^2<u<\exp(a)$ and $\exp(a)<u$, with $a=1$ for integrals involving
$\Delta^{-1}_{(\alpha)}$ and  $a=3-2\pi\kappa$ for integrals involving
$\Delta^{-1}_{(\theta)}$.  The integration
\begin{equation}
\int_{M^2/m_0^2}^{\exp(a)} \D u\,{\rm T}^0\hat F(u)
\end{equation}
is now trivial, and we are left with the task of evaluating
\begin{equation}
c \equiv 4\pi I^{\rm reg} = \lim_{\Lambda\to\infty}\left\{
\int_1^{\Lambda+2}\D t\left(1-{1\over t^2}\right)f\left(t+1\over t-1\right)
- \int_0^\Lambda\D u\,{\rm T}^{\rm(UV)}\hat F(u)
- \int_{\exp(a)}^\Lambda\D u\,{\rm T}^0\hat F(u)\right\}.
\end{equation}

Exact analytic results have been obtained in the following instances:
\begin{eqnarray}
f(\xi) &=& \ln\ln{\xi+1\over\xi-1}, \nonumber \\
c &=& \lim_{\Lambda\to\infty}\left\{
\int_1^{\Lambda+2}\D t\left(1-{1\over t^2}\right)\ln\ln t
- \int_0^\Lambda\D u \ln\left|\ln u\right|
- \int_e^\Lambda{\D u\over u}\,{2\over\ln u}\right\} \nonumber \\
&=& -\int_1^\infty {\D t\over t^2} \ln\ln t
- \int_0^1 \D u \ln\left|\ln u\right| =
-2 \int_1^\infty{\D t\over t^2} \ln\ln t \nonumber \\
&=& 2 \gamma_E\,;
\end{eqnarray}
\begin{eqnarray}
f(\xi) &=& {\displaystyle1-{1\over\xi}\over\displaystyle\ln{\xi+1\over\xi-1}},
	\nonumber \\
c &=& \lim_{\Lambda\to\infty}\left\{
\int_1^{\Lambda+2}\D t\left(t-1\over t^2\right)\,{2\over\ln t}
- \int_e^\Lambda{\D u\over u}\,{2\over\ln u}\right\} =
-2 \int_1^\infty{\D t\over t^2} \ln\ln t \nonumber \\
&=& 2 \gamma_E\,;
\end{eqnarray}
\begin{eqnarray}
f(\xi) &=& {\displaystyle{1\over\xi}-{4\xi\over3+\xi^2}
       \over\displaystyle\ln{\xi+1\over\xi-1}}, \nonumber \\
c &=& \lim_{\Lambda\to\infty}\left\{
- \int_1^{\Lambda+2}\D t\,{(t-1)^2\over t}\,{3\over\ln t}\,{1\over t^2-t+1}
+ \int_e^\Lambda{\D u\over u}\,{3\over\ln u}\right\} \nonumber \\
&=& -3 \int_1^\infty\D t\,{1-t^2\over\left(t^2-t+1\right)^2}\,\ln\ln t
= -3\left(\gamma_E-c_1\right),
\end{eqnarray}
where $c_1$ is given by Eq.~(\ref{c1});
\begin{eqnarray}
f(\xi) &=& {\displaystyle\sqrt{1-{1\over\xi^2}}
       \over\displaystyle2\ln{\xi+1\over\xi-1}},
	\nonumber \\
c &=& \lim_{\Lambda\to\infty}\left\{
\int_1^{\Lambda+2}\D t\,{t-1\over t\sqrt t}\,{1\over\ln t}
- \princint_0^\Lambda{\D u\over\sqrt u}\,{1\over\ln u}\right\} \nonumber \\
&=& \lim_{\varepsilon\to0} \left\{
- \int_{1+\varepsilon}^\infty{\D t\over t\sqrt t}\,{1\over\ln t}
- \int_0^{1-\varepsilon}{\D u\over\sqrt u}\,{1\over\ln u} \right\}
 \nonumber \\ &=& 0\,.
\end{eqnarray}

\section{Effective vertices in the continuum}
\label{effective vertices}

The effective vertices of the $1/N$ expansion are nothing but one-loop
integrals over the fundamental field propagators with appropriate couplings
to the external lines.  The problem of evaluating the most general
continuum one-loop integral in two dimension is solved in principle in
terms of elementary functions \cite{Kallen-Toll,Petersson,Schonfeld}.  It
is however convenient to derive explicit expressions for those special
kinematic configurations entering the actual computations we would like to
perform \cite{Campostrini-Rossi-CPN}.

One basic ingredient is the three-point scalar vertex
\begin{equation}
V_3(p_1,p_2) \equiv \intq\,{1\over q^2+m_0^2}
   \,{1\over (q+p_1)^2+m_0^2}\,{1\over (q+p_2)^2+m_0^2}\,,
\end{equation}
a symmetric function of $p_1$, $p_2$ and $p_1-p_2$.
Two-dimensional identities allow a reduction of the integrand to a
combination of terms involving only two fundamental field propagators.  The
integration is then straightforward, and the result is
\begin{eqnarray}
V_3(p_1,p_2) &=& D^{-1}(p_1,p_2)\Bigl[ p_1^2(p_2(p_2-p_1))
	\Delta_{(\alpha)}^{-1}(p_1)
\nonumber \\&&\qquad+\quad p_2^2(p_1(p_1-p_2))\Delta_{(\alpha)}^{-1}(p_2) +
   (p_1-p_2)^2(p_1p_2)\Delta_{(\alpha)}^{-1}(p_1-p_2)\Bigr],
\end{eqnarray}
where
\begin{equation}
D(p_1,p_2) = p_1^2p_2^2(p_1-p_2)^2 +
4m_0^2\left[p_1^2p_2^2-(p_1p_2)^2\right].
\end{equation}

Let us now consider the four-point vertices: the exceptional configurations
we are interested in are the cases when the external momenta are equal two
by two. Let us define
\begin{equation}
V_4^{(a)}(p_1,p_2) \equiv \intq\,{1\over \left[q^2+m_0^2\right]^2}
   \,{1\over (q+p_1)^2+m_0^2}\,{1\over (q+p_2)^2+m_0^2}\,.
\end{equation}
Again by applying algebraic identities we are led to an explicitly
integrable expression. The final result is
\begin{eqnarray}
V_4^{(a)}(p_1,p_2) &=& D^{-1}(p_1,p_2)
   \Biggl[{\left(p_2^2-(p_1p_2)\right)p_1^2\over p_1^2+4m_0^2}
      \left(\Delta_{(\alpha)}^{-1}(p_1)+\Delta_{(\alpha)}^{-1}(0)\right)
	\nonumber \\
&&\qquad\qquad+\quad{\left(p_1^2-(p_1p_2)\right)p_2^2\over p_2^2+4m_0^2}
      \left(\Delta_{(\alpha)}^{-1}(p_2)+\Delta_{(\alpha)}^{-1}(0)\right)
	\Biggr] \nonumber \\
&+&\quad D^{-2}(p_1,p_2) \Bigl\{
   \left[(p_1-p_2)^2(p_1p_2) + p_1^2p_2^2 - (p_1p_2)^2\right] \nonumber \\
&&\qquad\qquad\times\quad
   \left[\left(p_2^2-(p_1p_2)\right)p_1^2\,\Delta_{(\alpha)}^{-1}(p_1) +
    \left(p_1^2-(p_1p_2)\right)p_2^2\,\Delta_{(\alpha)}^{-1}(p_2)\right]
    \nonumber \\
&&\qquad\qquad+\quad \left[(p_1-p_2)^2(p_1p_2)\right]^2\,
	\Delta_{(\alpha)}^{-1}(p_1-p_2) \Bigr\} \nonumber \\
&-&\quad D^{-2}(p_1,p_2) \left[p_1^2p_2^2 - (p_1p_2)^2\right] \Bigl\{
 (p_1^2+4m_0^2)((p_1p_2)-p_1^2)\Delta_{(\alpha)}^{-1}(p_1) \nonumber \\
&&\qquad\qquad+\quad (p_2^2+4m_0^2)((p_1p_2)-p_2^2)
	\Delta_{(\alpha)}^{-1}(p_2) \nonumber \\
&&\qquad\qquad+\quad \left[(p_1-p_2)^2+4m_0^2\right](p_1-p_2)^2\,
	\Delta_{(\alpha)}^{-1}(p_1-p_2) \Bigr\}.
\end{eqnarray}

We must also evaluate
\begin{equation}
V_4^{(b)}(p_1,p_2) \equiv \intq\,
   {1\over q^2+m_0^2}\,{1\over (q+p_1)^2+m_0^2}\,
   {1\over (q+p_2)^2+m_0^2}\,{1\over (q+p_1+p_2)^2+m_0^2}\,.
\end{equation}
One can show that the result is expressible in the form
\begin{equation}
V_4^{(b)}(p_1,p_2) =
{1\over(p_1p_2)}\left[V_3(p_1,p_2) - V_3(p_1,-p_2)\right].
\end{equation}

In order to compute the correlation function of the composite operator
$P_{ij}(x)$, we also need the mixed four point scalar-vector vertices in
exceptional momentum configurations.  We quote here the definitions:
\begin{mathletters}
\begin{eqnarray}
   V_{\mu\nu}^{(a)}(p,k) &=& \intq\,{1\over(q^2+m_0^2)^2}\,
   {(2q_\mu+k_\mu)(2q_\nu+k_\nu) \over
	\left[(q+p)^2+m_0^2\right]\left[(q+k)^2+m_0^2\right]}\,,  \\
   V_{\mu\nu}^{(b)}(p,k) &=& \intq\,{1\over q^2+m_0^2}\,
   {2q_\mu+k_\mu\over(q+k)^2+m_0^2}\,{1\over (q+p)^2+m_0^2}\,
   {2q_\nu+2p_\nu+k_\nu\over(q+p+k)^2+m_0^2}\,.
\end{eqnarray}
\end{mathletters}

Actually we only need the combination of vertices appearing in
Fig.~\ref{Delta1} and this can be shown to be a transverse tensor.
Therefore we can limit ourselves to computing
\begin{mathletters}
\begin{eqnarray}
\left(\delta_{\mu\nu}-{k_\mu k_\nu\over k^2}\right)
	V_{\mu\nu}^{(a)}(p,k)
&=& - (k^2+4m_0^2)V_4^{(a)}(p,k) + 2V_3(p,k)
   + \left(1-{4m_0^2\over k^2}\,{(p\cdot k)\over p^2}\right)
   {\Delta_{(\alpha)}^{-1}(0)\over p^2+4m_0^2} \nonumber \\
&+&\,\left[(p+k)^2+4m_0^2\left(1+{(p\cdot k)\over p^2}
	\right)\right]
   {\Delta_{(\alpha)}^{-1}(p)\over k^2(p^2+4m_0^2)}
   - {1\over k^2}\,\Delta_{(\alpha)}^{-1}(p-k) \nonumber \\
\end{eqnarray}
and
\begin{eqnarray}
\left(\delta_{\mu\nu}-{k_\mu k_\nu\over k^2}\right)
	V_{\mu\nu}^{(b)}(p,k)
&=&   - (k^2+2p^2+4m_0^2) V_4^{(b)}(p,k)
   - {2\over k^2}\,\Delta_{(\alpha)}^{-1}(p)
   + {1\over k^2}\,\Delta_{(\alpha)}^{-1}(p-k) \nonumber \\
&+&\, {1\over k^2}\,\Delta_{(\alpha)}^{-1}(p+k) + 2V_3(p,k)+2V_3(p,-k)\,.
\end{eqnarray}
\end{mathletters}

\section{The bound state equation in the large-$\pbbox{N}$ limit}
\label{Schrodinger equation}
As discussed in Sect.~\ref{continuum 1/N}, we must solve the following
eigenvalue Schr\"odinger equation:
\begin{equation}
-{\D^2\psi\over\D \rho^2} + {6\pi\over N}\,A(x)\left(1-e^{-x\rho}\right)\psi
= \varepsilon\psi,
\label{Schrodinger-orig}
\end{equation}
where $\rho=m_0R$ and $x=m_\theta/m_0$.
The resulting bound state masses will be $m_B = m_0(2+\varepsilon)$.

It is convenient to introduce new ``natural'' variables: a rescaled
coordinate $y=x\rho$, a ``weak'' coupling
\begin{equation}
g = \left({6\pi\over N}\,xA(x)\right)^{1/3}
\goto_{x\to0}\left({6\pi\over N}\right)^{1/3},
\label{g-def}
\end{equation}
and a rescaled energy eigenvalue $\eta = \varepsilon/g^2$.
Eq.~(\ref{Schrodinger-orig}) now turns into
\begin{equation}
-{\D^2\psi\over\D y^2} + t^3\left(1-e^{-y}\right)\psi = \eta t^2\psi,
\label{Schrodinger-rescaled}
\end{equation}
where $t = g/x = g m_0/m_\theta$.

The general (unnormalized) solution of Eq.~(\ref{Schrodinger-rescaled})
is a Bessel function:
\begin{equation}
\psi_\eta(t,y) = J_{2t\sqrt{t-\eta}}\left(2 t^{3/2} e^{-y/2}\right),
\end{equation}
and the eigenvalue condition simply amounts to
\begin{mathletters}
\label{Schrodinger-eigenvalues}
\begin{eqnarray}
J'_{2t\sqrt{t-\eta}}\left(2t^{3/2}\right) &=& 0
\qquad\hbox{(even-parity levels),} \\
 J_{2t\sqrt{t-\eta}}\left(2t^{3/2}\right) &=& 0
\qquad\hbox{(odd-parity levels).}
\end{eqnarray}
\end{mathletters}
More specifically, denoting by $j_{\nu,k}$ the $k$th zero of $J_\nu(z)$ and
by $j'_{\nu,k}$ the $k$th zero of $J'_\nu(z)$, the $n$th energy level
$\eta_n$ is determined by solving
\begin{mathletters}
\begin{eqnarray}
j'_{2t\sqrt{t-\eta_n},\,(n+1)/2} &=& 2t^{3/2}\qquad\hbox{(odd $n$),} \\
 j_{2t\sqrt{t-\eta_n},\,n/2}   &=& 2t^{3/2}\qquad\hbox{(even $n$).}
\end{eqnarray}
\end{mathletters}
The rescaled energies $\eta_n$ are plotted as functions of $t$ in
Fig.~\ref{bound-states}. The energy eigenvalues $\varepsilon_n$ themselves
and their dependence on the mass parameters can be easily derived from
$\eta_n(t)$.

This kind of analysis is especially suitable for discussing the large-$t$
behavior, corresponding to the condition
\begin{equation}
m_\theta \ll gm_0 \cong \left(6\pi\over N\right)^{1/3}m_0.
\label{t-cond}
\end{equation}
{}From the asymptotic formula for $j'_{l,1}$ at large order $l$
\cite{Abramowitz}
\begin{equation}
j'_{l,1} \sim l\left[1+\sum_{k=1}^\infty \alpha_k l^{-2k/3}\right],
\end{equation}
one obtains an expansion in the form
\begin{equation}
\eta_1 = \sum_{k=0}^\infty \gamma_k t^{-k},
\end{equation}
where $\gamma_0 \cong 1.01879297$.
Similar expansions can be derived for higher energy levels.  As long as
$t\gg1$ these are good descriptions of the mass spectrum, and in particular
when $t\to\infty$ they reproduce known results for the bound states of the
$\CPN$ models.  In passing we notice that the condition (\ref{t-cond}) can
be rephrased in the form
\begin{equation}
m_\theta\left<R\right>_B \ll 1,
\label{R-cond}
\end{equation}
where $\left<R\right>_B$ is the semiclassical bound state radius.
Eq.~(\ref{R-cond}) is an obvious consistency condition for the
calculations.

Eqs.~(\ref{Schrodinger-eigenvalues}) indicate that, for sufficiently small
values of $t$, higher bound states may disappear from the spectrum.  More
precisely, defining $t_n$ by the condition
\begin{mathletters}
\begin{eqnarray}
j'_{0,\,(n+1)/2} &=& 2t_n^{3/2}\qquad\hbox{(odd $n$),} \\
 j_{0,\,n/2}     &=& 2t_n^{3/2}\qquad\hbox{(even $n$),}
\end{eqnarray}
\end{mathletters}
the corresponding state satisfying $\eta_n=t_n$, the
$n$th level disappears for all $t<t_n$.  In particular, when
$t<t_2 \cong 1.130756402$ all excited bound states disappear.
This phenomenon is illustrated in Fig.~\ref{bound-states}.

We may appreciate that, as long as $m_\theta \sim gm_0$, one may show
that the mass of the fundamental particle is well approximated by
\begin{equation}
m^2_F = m_0^2 + {1\over N}\,\Sigma_1(-m_0^2)
\cong m_0^2 + {1\over N}\,{6\pi m_0^3\over m_\theta} \,,
\end{equation}
implying
\begin{equation}
{m_F\over m_0} = 1 + {1\over2}\,g^2t.
\end{equation}
The threshold condition for the $n$th bound state in turn has the form
\begin{equation}
{m_{B(n)}\over m_0} = 2 + g^2\eta_n = 2 + g^2 t_n,
\end{equation}
and therefore it amounts to the condition
\begin{equation}
m_{B(n)}(t_n) = 2 m_F(t_n),
\end{equation}
as one may expect on physical grounds.

Finally, the threshold condition for the first bound state is simply $t=0$,
which corresponds to $g=0$, and from Eqs.~(\ref{g-def}) and (\ref{A-def})
this implies $m_\theta=2m_0$, a result known by independent arguments.

\section{Lattice integrals}
\label{lattice integr}
We know from general theorems that the perturbative expectation values of
quantities invariant under the full symmetry groups of the models must be
infrared-finite. Therefore in principle it must be possible to represent
finite lattice expectation values in terms of lattice integrals only.
However our regularization technique has led to the introduction of
continuum counterterms. There must therefore exist (infinitely many) simple
identities connecting lattice integrals and their continuum counterparts.

We have found all the identities that might be relevant in a three-loop
perturbative computation of lattice renormalization-group functions (when
$c_2=0$):
\begin{mathletters}
\begin{eqnarray}
 \intk A_0^{(\alpha)}(k) &-& \int_0^{32} {\D k^2\over4\pi}
    \,{1\over2\pi k^2} \ln{k^2\over32} = 0 \,,  \\
 \intk \hat k^2 A_1^{(\alpha)}(k) &+& \int_0^{32}
    {\D k^2\over4\pi}\, {1\over\pi k^2}\left(\ln{k^2\over32}-1\right)
= -{1\over4\pi^2} \,,  \\
 \intk {A_0^{(\theta)}(k)\over\hat k^2} &-& \int_0^{32}
    {\D k^2\over4\pi}\,{1\over2\pi k^2}
    \left(\ln{k^2\over32}-2\right) = -{1\over2\pi^2} \,, \\
 \intk A_1^{(\theta)}(k) &-& \int_0^{32} {\D k^2\over4\pi}
    \,{1\over\pi k^2} \left(\ln{k^2\over32}+1\right)
= -{1\over4\pi^2} \,.
\end{eqnarray}
\end{mathletters}
At the same order of approximation, a number of intrinsically finite
lattice integrals must be computed.  When $c_2=0$, some integrals can be
evaluated analytically:
\begin{mathletters}
\begin{eqnarray}
\intk \hat k^2 A_0^{(\alpha)}(k) &=& -{1\over4} \,, \\
\intk A_0^{(\theta)}(k) &=& -{3\over4} \,.
\end{eqnarray}
\end{mathletters}
However some computations can only be, to the best of our knowledge,
performed numerically \cite{Biscari-Campostrini-Rossi,Falcioni-Treves}:
\begin{mathletters}
\label{G1-at}
\begin{eqnarray}
G_1^{(\alpha)} &=& -{1\over4}\intk {\hat k^4\over\hat k^2}
\,A_0^{(\alpha)}(k) \cong 0.04616363, \\
G_1^{(\theta)} &=& -{1\over4}\intk {\hat k^4\over\left(\hat k^2\right)^2}
\,A_0^{(\theta)}(k) = G_1^{(\alpha)} + {1\over12}\,.
\end{eqnarray}
\end{mathletters}



\begin{figure}
\caption{Renormalization-group trajectories in the
$\{\beta_{\rm g},\beta_{\rm v}\}$ plane.} 
\label{kappa-plot}
\end{figure}

\begin{figure}
\caption{The large-$N$ mass ratio $m_\theta/m_0$, plotted as a function
of $\kappa$.}
\label{mu-plot}
\end{figure}

\begin{figure}
\caption{Feynman rules for the $1/N$ expansion in the continuum.}
\label{Feynman-rules-cont}
\end{figure}

\begin{figure}
\caption{The $O(1)$ (subleading) finite scaling part of the free energy
$c_F$, computed from Eq.~(\protect\ref{F-reg-cont}) (solid line) and from
Eq.~(\protect\ref{cF-1k}) (dot-dashed line); the $O(1/N)$ (subleading)
finite part of the mass gap $c_m$, computed from
Eq.~(\protect\ref{m1-cont}) (dashed line) and from
Eq.~(\protect\ref{cm-1/k}) (dotted line).}
\label{c-Fm}
\end{figure}

\begin{figure}
\caption{$O(1/N)$ contributions to the two-point function.}
\label{two-point}
\end{figure}

\begin{figure}
\caption{The $O(1/N)$ contribution to $\delta m^2_R$,
as a function of $\kappa$.}
\label{mR}
\end{figure}

\begin{figure}
\caption{Contributions to $\Delta_1^{-1}$, the correlation function of
the composite operator $P_{ij}(x)$.}
\label{Delta1}
\end{figure}

\begin{figure}
\caption{Effective vertices in the continuum. All momenta are entering in
the diagrams.}
\label{V3-V4}
\end{figure}

\begin{figure}
\caption{$c_Z$, the finite part of the renormalization constant $\tilde
Z_P$.}
\label{c-Z}
\end{figure}

\begin{figure}
\caption{The static potential $V(R)$, for several values of $\kappa$.}
\label{potential-fig}
\end{figure}

\begin{figure}
\caption{$O(1/N^2)$ contributions to the topological susceptibility.}
\label{topology-diagrams}
\end{figure}

\begin{figure}
\caption{Feynman rules for the $1/N$ expansion on the lattice.}
\label{Feynman-rules-latt}
\end{figure}

\begin{figure}
\caption{$M^2_L$ and $\alpha_1$ as functions of $c_2$ ($\alpha_1$ is
multiplied by $10^3$).}
\label{M2L}
\end{figure}

\begin{figure}
\caption{Contour plots of $A_0^{(\alpha)}(k)$ and
$A_0^{(\theta)}(k)+ 2/\pi$.  Contour lines are in logarithmic scale,
separated by a factor $\protect\sqrt2$.}
\label{A0-fig}
\end{figure}

\begin{figure}
\caption{$\mu_0^2(0)$ (solid line) and $\mu_0^2(\fourth\pi)$, (dashed
line), normalized to $m_0^2$.}
\label{large-N-fig}
\end{figure}

\begin{figure}
\caption{$O(1/N)$ contributions to the self-energy.}
\label{Feynman-diagrams-fig}
\end{figure}

\begin{figure}
\caption{The renormalization-group flow trajectories of the lattice models,
as defined by Eq.~(\protect\ref{kappa-redef}).}
\label{RG-lattice-fig}
\end{figure}

\begin{figure}
\caption{$\delta m_1^2$ as a function of $f$, for $c_2=0$ and
$\kappatilde = 0.01$, $\kappatilde = 0.1$, $\kappatilde = 1/\pi$, and
$\kappatilde = 1$. Solid and dashed lines are the results of
Eq.~(\protect\ref{mu1-c0}) for $\theta=0$ and $\theta=\pi/4$ respectively;
dot-dashed lines are the results of Eq.~(\protect\ref{mu10-c0}); dotted
lines are the results of a power series expansion of
Eq.~(\protect\ref{mu10-c0}) to 3rd and 6th order.}
\label{delta-m}
\end{figure}

\begin{figure}
\caption{$\delta Z_1$ as a function of $f$, for $c_2=0$ and
$\kappatilde = 0.01$, $\kappatilde = 0.1$, $\kappatilde = 1/\pi$, and
$\kappatilde = 1$. Solid lines are the results of
Eq.~(\protect\ref{Z10-c0}); dotted lines are the results of a power series
expansion of Eq.~(\protect\ref{Z10-c0}) to 3rd and 6th order.}
\label{delta-Z}
\end{figure}

\begin{figure}
\caption{$\delta m_1^2$ as a function of $f$, for the ${\rm O}(2N)$ model.
Solid and dashed lines are the results of Eq.~(\protect\ref{mu1-c0-ON}) for
$\theta=0$ and $\theta=\pi/4$ respectively; dot-dashed lines are the
results of Eq.~(\protect\ref{m-Z-scaling-ON}a); dotted lines are the
results of a power series expansion of Eq.~(\protect\ref{m-Z-scaling-ON}a)
to 3rd and 6th order.}
\label{delta-m-ON}
\end{figure}

\begin{figure}
\caption{$\delta m_1^2$ as a function of $f$, for the $\CPN$ model. The
solid line is the result Eq.~(\protect\ref{m1-scaling-CPN}); dotted lines
are the results of a power series expansion of
Eq.~(\protect\ref{m1-scaling-CPN}) to 3rd and 6th order.}
\label{delta-m-CPN}
\end{figure}

\begin{figure}
\caption{The leading contribution to the difference $\chi_t^{(1)}-\chi_t$.}
\label{chi-diff}
\end{figure}

\begin{figure}
\caption{The tadpole graph contributing to the difference between
Eq.~(\protect\ref{Lambda-conv3}) and Eq.~(\protect\ref{Lambda-conv4}).}
\label{tadpole-graph}
\end{figure}

\begin{figure}
\caption{The graphical form of the Schwinger--Dyson equations truncated to
$O(1/N^2)$. Open circles indicate dressed propagators; full circles
indicate dressed vertices}
\label{SD-graph}
\end{figure}

\begin{figure}
\caption{The rescaled energy levels $\eta_n$ as functions of $t$ (solid
lines).  The dashed line corresponds to $\eta=t$, where the excited levels
disappear.}
\label{bound-states}
\end{figure}


\begin{thebibliography}{100}

\bibitem{Coleman-aspects}
{\sc S. Coleman}, {\em Aspects of Symmetry} (Cambridge University Press,
  Cambridge, 1985), p.\ 351, and references therein.

\bibitem{DeVega}
{\it {\sc H. de~Vega}, Phys. Lett.} {\bf 98B},  280  (1981).

\bibitem{Avan-DeVega-I}
{\it {\sc J. Avan {\rm and} H. de~Vega}, Phys. Rev.} {\bf D29},  2891  (1984).

\bibitem{Avan-DeVega-II}
{\it {\sc J. Avan {\rm and} H. de~Vega}, Phys. Rev.} {\bf D29},  2904  (1984).

\bibitem{Campostrini-Rossi-dim}
{\it {\sc M. Campostrini {\rm and} P. Rossi}, Int. Jour. Mod. Phys.} {\bf A7},
  3265  (1992).

\bibitem{Campostrini-Rossi-condensates}
{\it {\sc M. Campostrini {\rm and} P. Rossi}, Phys. Lett.} {\bf 242B},  81
  (1990).

\bibitem{Samuel}
{\it {\sc S. Samuel}, Phys. Rev.} {\bf D28},  2628  (1983).

\bibitem{GellMann-Levy}
{\it {\sc M. Gell-Mann {\rm and} M. Levy}, Nuovo Cim.} {\bf 16},  705  (1960).

\bibitem{Stanley-I}
{\it {\sc H.~E. Stanley}, Phys. Rev. Lett.} {\bf 20},  589  (1968).

\bibitem{Stanley-II}
{\it {\sc H.~E. Stanley}, Phys. Rev.} {\bf 176},  718  (1968).

\bibitem{Golo-Perelomov}
{\it {\sc V.~L. Golo {\rm and} A.~M. Perelomov}, Phys. Lett.} {\bf 79B},  112
  (1978).

\bibitem{Eichenherr}
{\it {\sc H. Eichenherr}, Nucl. Phys.} {\bf B146},  215  (1978).

\bibitem{Abdalla-Abdalla-Alves-Carneiro}
{\it {\sc E. Abdalla, M.~C.~B. Abdalla, N.~A. Alves, {\rm and} C.~E.~I.
  Carneiro}, Phys. Rev.} {\bf D41},  571  (1990).

\bibitem{Brezin-ZinnJustin-breakdown}
{\it {\sc E. Brezin {\rm and} J. Zinn-Justin}, Phys. Rev.} {\bf B14},  3110
  (1976).

\bibitem{Amit-Ma-Zia}
{\it {\sc D.~J. Amit, {S.-k. Ma}, {\rm and} R.~K.~P. Zia}, Nucl. Phys.} {\bf
  B180 [FS2]},  157  (1981).

\bibitem{Mermin-Wagner}
{\it {\sc N.~D. Mermin {\rm and} H. Wagner}, Phys. Rev. Lett.} {\bf 17},  1133
  (1966).

\bibitem{Coleman}
{\it {\sc S. Coleman}, Comm. Math. Phys.} {\bf 31},  259  (1973).

\bibitem{Polyakov-goldstone}
{\it {\sc A.~M. Polyakov}, Phys. Lett.} {\bf 59B},  79  (1975).

\bibitem{Brezin-ZinnJustin-renormalization}
{\it {\sc E. Brezin {\rm and} J. Zinn-Justin}, Phys. Rev. Lett.} {\bf 36},
   691 (1975).

\bibitem{Brezin-ZinnJustin-LeGuillou}
{\it {\sc E. Brezin, J. Zinn-Justin, {\rm and} J.~C.~L. Guillou}, Phys. Rev.}
  {\bf D14},  2615  (1976).

\bibitem{Bardeen-Lee-Shrock}
{\it {\sc W.~A. Bardeen, B.~W. Lee, {\rm and} R.~E. Shrock}, Phys. Rev.} {\bf
  D14},  985  (1976).

\bibitem{Valent}
{\it {\sc G. Valent}, Nucl. Phys.} {\bf B238},  142  (1984).

\bibitem{Eichenherr-Forger}
{\it {\sc H. Eichenherr {\rm and} M. Forger}, Nucl. Phys.} {\bf B155},  381
  (1979).

\bibitem{Eichenherr-Forger-II}
{\it {\sc H. Eichenherr {\rm and} M. Forger}, Nucl. Phys.} {\bf B164},  528
  (1980).

\bibitem{Pisarski}
{\it {\sc R.~D. Pisarski}, Phys. Rev.} {\bf D20},  3358  (1979).

\bibitem{Duane}
{\it {\sc S. Duane}, Nucl. Phys.} {\bf B168},  32  (1980).

\bibitem{Brezin-Hikami-ZinnJustin}
{\it {\sc E. Brezin, S. Hikami, {\rm and} J. Zinn-Justin}, Nucl. Phys.} {\bf
  B165},  528  (1980).

\bibitem{Jevicki-ground-state}
{\it {\sc A. Jevicki}, Phys. Lett.} {\bf 71B},  327  (1977).

\bibitem{Elitzur}
{\it {\sc S. Elitzur}, Nucl. Phys.} {\bf B212},  501  (1983).

\bibitem{McKane-Stone}
{\it {\sc A. McKane {\rm and} M. Stone}, Nucl. Phys.} {\bf B163},  169  (1980).

\bibitem{Amit-Kotliar}
{\it {\sc D.~J. Amit {\rm and} G.~B. Kotliar}, Nucl. Phys.} {\bf B170 [FS1]},
  187  (1980).

\bibitem{Wolff-O3-mass}
{\it {\sc U. Wolff}, Nucl. Phys.} {\bf B334},  581  (1990).

\bibitem{David-infrared}
{\it {\sc F. David}, Comm. Math. Phys.} {\bf 81},  149  (1981).

\bibitem{David-infrared-letter}
{\it {\sc F. David}, Phys. Lett.} {\bf 96B},  371  (1980).

\bibitem{David-global-constraint}
{\it {\sc F. David}, Nucl. Phys.} {\bf B190 [FS3]},  205  (1981).

\bibitem{Hikami-Brezin}
{\it {\sc S. Hikami {\rm and} E. Brezin}, J. Phys. {\bf A}{\rm:} Math. Gen.}
  {\bf 11},  1141  (1978).

\bibitem{Hikami-CPN}
{\it {\sc S. Hikami}, Prog. Theor. Phys.} {\bf 62},  226  (1979).

\bibitem{Hikami-3loop}
{\it {\sc S. Hikami}, Phys. Lett.} {\bf 98B},  208  (1981).

\bibitem{Hikami-sigma}
{\it {\sc S. Hikami}, Nucl. Phys.} {\bf B215 [FS7]},  555  (1983).

\bibitem{Hof-Wegner}
{\it {\sc D. Hof {\rm and} F. Wegner}, Nucl. Phys.} {\bf B275},  561  (1986).

\bibitem{Wegner-1}
{\it {\sc F. Wegner}, Nucl. Phys.} {\bf B280 [FS18]},  193  (1987).

\bibitem{Wegner-2}
{\it {\sc F. Wegner}, Nucl. Phys.} {\bf B280 [FS18]},  210  (1987).

\bibitem{Bernreuther-Wegner}
{\it {\sc W. Bernreuther {\rm and} F.~J. Wegner}, Phys. Rev. Lett.} {\bf 57},
  1383  (1986).

\bibitem{Wegner}
{\it {\sc F. Wegner}, Nucl. Phys.} {\bf B316},  663  (1989).

\bibitem{Jolicoeur-Niel-mass}
{\it {\sc T. Jolicoeur {\rm and} J.~C. Niel}, Nucl. Phys.} {\bf B300 [FS22]},
  517  (1988).

\bibitem{Jolicoeur-Niel-magnetic}
{\it {\sc T. Jolicoeur {\rm and} J.~C. Niel}, Phys. Lett.} {\bf 215B},  735
  (1988).

\bibitem{Abe-II}
{\it {\sc R. Abe}, Prog. Theor. Phys.} {\bf 49},  113  (1973).

\bibitem{Abe-eta-II}
{\it {\sc R. Abe}, Prog. Theor. Phys.} {\bf 49},  1877  (1973).

\bibitem{Abe-Hikami-scaling}
{\it {\sc R. Abe {\rm and} S. Hikami}, Prog. Theor. Phys.} {\bf 49},  442
  (1973).

\bibitem{Brezin-Wallace}
{\it {\sc E. Brezin {\rm and} D.~J. Wallace}, Phys. Rev.} {\bf B7},  1967
  (1973).

\bibitem{Ma-exponents}
{\it {\sc {S.-k. Ma}}, Phys. Rev.} {\bf A7},  2172  (1973).

\bibitem{Ma-largeN}
{\it {\sc {S.-k. Ma}}, J. Math. Phys.} {\bf 15},  1866  (1974).

\bibitem{Vasilev-Pismak-Khonkonen-indices}
{\it {\sc A.~N. Vasil'ev, {Yu. M. Pis'mak}, {\rm and} {Yu. R. Khonkonen}},
  Theor. Math. Phys.} {\bf 46},  157  (1981).

\bibitem{Vasilev-Pismak-Khonkonen-exponents}
{\it {\sc A.~N. Vasil'ev, {Yu. M. Pis'mak}, {\rm and} {Yu. R. Khonkonen}},
  Theor. Math. Phys.} {\bf 47},  291  (1981).

\bibitem{Vasilev-Pismak-Khonkonen-bootstrap}
{\it {\sc A.~N. Vasil'ev, {Yu. M. Pis'mak}, {\rm and} {Yu. R. Khonkonen}},
  Theor. Math. Phys.} {\bf 50},  195  (1982).

\bibitem{Vasilev-Nalimov-CPN}
{\it {\sc A.~N. Vasil'ev {\rm and} {M. Yu. Nalimov}}, Theor. Math. Phys.} {\bf
  56},  15  (1983).

\bibitem{Symanzik-ON}
{\sc K. Symanzik}, {\em $1/N$ expansions in $P(\phi^2)_{4-\epsilon}$ theory. 1.
  massless theory, $0<\epsilon<2$}, preprint DESY 77/05, unpublished, 1977.

\bibitem{Symanzik-Pphi2}
{\sc K. Symanzik},  in {\em Symposium on Quantum Fields -- Algebras, Processes,
  Bielefeld 1978}, edited by {\sc L. Streit} (Springer, Wien, 1980), p.\ 379.

\bibitem{Arefeva-I}
{\it {\sc {I. Ya. Aref'eva}}, Theor. Math. Phys.} {\bf 29},  147  (1976).

\bibitem{Arefeva-II}
{\it {\sc {I. Ya. Aref'eva}}, Theor. Math. Phys.} {\bf 31},  3  (1977).

\bibitem{Arefeva-III}
{\it {\sc {I. Ya. Aref'eva}}, Theor. Math. Phys.} {\bf 36},  24  (1978).

\bibitem{Arefeva-IV}
{\it {\sc {I. Ya. Aref'eva}}, Theor. Math. Phys.} {\bf 36},  159  (1978).

\bibitem{Arefeva-AP}
{\it {\sc {I. Ya. Aref'eva}}, Ann. Phys. (N. Y.)} {\bf 117},  393  (1979).

\bibitem{Arefeva-Azakov}
{\it {\sc {I. Ya. Aref'eva} {\rm and} S.~I. Azakov}, Nucl. Phys.} {\bf B162},
  298  (1980).

\bibitem{Arefeva-Nissimov-Pacheva}
{\it {\sc {I. Ya. Aref'eva}, E.~R. Nissimov, {\rm and} S.~J. Pacheva}, Comm.
  Math. Phys.} {\bf 71},  213  (1980).

\bibitem{Vasilev-Nalimov-dimension}
{\it {\sc A.~N. Vasil'ev {\rm and} {M. Yu. Nalimov}}, Theor. Math. Phys.} {\bf
  55},  163  (1983).

\bibitem{Rim-Weisberger}
{\it {\sc C. Rim {\rm and} W.~I. Weisberger}, Phys. Rev.} {\bf D30},  1763
  (1984).

\bibitem{Jevicki-instantons}
{\it {\sc A. Jevicki}, Phys. Rev.} {\bf D20},  3331  (1979).

\bibitem{David}
{\it {\sc F. David}, Phys. Lett.} {\bf 138B},  139  (1984).

\bibitem{Fateev-Frolov-Schwarz}
{\it {\sc V.~A. Fateev, I.~V. Frolov, {\rm and} A.~S. Schwarz}, Nucl. Phys.}
  {\bf B154},  1  (1979).

\bibitem{Berg-Luscher-CPN}
{\it {\sc B. Berg {\rm and} M. {L\"uscher}}, Comm. Math. Phys.} {\bf 69},  57
  (1979).

\bibitem{Affleck-I}
{\it {\sc I. Affleck}, Nucl. Phys.} {\bf B162},  461  (1980).

\bibitem{Affleck-II}
{\it {\sc I. Affleck}, Nucl. Phys.} {\bf B171},  420  (1980).

\bibitem{Actor}
{\it {\sc A. Actor}, Fortschr. Phys.} {\bf 33},  333  (1985).

\bibitem{Davis-Matheson-chiral}
{\it {\sc A.~C. Davis {\rm and} A.~M. Matheson}, Nucl. Phys.} {\bf B258},  373
  (1985).

\bibitem{Zhitnitsky-CPN}
{\it {\sc A.~R. Zhitnitsky}, Nucl. Phys.} {\bf B374},  183  (1992).

\bibitem{Parisi-infrared}
{\it {\sc G. Parisi}, Nucl. Phys.} {\bf B150},  163  (1979).

\bibitem{Shifman-Vainshtein-Zakharov}
{\it {\sc M.~A. Shifman, A.~I. Vainshtein, {\rm and} V.~I. Zakharov}, Nucl.
  Phys.} {\bf B147},  385, 448, 519  (1979).

\bibitem{David-renormalons}
{\it {\sc F. David}, Nucl. Phys.} {\bf B209},  433  (1982).

\bibitem{David-ambiguity}
{\it {\sc F. David}, Nucl. Phys.} {\bf B234},  237  (1984).

\bibitem{David-OPE}
{\it {\sc F. David}, Nucl. Phys.} {\bf B263},  637  (1986).

\bibitem{Brunelli-Gomes}
{\it {\sc J.~C. Brunelli {\rm and} M. Gomes}, Zeit. Phys.} {\bf C42},  649
  (1989).

\bibitem{Novikov-Shifman-Vainshtein-Zakharov-sigma}
{\it {\sc V.~A. Novikov, M.~A. Shifman, A.~I. Vainshtein, {\rm and} V.~I.
  Zakharov}, Phys. Rep.} {\bf C116},  103  (1984).

\bibitem{Novikov-Shifman-Vainshtein-Zakharov-OPE}
{\it {\sc V.~A. Novikov, M.~A. Shifman, A.~I. Vainshtein, {\rm and} V.~I.
  Zakharov}, Nucl. Phys.} {\bf B249},  445  (1985).

\bibitem{Lang-Ruhl-field}
{\it {\sc K. Lang {\rm and} W. Ruhl}, Zeit. Phys.} {\bf C50},  285  (1991).

\bibitem{Lang-Ruhl-tensor}
{\it {\sc K. Lang {\rm and} W. Ruhl}, Zeit. Phys.} {\bf C51},  127  (1991).

\bibitem{Lang-Ruhl-OPE}
{\it {\sc K. Lang {\rm and} W. Ruhl}, Nucl. Phys.} {\bf B377},  371  (1992).

\bibitem{Lang-Ruhl-ancestor}
{\it {\sc K. Lang {\rm and} W. Ruhl}, Phys. Lett.} {\bf B275},  93  (1992).

\bibitem{Lang-Ruhl-quasiprimary}
{\sc K. Lang {\rm and} W. Ruhl}, {\em The critical ${\rm O}(N)$ sigma model at
  dimensions $2<d<4$: A list of quasiprimary fields}, Kaiserslautern preprint
  KL-TH-92/7, 1992.

\bibitem{Lang-Ruhl-fusion}
{\sc K. Lang {\rm and} W. Ruhl}, {\em The critical ${\rm O}(N)$ sigma model at
  dimensions $2<d<4$: fusion coefficients and anomalous dimensions},
  Kaiserslautern preprint KL-TH-92/15, 1992.

\bibitem{Davis-Nahm-sigma}
{\it {\sc A.~C. Davis {\rm and} W. Nahm}, Phys. Lett.} {\bf 155B},  404
  (1985).

\bibitem{Davis-Nahm-CPN}
{\it {\sc A.~C. Davis {\rm and} W. Nahm}, Phys. Lett.} {\bf 159B},  294
  (1985).

\bibitem{Davis-Mayger}
{\it {\sc A.~C. Davis {\rm and} E.~M. Mayger}, Nucl. Phys.} {\bf B306},  199
  (1988).

\bibitem{DelrioGatzelurrutia-Davis}
{\it {\sc T. del Rio~Gatzelurrutia {\rm and} A.~C. Davis}, Nucl. Phys.} {\bf
  B347},  319  (1990).

\bibitem{Luscher-secret}
{\it {\sc M. {L\"uscher}}, Phys. Lett.} {\bf 78B},  465  (1978).

\bibitem{DAdda-Luscher-DiVecchia-1/N}
{\it {\sc A. D'Adda, M. {L\"uscher}, {\rm and} P. {Di Vecchia}}, Nucl. Phys.}
  {\bf B146},  63  (1978).

\bibitem{Witten-CPN}
{\it {\sc E. Witten}, Nucl. Phys.} {\bf B149},  285  (1979).

\bibitem{Abdalla-LimaSantos-CPN}
{\it {\sc E. Abdalla {\rm and} A. Lima-Santos}, Phys. Rev.} {\bf D29},  1851
  (1984).

\bibitem{DAdda-DiVecchia-Luscher}
{\it {\sc A. D'Adda, P. {Di Vecchia}, {\rm and} M. {L\"uscher}}, Nucl. Phys.}
  {\bf B152},  125  (1979).

\bibitem{Iagolnitzer}
{\it {\sc D. Iagolnitzer}, Phys. Lett.} {\bf 76B},  207  (1978).

\bibitem{Luscher-Pohlmeyer}
{\it {\sc M. {L\"uscher} {\rm and} K. Pohlmeyer}, Nucl. Phys.} {\bf B137},  46
  (1978).

\bibitem{Luscher-charge}
{\it {\sc M. {L\"uscher}}, Nucl. Phys.} {\bf B135},  1  (1978).

\bibitem{Abdalla-Abdalla-Gomes-I}
{\it {\sc E. Abdalla, M.~C.~B. Abdalla, {\rm and} M. Gomes}, Phys. Rev.} {\bf
  D23},  1800  (1981).

\bibitem{Abdalla-Abdalla-Gomes-III}
{\it {\sc E. Abdalla, M.~C.~B. Abdalla, {\rm and} M. Gomes}, Phys. Rev.} {\bf
  D27},  825  (1983).

\bibitem{Abdalla-Abdalla-Gomes-II}
{\it {\sc E. Abdalla, M.~C.~B. Abdalla, {\rm and} M. Gomes}, Phys. Rev.} {\bf
  D25},  452  (1982).

\bibitem{Zamolodchikov-Zamolodchikov-ON}
{\it {\sc A.~B. Zamolodchikov {\rm and} {Al. B. Zamolodchikov}}, Nucl. Phys.}
  {\bf B133},  525  (1978).

\bibitem{Banks-Zaks}
{\it {\sc T. Banks {\rm and} A. Zaks}, Nucl. Phys.} {\bf B128},  333  (1977).

\bibitem{Berg-Karowski-Kurak-Weisz}
{\it {\sc B. Berg, M. Karowski, V. Kurak, {\rm and} P. Weisz}, Phys. Lett.}
  {\bf 76B},  502  (1978).

\bibitem{Berg-Karowski-Weisz-Kurak}
{\it {\sc B. Berg, M. Karowski, P. Weisz, {\rm and} V. Kurak}, Nucl. Phys.}
  {\bf B134},  125  (1978).

\bibitem{Koberle-Kurak-CPN}
{\it {\sc R. {K\"oberle} {\rm and} V. Kurak}, Phys. Rev. Lett.} {\bf 58},  627
  (1987).

\bibitem{Karowski-Weisz}
{\it {\sc M. Karowski {\rm and} P. Weisz}, Nucl. Phys.} {\bf B139},  455
  (1978).

\bibitem{Hasenfratz-Maggiore-Niedermayer}
{\it {\sc P. Hasenfratz, M. Maggiore, {\rm and} F. Niedermayer}, Phys. Lett.}
  {\bf 245B},  522  (1990).

\bibitem{Hasenfratz-Niedermayer-ON}
{\it {\sc P. Hasenfratz {\rm and} F. Niedermayer}, Phys. Lett.} {\bf 245B},
  529  (1990).

\bibitem{Campostrini-Rossi-CPN}
{\it {\sc M. Campostrini {\rm and} P. Rossi}, Phys. Rev.} {\bf D45},  618
  (1992); {\bf D46}, 2741 (1992) (E).

\bibitem{Orloff-Brout}
{\it {\sc J. Orloff {\rm and} R. Brout}, Nucl. Phys.} {\bf B270 [FS16]},  273
  (1986).

\bibitem{Biscari-Campostrini-Rossi}
{\it {\sc P. Biscari, M. Campostrini, {\rm and} P. Rossi}, Phys. Lett.} {\bf
  242B},  225  (1990).

\bibitem{Flyvbjerg-scaling}
{\it {\sc H. Flyvbjerg}, Phys. Lett.} {\bf 245B},  533  (1990).

\bibitem{Flyvbjerg}
{\it {\sc H. Flyvbjerg}, Nucl. Phys.} {\bf B348},  714  (1991).

\bibitem{Haber-Hinchliffe-Rabinovici}
{\it {\sc H.~E. Haber, I. Hinchliffe, {\rm and} E. Rabinovici}, Nucl. Phys.}
  {\bf B172},  458  (1980).

\bibitem{Campostrini-Rossi-CPN-topology}
{\it {\sc M. Campostrini {\rm and} P. Rossi}, Phys. Lett.} {\bf B272},  305
  (1991).

\bibitem{Stone}
{\it {\sc M. Stone}, Nucl. Phys.} {\bf B152},  97  (1979).

\bibitem{Rossi-Brihaye}
{\it {\sc P. Rossi {\rm and} Y. Brihaye}, Physica} {\bf 126A},  237  (1984).

\bibitem{Hamer-Kogut-Susskind}
{\it {\sc C.~J. Hamer, J.~B. Kogut, {\rm and} L. Susskind}, Phys. Rev.} {\bf
  D19},  3091  (1979).

\bibitem{Srednicki}
{\it {\sc M. Srednicki}, Phys. Rev.} {\bf B20},  3783  (1979).

\bibitem{Banks}
{\it {\sc J.~L. Banks}, Phys. Lett.} {\bf 93B},  161  (1980).

\bibitem{Guha-Sakita}
{\it {\sc A. Guha {\rm and} B. Sakita}, Phys. Lett.} {\bf 100B},  489  (1981).

\bibitem{Campostrini-Rossi-Vicari-I}
{\it {\sc M. Campostrini, P. Rossi, {\rm and} E. Vicari}, Phys. Rev.} {\bf
  D46},  2647  (1992).

\bibitem{Campostrini-Rossi-Vicari-II}
{\it {\sc M. Campostrini, P. Rossi, {\rm and} E. Vicari}, Phys. Rev.} {\bf
  D46},  4643  (1992).

\bibitem{Irving-Michael-potential}
{\it {\sc A.~C. Irving {\rm and} C. Michael}, Nucl. Phys.} {\bf B371},  521
  (1992).

\bibitem{Vicari}
{\sc E. Vicari}, {\em Monte Carlo simulation of lattice ${\rm CP}^{N-1}$ models
  at large $N$}, preprint IFUP-TH 35/92, 1992.

\bibitem{Rabinovici-Samuel}
{\it {\sc E. Rabinovici {\rm and} S. Samuel}, Phys. Lett.} {\bf 101B},  323
  (1981).

\bibitem{DiVecchia-Holtkamp-Musto-Nicodemi-Pettorino}
{\it {\sc P. {Di Vecchia}, A. Holtkamp, R. Musto, F. Nicodemi, {\rm and} R.
  Pettorino}, Nucl. Phys.} {\bf B190},  719  (1981).

\bibitem{Duane-Green}
{\it {\sc S. Duane {\rm and} M.~B. Green}, Phys. Lett.} {\bf 103B},  359
  (1981).

\bibitem{DiVecchia-Musto-Nicodemi-Pettorino-Rossi}
{\it {\sc P. {Di Vecchia}, R. Musto, F. Nicodemi, R. Pettorino, {\rm and} P.
  Rossi}, Nucl. Phys.} {\bf B235 [FS11]},  478  (1984).

\bibitem{Ruhl-mean}
{\it {\sc W. {R\"uhl}}, Zeit. Phys.} {\bf C32},  265  (1986).

\bibitem{Symanzik-sigma}
{\it {\sc K. Symanzik}, Nucl. Phys.} {\bf B226},  205  (1983).

\bibitem{Musto-Nicodemi-Pettorino}
{\it {\sc R. Musto, F. Nicodemi, {\rm and} R. Pettorino}, Phys. Lett.} {\bf
  129B},  95  (1983).

\bibitem{Gradstein}
{\sc I.~S. {Grad\v stein} {\rm and} I.~M. {Ry\v zik}}, {\em Table of Integrals,
  Series and Products} (Academic Press, Orlando, 1980).

\bibitem{Muller-Raddatz-Ruhl}
{\it {\sc V.~F. {M\"uller}, T. Raddatz, {\rm and} W. {R\"uhl}}, Nucl. Phys.}
  {\bf B251 [FS13]},  212  (1985), {\bf B259}; 745 (1985) (E).

\bibitem{Cristofano-Musto-Nicodemi-Pettorino-Pezzella}
{\it {\sc G. Cristofano, R. Musto, F. Nicodemi, R. Pettorino, {\rm and} F.
  Pezzella}, Nucl. Phys.} {\bf B257 [FS14]},  505  (1985).

\bibitem{Affleck-Levine}
{\it {\sc I.~K. Affleck {\rm and} H. Levine}, Nucl. Phys.} {\bf B195},  493
  (1982).

\bibitem{Musto-Nicodemi-Pettorino-Clarizia}
{\it {\sc R. Musto, F. Nicodemi, R. Pettorino, {\rm and} A. Clarizia}, Nucl.
  Phys.} {\bf B210 [FS6]},  263  (1982).

\bibitem{Butera-Comi-Marchesini-sigma}
{\it {\sc P. Butera, M. Comi, {\rm and} G. Marchesini}, Nucl. Phys.} {\bf B300
  [FS22]},  1  (1988).

\bibitem{Butera-Comi-Marchesini-Onofri}
{\it {\sc P. Butera, M. Comi, G. Marchesini, {\rm and} E. Onofri}, Nucl. Phys.}
  {\bf B326},  758  (1989).

\bibitem{Bonnier-Hontebeyrie}
{\it {\sc B. Bonnier {\rm and} M. Hontebeyrie}, Phys. Lett.} {\bf 226B},  361
  (1989).

\bibitem{Butera-Comi-Marchesini-series}
{\it {\sc P. Butera, M. Comi, {\rm and} G. Marchesini}, Phys. Rev.} {\bf B41},
  11494  (1990).

\bibitem{Clarizia-Cristofano-Musto-Nicodemi-Pettorino}
{\it {\sc A. Clarizia, G. Cristofano, R. Musto, F. Nicodemi, {\rm and} R.
  Pettorino}, Phys. Lett.} {\bf 148B},  323  (1984).

\bibitem{Falcioni-Treves}
{\it {\sc M. Falcioni {\rm and} A. Treves}, Nucl. Phys.} {\bf B265 [FS15]},
  671  (1986).

\bibitem{Luscher-Weisz-Wolff}
{\it {\sc M. {L\"uscher}, P. Weisz, {\rm and} U. Wolff}, Nucl. Phys.} {\bf
  B359},  221  (1991).

\bibitem{Berg-Luscher-O3}
{\it {\sc B. Berg {\rm and} M. {L\"uscher}}, Nucl. Phys.} {\bf B190 [FS3]},
  412  (1981).

\bibitem{Berg-letter}
{\it {\sc B. Berg}, Phys. Lett.} {\bf 104B},  475  (1981).

\bibitem{Luscher-lattice-chi}
{\it {\sc M. {L\"uscher}}, Nucl. Phys.} {\bf B200 [FS4]},  61  (1982).

\bibitem{Parisi-coupling}
{\it {\sc G. Parisi}, Phys. Lett.} {\bf 92B},  133  (1980).

\bibitem{Berg-Lambda}
{\it {\sc B. Berg}, Zeit. Phys.} {\bf C20},  243  (1983).

\bibitem{Brezin-FSS}
{\it {\sc E. Brezin}, Journ. de Phys.} {\bf 43},  15  (1982).

\bibitem{Brezin-Korutcheva-Jolicoeur-ZinnJustin}
{\sc E. Brezin, E. Korutcheva, T. Jolicoeur, {\rm and} J. Zinn-Justin}, {\em
  ${\rm O}(N)$ vector-model with twisted boundary conditions}, preprint
  SPhT/92/039, 1992.

\bibitem{Luscher-low-lying-states}
{\it {\sc M. {L\"uscher}}, Phys. Lett.} {\bf 118B},  391  (1982).

\bibitem{Floratos-Petcher-I}
{\it {\sc E.~G. Floratos {\rm and} D. Petcher}, Phys. Lett.} {\bf 133B},  206
  (1983).

\bibitem{Floratos-Petcher-II}
{\it {\sc E.~G. Floratos {\rm and} D. Petcher}, Nucl. Phys.} {\bf B252},  689
  (1985).

\bibitem{Floratos-Vlachos-CPN}
{\it {\sc E.~G. Floratos {\rm and} N.~D. Vlachos}, Nucl. Phys.} {\bf B278},
  170  (1986).

\bibitem{Campostrini-Rossi-fss}
{\it {\sc M. Campostrini {\rm and} P. Rossi}, Phys. Lett.} {\bf 255B},  89
  (1991).

\bibitem{Rossi-Vicari}
{\sc P. Rossi {\rm and} E. Vicari}, {\em Finite size scaling in ${\rm
  CP}^{N-1}$ models}, Pisa preprint IFUP-TH 59/92, 1992.

\bibitem{Hasenfratz-zero-modes}
{\it {\sc P. Hasenfratz}, Phys. Lett.} {\bf 141B},  385  (1984).

\bibitem{Brihaye-Spindel}
{\it {\sc Y. Brihaye {\rm and} P. Spindel}, Nucl. Phys.} {\bf B280},  466
  (1987).

\bibitem{Flyvbjerg-DS}
{\it {\sc H. Flyvbjerg}, J. Phys. {\bf A}{\rm:} Math. Gen.} {\bf 22},  3393
  (1989).

\bibitem{Luscher-volume-dependence}
{\it {\sc M. {L\"uscher}}, Comm. Math. Phys.} {\bf 104},  177  (1986).

\bibitem{Schultka-MullerPreussker}
{\it {\sc N. Schultka {\rm and} M. {M\"uller-Preu\ss ker}}, Nucl. Phys.} {\bf
  B386},  214  (1992).

\bibitem{Ruhl-critical}
{\it {\sc W. {R\"uhl}}, Fortschr. Phys.} {\bf 35},  707  (1987).

\bibitem{Muller-Ruhl}
{\it {\sc V.~F. {M\"uller} {\rm and} W. {R\"uhl}}, Ann. Phys. (N. Y.)} {\bf
  168},  425  (1986).

\bibitem{Hasenfratz-Leutwyler}
{\it {\sc P. Hasenfratz {\rm and} H. Leutwyler}, Nucl. Phys.} {\bf B343},  241
  (1990).

\bibitem{Flyvbjerg-1/N}
{\it {\sc H. Flyvbjerg}, Phys. Lett.} {\bf 219B},  323  (1989).

\bibitem{Flyvbjerg-Varsted}
{\it {\sc H. Flyvbjerg {\rm and} S. Varsted}, Nucl. Phys.} {\bf B344},  646
  (1990).

\bibitem{Flyvbjerg-Larsen-Kristjansen}
{\sc H. Flyvbjerg, F. Larsen, {\rm and} C. Kristjansen},  in {\em Lattice '90,
  International Symposium on Lattice Field Theory}, Vol.~20 of {\em Nucl. Phys.
  B (Proc. Suppl.)}, edited by {\sc U.~M. Heller {\it et~al.}} (North Holland,
  Amsterdam, 1991), p.\ 44.

\bibitem{Flyvbjerg-Larsen-support}
{\it {\sc H. Flyvbjerg {\rm and} F. Larsen}, Phys. Lett.} {\bf B266},  92
  (1991).

\bibitem{Flyvbjerg-Larsen-susceptibility}
{\it {\sc H. Flyvbjerg {\rm and} F. Larsen}, Phys. Lett.} {\bf B266},  99
  (1991).

\bibitem{Drouffe-Flyvbjerg-letter}
{\it {\sc J.-M. Drouffe {\rm and} H. Flyvbjerg}, Phys. Lett.} {\bf 206B},  285
  (1988).

\bibitem{Drouffe-Flyvbjerg}
{\it {\sc J.-M. Drouffe {\rm and} H. Flyvbjerg}, Nucl. Phys.} {\bf B332},  687
  (1990).

\bibitem{Stanley}
{\it {\sc H.~E. Stanley}, Phys. Rev.} {\bf 179},  570  (1969).

\bibitem{Rosenstein-Warr-Park-renorm}
{\it {\sc B. Rosenstein, B.~J. Warr, {\rm and} S.~H. Park}, Phys. Rev. Lett.}
  {\bf 62},  1433  (1989).

\bibitem{Gat-Kovner-Rosenstein-Warr}
{\it {\sc G. Gat, A. Kovner, B. Rosenstein, {\rm and} B.~J. Warr}, Phys. Lett.}
  {\bf 240B},  158  (1990).

\bibitem{Rosenstein-Warr-Park-breaking}
{\it {\sc B. Rosenstein, B.~J. Warr, {\rm and} S.~H. Park}, Phys. Rep.} {\bf
  205},  59  (1991).

\bibitem{Kikukawa-Yamawaki}
{\it {\sc Y. Kikukawa {\rm and} K. Yamawaki}, Phys. Lett.} {\bf 234B},  497
  (1990).

\bibitem{He-Kuang-Wang-Yi}
{\it {\sc {H.-J. He}, {Y.-P. Kuang}, {Q. Wang}, {\rm and} {Y.-P. Yi}}, Phys.
  Rev.} {\bf D45},  4610  (1992).

\bibitem{ZinnJustin}
{\it {\sc J. Zinn-Justin}, Nucl. Phys.} {\bf B367},  105  (1991).

\bibitem{Hands-Kocic-Kogut-I}
{\it {\sc S. Hands, A. Kocic, {\rm and} J.~B. Kogut}, Phys. Lett.} {\bf B273},
  111  (1991).

\bibitem{Hands-Kocic-Kogut-II}
{\sc S. Hands, A. Kocic, {\rm and} J.~B. Kogut}, {\em Four-fermi theories in
  fewer than four dimensions}, preprint CERN-TH 6557/92, 1992.

\bibitem{Gracey-GN-eta}
{\it {\sc J.~A. Gracey}, Int. Jour. Mod. Phys.} {\bf A6},  395  (1991).

\bibitem{Bondi-Curci-Paffuti-Rossi-central-charge}
{\it {\sc A. Bondi, G. Curci, G. Paffuti, {\rm and} P. Rossi}, Ann. Phys. (N.
  Y.)} {\bf 199},  268  (1990).

\bibitem{Abdalla-Abdalla-Rothe}
{\sc E. Abdalla, M.~C.~B. Abdalla, {\rm and} K.~D. Rothe}, {\em
  Non-Perturbative Methods in 2 Dimensional Quantum Field Theory} (World
  Scientific, Singapore, 1991).

\bibitem{Wilson-lecture}
{\sc K.~G. Wilson},  in {\em New Phenomena in Subnuclear Physics}, edited by
  {\sc A. Zichichi} (Plenum, New York, 1977), p.\ 69.

\bibitem{Cohen-Elitzur-Rabinovici}
{\it {\sc Y. Cohen, S. Elitzur, {\rm and} E. Rabinovici}, Nucl. Phys.} {\bf
  B220 [FS8]},  102  (1983).

\bibitem{Susskind}
{\it {\sc L. Susskind}, Phys. Rev.} {\bf D16},  3031  (1977).

\bibitem{Jolicoeur-Morel-Petersson}
{\it {\sc T. Jolicoeur, A. Morel, {\rm and} B. Petersson}, Nucl. Phys.} {\bf
  B274},  225  (1986).

\bibitem{Eguchi-Nakayama}
{\it {\sc T. Eguchi {\rm and} R. Nakayama}, Phys. Lett.} {\bf 126B},  89
  (1983).

\bibitem{Aoki-Higashijima}
{\it {\sc S. Aoki {\rm and} K. Higashijima}, Prog. Theor. Phys.} {\bf 76},  521
   (1981).

\bibitem{David-Hamber}
{\it {\sc F. David {\rm and} H.~W. Hamber}, Nucl. Phys.} {\bf B248},  381
  (1984).

\bibitem{Wetzel-impr}
{\it {\sc W. Wetzel}, Nucl. Phys.} {\bf B255},  659  (1985).

\bibitem{Ma-Wetzel}
{\it {\sc J.~P. Ma {\rm and} W. Wetzel}, Phys. Lett.} {\bf 176B},  441  (1986).

\bibitem{Belanger-Lacaze-Morel-Attig-Petersson-Wolff}
{\it {\sc L. Belanger, R. Lacaze, A. Morel, N. Attig, B. Petersson, {\rm and}
  M. Wolff}, Nucl. Phys.} {\bf B340},  245  (1990).

\bibitem{Campostrini-Curci-Rossi}
{\it {\sc M. Campostrini, G. Curci, {\rm and} P. Rossi}, Nucl. Phys.} {\bf
  B314},  467  (1989).

\bibitem{Ichinose}
{\it {\sc I. Ichinose}, Ann. Phys. (N. Y.)} {\bf 152},  451  (1984).

\bibitem{Abdalla-Abdalla-Kawamoto}
{\it {\sc E. Abdalla, M.~C.~B. Abdalla, {\rm and} N. Kawamoto}, Phys. Rev.}
  {\bf D31},  3213  (1985).

\bibitem{Kallen-Toll}
{\it {\sc G. {K\"allen} {\rm and} J. Toll}, J. Math. Phys.} {\bf 6},  299
  (1965).

\bibitem{Petersson}
{\it {\sc B. Petersson}, J. Math. Phys.} {\bf 6},  1955  (1965).

\bibitem{Schonfeld}
{\it {\sc J.~F. Schonfeld}, Nucl. Phys.} {\bf B95},  148  (1975).

\bibitem{Abramowitz}
{\sc M. Abramowitz {\rm and} I.~A. Stegun}, {\em Handbook of Mathematical
  Functions} (Dover, New York, 1970).

\end{thebibliography}
\end{document}